\documentclass[a4paper,11pt]{article}
\usepackage{jheppub}
\usepackage[utf8x]{inputenc}
\usepackage[T1]{fontenc}
\usepackage{braket}
\usepackage{physics}
\usepackage{color}
\usepackage{amsfonts}
\usepackage{amssymb}
\usepackage{cancel}
\usepackage{amsmath}
\usepackage[normalem]{ulem}
\usepackage[font={footnotesize}]{caption}
\usepackage{graphicx}
\usepackage{epstopdf}
\usepackage{enumitem}
\usepackage{subcaption}
\usepackage{comment}
\usepackage{mathtools}
\usepackage{ragged2e}
\usepackage{hyperref}
\hypersetup{
	colorlinks=true,
	linkcolor=blue,
	filecolor=cyan, 
	urlcolor=blue,
	citecolor=red,
} 
\usepackage[titletoc,toc,title]{appendix}
\numberwithin{equation}{section}
\usepackage[nameinlink]{cleveref}
\usepackage{float}

\makeatletter
\newcommand{\vast}{\bBigg@{3}}
\newcommand{\Vast}{\bBigg@{5}}
\makeatother

\newcommand{\zbar}{\raisebox{0.2ex}{--}\kern-0.6em Z}

\def\d{\textrm{d}}

\hypersetup{
	colorlinks=true,
	linkcolor=blue,
	filecolor=red,      
	urlcolor=blue,
	citecolor=red
} 

\title{Reflected entropy and islands in JT gravity from partial dimensional reduction}

\author[a]{Debarshi Basu}

\affiliation[a]{Shing-Tung Yau Center and School of Physics, Southeast University, Nanjing 210096, China}

\emailAdd{debarshi.128@gmail.com}

\author[b,c]{Himanshu Chourasiya}

\affiliation[b]{Department of Physics, Indian Institute of Technology Kanpur, 208016, India}

\affiliation[c]{ MAEER's Maharashtra Institute of Technology, Mumbai, 401107, India}

\emailAdd{himanshu.hep.th@gmail.com}

\abstract{
We study how mixed-state correlations reorganize across island transitions in Jackiw--Teitelboim gravity coupled to a non-gravitating bath. Using the partial-dimensional-reduction realization of double holography, we compute reflected entropy for adjacent and disjoint radiation regions in the vacuum and in the thermofield-double state of an eternal black hole. Reflected entropy probes the internal partition of the entanglement wedge and therefore resolves structure that is invisible to the ordinary Page curve: its dominant saddle can change even when the entropy saddle of the union remains fixed, and it vanishes at leading order when the entanglement wedge disconnects. We show that the lower-dimensional reflected-island description agrees, to leading order in the dynamical brane fluctuation, with the higher-dimensional defect-extremal-surface picture. The finite-angle comparison further suggests a projection-covariant identification of island and brane endpoints at linear order. The resulting Page-like curves and Markov gaps provide a geometric probe of the reorganization of the purification and of non-Markovian, multipartite correlation structure during island transitions.
}

\begin{document} 
	\maketitle
	\flushbottom
	
	\section{Introduction}
The black-hole information problem asks how an apparently thermal Hawking flux can be compatible with unitary quantum evolution.  In its modern formulation the sharpest diagnostic is the fine-grained entropy of the Hawking radiation: a unitary system should follow the Page curve, whereas a calculation based on local quantum field theory on a fixed evaporating background gives the monotonically increasing Hawking result \cite{Page:1993wv,Page:1993df,Page:2013dx,Hawking:1976ra,Almheiri:2019hni,Penington:2019npb,Almheiri:2020cfm}.  Holography turns this tension into a question about competing extremal geometries.  The Ryu--Takayanagi and Hubeny--Rangamani--Takayanagi prescriptions, their replica derivations, and their quantum generalizations identify the entropy of a boundary region with an extremization problem in the bulk \cite{Ryu:2006bv,Ryu:2006ef,Hubeny:2007xt,Fursaev:2006ih,Casini:2011kv,Lewkowycz:2013nqa,Faulkner:2013ana,Engelhardt:2014gca,Dong:2016hjy}.  The time-dependent competition between exterior and wormhole-crossing surfaces is already visible in the HM geometry of the thermofield-double state \cite{Hartman:2013qma,Hartman:2013mia}.

Replica wormholes and quantum extremal surfaces provide the semiclassical mechanism that restores the Page curve \cite{Penington:2019kki,Almheiri:2019psf,Almheiri:2019yqk,Almheiri:2019qdq,Chen:2020uac,Chen:2020hmv,Hartman:2020swn,Gautason:2020tmk,Goto:2020wnk}.  The crucial new ingredient is the island: after the Page time, a region in the gravitating spacetime can become part of the entanglement wedge of the non-gravitating radiation.  Since the original derivations, the island mechanism has been tested in asymptotically flat and Schwarzschild geometries, charged and rotating black holes, dilaton and Kaluza--Klein models, higher-derivative and massive-gravity theories, and configurations with grey-body factors or multiple competing quantum extremal surfaces \cite{Anegawa:2020ezn,Hashimoto:2020cas,Krishnan:2020oun,Ling:2020laa,Ahn:2021chg,Yu:2021cgi,Yu:2021rfg,Sun:2021dfl,Lu:2021gmv,Alishahiha:2020qza,Cao:2021ujs,Geng:2020qvw,Grimaldi:2022suv,Hollowood:2021lsw,Hollowood:2021wkw,Arefeva:2021kfx,Karananas:2020fwx}.  Related analyses have explored gravitating and deformed baths, moving mirrors, thermalization, complete evaporation, and microscopic or information-theoretic interpretations of the Page curve \cite{Geng:2020fxl,Anderson:2020vwi,Chen:2020jvn,Caceres:2021fuw,Balasubramanian:2021xcm,Akal:2020twv,Agon:2020fqs,Arefeva:2022guf,Chou:2021boq,Okuyama:2021bqg,Chen:2019iro,Chen:2020wiq,Urbach:2021zil,Marolf:2020rpm,Marolf:2021ghr,Renner:2021qbe,Wang:2021fza,Krishnan:2021faa,Pedraza:2021ssc,Bousso:2021sji,He:2021mst,Krishnan:2020fer}.  Islands have also been studied in cosmological and de Sitter settings, where the interpretation of the gravitating region and even the existence of a Page transition are more subtle \cite{Hartman:2020khs,Balasubramanian:2020xqf,Sybesma:2020fxg,Azarnia:2021uch,Aguilar-Gutierrez:2021bns,Teresi:2021qff,Kames-King:2021etp,Bousso:2022gth}.  Other useful tests include higher-dimensional islands, Janus and warped-CFT settings, stretched-horizon descriptions and alternative interpretations of the bulk Page curve \cite{Almheiri:2019psy,Bak:2020enw,Matsuo:2020ypv,Caceres:2020jcn,Krishnan:2021ffb,Cadoni:2021ypx}.

A complementary line of work has made the island mechanism geometric through AdS/BCFT, braneworld and double-holographic constructions.  In these models the quantum extremal surface in the lower-dimensional theory is represented by an ordinary higher-dimensional extremal surface ending on, or interacting with, a brane.  This viewpoint has been developed in two and higher dimensions, for dynamical and Planck branes, in string-theory embeddings, and for higher-curvature gravity \cite{Suzuki:2022xwv,Anous:2022wqh,Chu:2021gdb,Hu:2022ymx,Chang:2023gkt,Geng:2024xpj,Uhlemann:2021nhu,Hernandez:2020nem,Ageev:2021ipd,Bhattacharya:2021jrn,Neuenfeld:2021bsb,Karch:2022rvr,Verheijden:2021yrb,Deng:2020ent,Deng:2022yll}.  The same geometric language has exposed limitations and consistency conditions of the island prescription, including questions of long-range gravity, homology, monogamy and coarse versus fine graining \cite{Geng:2021hlu,Karlsson:2020uga,Li:2021lfo,Ghosh:2021axl,Neuenfeld:2021bsb}.

These developments have made the fine-grained entropy of Hawking radiation calculable in a broad class of semiclassical models.  At the same time, they raise a more detailed question.  The ordinary entropy of a radiation region only measures the total amount of entanglement across one bipartition; it does not reveal how correlations are distributed among different parts of the radiation, how much of those correlations are intrinsically mixed-state, or how the internal organization of the entanglement wedge changes when an island appears.  A complete information-theoretic picture of the Page transition should therefore go beyond the entropy of a single region.

Mixed-state correlation measures provide precisely this additional resolution.  Entanglement of purification, logarithmic negativity, odd entropy, balanced partial entanglement and related quantities have all been used to probe structures invisible to the von Neumann entropy \cite{Takayanagi:2017knl,Caputa:2018xuf,Tamaoka:2018ned,Kudler-Flam:2018qjo,Jeong:2019xdr,Wen:2021qgx,Sahraei:2021wqn,BabaeiVelni:2019pkw,BabaeiVelni:2020wfl}.  Reflected entropy is particularly suitable for the present purpose because it combines three useful properties.  It has a precise field-theoretic definition in terms of the canonical purification, it is computable by a controlled replica construction, and in holographic states its leading contribution is twice the entanglement-wedge cross section \cite{Dutta:2019gen,Jeong:2019xdr,Bueno:2020fle,Bueno:2020vnx,Moosa:2020vcs,Bao:2019zqc,Chu:2019etd,Kudler-Flam:2020url,Kusuki:2019evw}.  Consequently, reflected entropy does not merely ask whether the entanglement wedge of $A\cup B$ is connected; it probes how that wedge is partitioned between $A$ and $B$.  In an island phase this is a genuinely stronger question, because the optimal cross section can terminate on an ordinary RT surface, on an island surface, or directly on an end-of-the-world brane.

This richer endpoint structure is the central motivation for the present work.  The island formula for reflected entropy contains an internal island cross section separating the portions assigned to $A$ and $B$, while the higher-dimensional description contains the corresponding defect extremal cross section \cite{Li:2020ceg,Chandrasekaran:2020qtn,Li:2021dmf,Ling:2021vxe,Akers:2022max,Basak:2023bnc,Basu:2023jtf,Basu:2025sqk,Dey:2025dlj}.  More generally, mixed-state and multipartite probes have already revealed information that is invisible to the ordinary Page curve: islands have been incorporated into entanglement negativity and replica-negativity constructions, multipartite purification and geometric secret-sharing models, partial-entanglement and complexity diagnostics, and random-tensor-network realizations \cite{KumarBasak:2020ams,KumarBasak:2021rrx,Dong:2021oad,Afrasiar:2022ebi,Shao:2022wrm,Basu:2022reu,Basu:2023wmv,Bhattacharya:2020ymw,Bhattacharya:2021dnd,Bhattacharya:2020uun,Balasubramanian:2020hfs,Saha:2021ohr,Akers:2021pvd}.  Related reflected-entropy studies in topological, anomalous, moving-mirror and irrelevant-deformation settings further illustrate how sensitively $S_R$ responds to the structure of the purification \cite{Berthiere:2020ihq,Basu:2022nds,BasakKumar:2022stg,Deng:2023pjs,Basu:2024bal}.  Agreement of the island and defect descriptions is therefore a substantially more stringent test of double holography than agreement of a single entanglement entropy.  It tests not only the generalized entropy of an island endpoint, but also the geometry of the purification inside the entanglement wedge and the way the island itself is partitioned between two radiation subsystems.

A particularly clean arena for such a test is the partial-dimensional-reduction realization of Jackiw--Teitelboim (JT) gravity.  In this construction an AdS$_3$/BCFT$_2$ geometry is truncated by an end-of-the-world brane and only the finite transverse wedge between the brane and a reference slice is reduced.  The background brane position sets the two-dimensional Newton constant, while a transverse fluctuation of the brane becomes the JT dilaton \cite{Takayanagi:2011zk,Fujita:2011fp,Cardy:2004hm,Sully:2020pza,Deng:2020ent,Deng:2022yll,Verheijden:2021yrb,Neuenfeld:2021bsb,Karch:2022rvr}.  The ordinary island/DES correspondence in this setup has already provided a geometric interpretation of the JT area term.  Reflected entropy asks whether the same dictionary continues to work for an observable whose extremal object can move inside the entanglement wedge and can change its endpoint topology as the state evolves.

There are several physical reasons to expect this extension to be informative.  First, the brane fluctuation gives a controlled gravitational correction to the cost of island-supported cross sections.  Its effect on reflected entropy therefore measures how a dynamical gravitational degree of freedom redistributes correlations inside the radiation, rather than merely shifting the entropy of the radiation as a whole.  Second, the transition of the union entropy and the transition of the optimal entanglement-wedge cross section need not coincide.  The resulting hierarchy of transition times can distinguish the onset of island dominance from the later reorganization of mixed-state correlations.  Third, the Markov gap, defined by the excess of reflected entropy over mutual information, is sensitive to the number and character of nontrivial cross-section endpoints \cite{Hayden:2021gno,Lu:2022cgq}.  Information-theoretically, a nonzero gap is an obstruction to exact Markov recovery; in a global pure-state partition into $A$, $B$ and the purifying remainder $C$, an order-$c$ holographic gap is therefore naturally sensitive to multipartite organization that cannot be represented solely by independent pairwise links.  We will use this as a diagnostic of tripartite organization rather than as a standalone tripartite entanglement monotone.  In the present setting the third party $C$ can include the complementary bath, black-hole interior and island degrees of freedom, so a kink in $\mathcal M$ can diagnose a reorganization of the purification when the purification moves from a wormhole-supported configuration to one controlled by ordinary RT surfaces, island surfaces or the brane \cite{Chandrasekaran:2020qtn,Hayden:2021gno,Lu:2022cgq}.  Finally, asymmetric left/right intervals require the full HM geometry rather than the special constant-position profile and thereby make explicit normalization and channel assignments that are hidden in symmetric examples.

These questions are closely related to, but distinct from, earlier analyses of reflected entropy in boundary and interface CFTs and in black-hole backgrounds.  Reflected-entropy Page curves and Markov gaps have been studied in rigid Karch--Randall and braneworld geometries, where the cross-section topology already exhibits a rich sequence of time-dependent phases \cite{Basu:2023jtf,Basak:2023bnc,Basu:2025sqk,Dey:2025dlj}.  In those constructions, the brane itself is usually treated as a fixed geometric ingredient.  The partial-reduction setup instead allows us to turn on the leading dynamical JT correction and to ask how the phase structure is deformed by the brane fluctuation.  This distinction is important: the aim is not simply to reproduce the rigid-brane phase catalogue, but to identify which parts of the mixed-state correlation structure are robust and which are sensitive to the JT mode.

In this paper we carry out this program for adjacent and disjoint radiation regions, both in the vacuum and in the thermofield-double state dual to an eternal BTZ black hole. On the two-dimensional side we evaluate the appropriate replica correlators and include the island contribution; on the three-dimensional side we extremize the corresponding defect cross sections. The comparison is performed phase by phase, with connected reflected-entropy formulas restricted to the domain in which the corresponding connected entanglement wedge is semiclassically dominant. We keep the leading correction in the brane fluctuation and use a perturbative-extremization identity to separate on-shell information from the coordinate motion of an extremal endpoint. At finite brane angle the minimal reduced and three-dimensional functionals agree in value and first variation at the rigid saddle but differ in their off-shell curvature by a term controlled by $e^{-2\rho_0/\ell}$. The apparent linear mismatch between their raw AdS$_2$-slicing endpoint coordinates has a more refined interpretation: once the endpoint on the bent brane is compared using the background projection onto the reduced coordinate, a compensating tangential (St\"uckelberg) shift restores the QES/DES location through $O(\Phi_r)$ at arbitrary finite angle. We regard this as a natural projection-covariant resolution at linear order, supported by the vacuum, disjoint and thermal geometries, rather than as a complete nonlinear theorem derived from the reduced action. Genuine finite-angle corrections to the off-shell functional begin quadratically around the rigid saddle and are suppressed in the large-angle limit. At finite temperature we also use the general embedding-space parametrization of HM geodesics, which is essential when the left and right endpoints are not spatially aligned.

The resulting picture is that the ordinary Page transition and the reflected-entropy transition encode complementary pieces of information.  The former determines when a new homology class becomes favored for the entropy of the union, whereas the latter records how the purification reorganizes inside that homology region.  In several of the examples below the reflected entropy changes slope, changes its limiting value, or switches endpoint topology without being reducible to a simple copy of the union-entropy transition.  The accompanying Markov-gap curves make this distinction especially transparent: they track the part of the correlation structure that is not already captured by mutual information and hence offer a geometric diagnostic of the internal connectivity of the entanglement wedge.  These features are the main physical output of the analysis, while the detailed agreement between the island and defect descriptions extends the partial-dimensional-reduction dictionary beyond ordinary entanglement entropy.

The technical ingredients are standard but their combination is somewhat intricate.  We use vacuum and thermal twist-field correlators, large-$c$ factorization of the relevant conformal blocks, and explicit geodesic extremization in Poincar\'e AdS$_3$ and BTZ \cite{Calabrese:2004eu,Calabrese:2009ez,Calabrese:2009qy,Hartman:2013mia,Fitzpatrick:2014vua,Banerjee:2016qca}.  Section~\ref{sec:partial-reduction-review} reviews partial dimensional reduction and the island/DES dictionary, while section~\ref{sec:RE-review} summarizes reflected entropy, the replica construction and the Markov gap.  Section~\ref{sec:vacuum-reflected} treats adjacent and disjoint intervals in the vacuum.  Section~\ref{sec:BTZ-review} introduces the eternal BTZ geometry and the competing island and HM entanglement saddles.  Section~\ref{sec:thermal-reflected} develops the finite-temperature reflected-entropy phase structure and the associated Page-like curves.  We conclude with the physical implications and limitations of the construction, and collect the embedding-space geodesic formulas in the appendix.

\section{Review and setup}

The calculations in this paper repeatedly move between three descriptions of the same physics: a three-dimensional AdS/BCFT geometry with an end-of-the-world brane, the partially reduced two-dimensional JT theory with islands, and the replica construction of reflected entropy.  It is therefore useful to organize the review as a single chain of dictionaries rather than as a list of separate formulas.  We first explain how the brane geometry reduces to JT gravity and how the ordinary island/DES correspondence emerges.  A one-interval warm-up then shows how the dictionary works at the level of an extremization problem.  Finally, we extend the same logic from ordinary entropy to reflected entropy, where the new ingredient is an internal cross section of the entanglement wedge.

\subsection{Partial dimensional reduction and JT gravity}\label{sec:partial-reduction-review}
The starting point is AdS$_3$/BCFT$_2$: an asymptotically AdS$_3$ spacetime $N$ is bounded by the ordinary conformal boundary $M$ and by an end-of-the-world (EOW) brane $Q$.  In the simplest constant-tension model the action is \cite{Takayanagi:2011zk,Fujita:2011fp,Sully:2020pza}
\begin{align}
I=\frac{1}{16\pi G_N}\int_N\!\sqrt{-g}\,(R-2\Lambda)
+\frac{1}{8\pi G_N}\int_Q\!\sqrt{-h}\,(K-T)+I_{\rm matter}[Q] .
\end{align}
The variational principle gives the Neumann condition $K_{ij}-(K-T)h_{ij}=0$, which fixes the embedding of the brane.  On the Poincar\'e patch it is convenient to describe the same geometry by AdS$_2$ slices,
\begin{align}
\d s^2&=\ell^2\frac{-\d t^2+\d x^2+\d z^2}{z^2}
=\d \rho^2+\ell^2\cosh^2\!\left(\frac{\rho}{\ell}\right)\frac{-\d t^2+\d y^2}{y^2},\notag\\
z&=y\operatorname{sech}\!\left(\frac{\rho}{\ell}\right),\qquad
x=-y\tanh\!\left(\frac{\rho}{\ell}\right).\label{AdS2-slicing-map-Poincare}
\end{align}
The undeformed EOW brane is the constant-$\rho$ surface $\rho=\rho_0$.  Its position can equivalently be encoded in the Poincar\'e brane angle $\theta_0$,
\begin{align}
\cos\theta_0=\operatorname{sech}\!\left(\frac{\rho_0}{\ell}\right)\,.\label{brane-angle-dictionary}
\end{align}
In the present constant-tension convention one also has $T\ell=\tanh(\rho_0/\ell)=\sin\theta_0$, so the large-angle limit $\rho_0/\ell\to\infty$ is equivalently the near-critical-tension limit $T\ell\to1^-$. 

Partial dimensional reduction now amounts to integrating only over the finite transverse wedge between a reference AdS$_2$ slice and the EOW brane, rather than compactifying the full three-dimensional geometry.  A small local displacement of the upper end of this wedge,
$\rho_0\rightarrow\rho_0+\Delta\rho(x)$, becomes a genuine two-dimensional degree of freedom.  To first order in this displacement the reduced action is \cite{Deng:2020ent,Deng:2022yll,Verheijden:2021yrb}
\begin{align}
I_\textrm{JT}={}&\frac{1}{16\pi G_N^{(2)}}\int\d^2x\sqrt{-g^{(2)}}\,R^{(2)}
+\frac{1}{16\pi G_N^{(2)}}\int\d^2x\sqrt{-g^{(2)}}\,
\frac{\Delta\rho(x)}{\rho_0}\notag\\
&\times\left(R^{(2)}+\frac{2}{\ell^2\cosh^2(\rho_0/\ell)}\right)
+I_\textrm{CFT}\, .\label{2d-action}
\end{align}
The first term is topological.  The second is precisely the JT dilaton coupling on an AdS$_2$ background of radius $\ell\cosh(\rho_0/\ell)$.  The reduction therefore identifies
\begin{align}
G_N^{(2)}=\frac{G_N}{\rho_0},
\qquad
\Phi(x)=\frac{\Delta\rho(x)}{\rho_0}.\label{dilaton-identifier}
\end{align}
This is the central geometric statement of the construction: the background width of the wedge fixes the effective Newton constant, while the transverse motion of the brane is the JT dilaton.  The construction has close relatives in braneworld and double-holographic realizations of islands, where an apparently quantum extremal surface in the reduced theory becomes an ordinary higher-dimensional extremal surface \cite{Chen:2020uac,Chen:2020hmv,Chu:2021gdb,Neuenfeld:2021bsb,Karch:2022rvr,Chang:2023gkt,Geng:2024xpj,Uhlemann:2021nhu}.
\begin{figure}[t]
\centering
\includegraphics[width=0.68\linewidth]{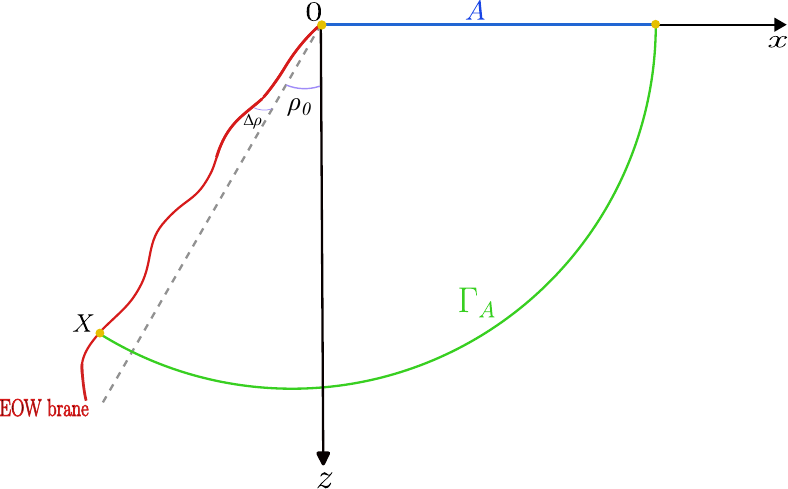}
\caption{Partial-dimensional-reduction dictionary.  The AdS$_3$/BCFT$_2$ description contains an asymptotic bath boundary $M$ and an EOW brane $Q$ at $\rho=\rho_0+\Delta\rho$.  Reducing only the finite transverse wedge between a reference AdS$_2$ slice and $Q$ gives JT gravity coupled to a non-gravitating bath.  The background brane position determines $G_N^{(2)}$, while the local displacement $\Delta\rho$ becomes the JT dilaton.  A three-dimensional surface ending on $Q$ is therefore the higher-dimensional representative of a QES ending at the corresponding island point.}
\label{fig:partial-reduction-review}
\end{figure}
The geometric content of these identifications is summarized in \cref{fig:partial-reduction-review}.  The figure should be read from left to right: a motion of the brane endpoint changes the length of a bulk geodesic in three dimensions, while the same motion changes the generalized-area term through the dilaton in two dimensions.  The equality of these two responses is the local statement underlying the partial-reduction dictionary used throughout the paper.

The entropy dictionary follows immediately.  For a bath region $A$, the two-dimensional description asks us to enlarge $A$ by a candidate island $I_A$ and extremize the generalized entropy,
\begin{align}
S_A=\min_{I_A}\,\underset{I_A}{\operatorname{ext}}\left[\frac{\operatorname{Area}(\partial I_A)}{4G_N}+S^{\mathrm{eff}}(A\cup I_A)\right].\label{Island-prescription}
\end{align}
In the three-dimensional picture the same competition is expressed by the defect-extremal-surface (DES) prescription \cite{Deng:2020ent,Deng:2022yll,Li:2021dmf,Ling:2021vxe},
\begin{align}
S_{\rm DES}(A)=\min_X\,\operatorname{ext}_X\left[\frac{{\rm Area}(\Gamma_A)}{4G_N}+S_{\rm defect}(D_X)\right],
\label{DES-review-formula}
\end{align}
where $X=\Gamma_A\cap Q$.  The two formulas are not independent prescriptions: the DES endpoint $X$ is the geometric representative of the QES endpoint $\partial I_A$, and the defect matter term is the higher-dimensional image of the effective matter entropy.  This is why partial reduction is especially useful for the present problem: it turns the island formula into a directly checkable geometric statement.

\subsubsection*{Warm-up: single-interval entanglement entropy}
We now review the single interval entanglement entropy computation in this model \cite{Deng:2022yll}. Consider $A=[0,L]$ and a candidate island $I_A=[-a,0]$.  The vacuum JT solution has dilaton profile \cite{Almheiri:2019yqk}
\begin{align}
\Phi(y)=\frac{2\Phi_r}{y^+-y^-}=\frac{\Phi_r}{y},\qquad y^\pm=t\pm y,
\end{align}
so that the static solution has $\Phi=\Phi_r/y$ and $\Phi|_{\rm bdy}=\Phi_r/\epsilon$. The induced brane metric is AdS$_2$,
\begin{align}
\d s^2_\textrm{brane}=\ell^2\cosh^2\!\left(\frac{\rho_0}{\ell}\right)\frac{-\d t^2+\d y^2}{y^2}
=-\Omega(y)^{-2}\d y^+\d y^- ,
\end{align}
and the Weyl factor that brings the matter correlator to the flat frame is
\begin{align}
\Omega(y)=\left|\frac{y}{\ell\cosh(\rho_0/\ell)}\right|.\label{Omega-y}
\end{align}
Using the two-point twist correlator for $A\cup I_A$, the generalized entropy becomes
\begin{align}
S_\textrm{gen}(a)&=\frac{1}{4G_N^{(2)}}\left(1+\frac{\Delta \rho}{\rho_0}\right)
+\lim_{n\to 1}\frac{1}{1-n}\log\!\left[\Omega(-a)^{\Delta_n}\left<\sigma_n(L)\bar\sigma_n(-a)\right>\right]\notag\\
&=\frac{c}{6}\frac{\rho_0}{\ell}\left(1+\frac{\Phi_r}{a}\right)
+\frac{c}{6}\log\!\left[\frac{(L+a)^2}{\epsilon\,a(\epsilon_y\cos\theta_0/\ell)}\right].\label{Sgen}
\end{align}
The two terms have a transparent interpretation: moving the candidate QES changes the dilaton cost and simultaneously changes the matter interval joining the bath endpoint to the island endpoint.  Defining
\begin{align}
\mu\equiv\frac{\rho_0\Phi_r}{\ell L},
\end{align}
the stationarity equation is
\begin{align}
a^2-a(1+\mu)L-\mu L^2=0,
\end{align}
and the root continuously connected to the rigid-brane saddle is
\begin{align}
a=\frac{L}{2}\left(1+\mu+\sqrt{1+6\mu+\mu^2}\right).\label{extremal-soln}
\end{align}
Hence $a=L+2L\mu+\mathcal O(\mu^2)$, and the on-shell island entropy is
\begin{align}\label{EE_island}
S_\textrm{QES}(A)=\frac{c}{6}\log\!\left(\frac{2L}{\epsilon}\right)
+\frac{c}{6}\log\!\left(\frac{2\ell}{\epsilon_y\cos\theta_0}\right)
+\frac{c}{6}\log(\sec\theta_0+\tan\theta_0)
+\frac{c\mu}{6}+\mathcal O(\mu^2).
\end{align}
The first term is the bath contribution, the next two reproduce the rigid-brane boundary/defect entropy, and the final term is the leading dynamical JT correction.

On the doubly-holographic side, the same candidate endpoint is represented by a geodesic $\Gamma_A$ from $(L,\epsilon)$ to $(-a\sin\theta,a\cos\theta)$ on the fluctuating brane.  Its regulated length is
\begin{align}
\frac{\operatorname{Area}(\Gamma_A)}{4G_N}=\frac{\ell}{4G_N}\operatorname{arccosh}\!\left[\frac{(L+a\sin\theta)^2+a^2\cos^2\theta}{2a\epsilon\cos\theta}\right].
\end{align}
The brane displacement changes the opening angle according to
\begin{align}
\theta=\arccos\!\left[\operatorname{sech}\!\left(\frac{\rho_0+\Delta\rho}{\ell}\right)\right].
\end{align}
Expanding the geodesic length at fixed $a$ isolates the normal response of the DES,
\begin{align}
\frac{\operatorname{Area}(\Gamma_A)}{4G_N}=\frac{\ell}{4G_N}\left\{\log\!\left[\frac{a^2+L^2+2aL\sin\theta_0}{a\epsilon\cos\theta_0}\right]
+\frac{\Delta\rho}{\ell}\frac{(a^2+L^2)\sin\theta_0+2aL}{a^2+L^2+2aL\sin\theta_0}\right\}+\mathcal O(\Delta\rho^2).
\end{align}
At this order the defect contribution is independent of $a$,
\begin{align}
S_\textrm{defect}=\frac{c}{6}\log\frac{2\ell}{\epsilon_y\cos\theta_0}.
\end{align}
The normal-response factor multiplying $\Delta\rho$ in the preceding equation equals one at the rigid saddle $a=L$ and has vanishing first derivative there.  Thus, for the purpose of determining the endpoint shift and the on-shell entropy through $O(\Phi_r)$, it may be replaced by unity in a neighborhood of the rigid saddle.  A convenient near-saddle representative is
\begin{align}
S_\textrm{DES}^{\rm loc}(a)=\frac{\ell}{4G_N}\log\!\left[\frac{a^2+L^2+2aL\sin\theta_0}{a\epsilon\cos\theta_0}\right]
+\frac{\rho_0}{4G_N}\frac{\Phi_r}{a}
+\frac{c}{6}\log\frac{2\ell}{\epsilon_y\cos\theta_0},\label{S-des}
\end{align}
where $S_\textrm{DES}^{\rm loc}$ is not intended as the complete off-shell functional away from $a=L$.  Expanding its stationarity condition about the rigid saddle gives
\begin{align}
\frac{a(a^2-L^2)}{a^2+2aL\sin\theta_0+L^2}-\mu L=0+\mathcal O(\mu^2),
\end{align}
and hence $a=L+L(1+\sin\theta_0)\mu+\mathcal O(\mu^2)$.  Although this coordinate shift differs from the QES shift, the extremized entropy is
\begin{align}
S_\textrm{DES}(A)=\frac{c}{6}\left[\log\!\left(\frac{2L}{\epsilon}\right)
+\log(\sec\theta_0+\tan\theta_0)
+\log\frac{2\ell}{\epsilon_y\cos\theta_0}+\mu\right]+\mathcal O(\mu^2),
\end{align}
which agrees with \cref{EE_island}.  The equality of the on-shell entropies, however, should not be confused with equality of the two off-shell endpoint functionals.  The distinction is important because it separates the role of a large brane angle in off-shell matching from the coordinate identification of the perturbed endpoint.

\subsubsection*{Finite-angle projection and the off-shell functional}\label{sec:large-angle-matching}
The finite-angle comparison is most transparent if one distinguishes the off-shell entropy functional from the coordinate used to label a point on the fluctuating brane. Set
\begin{align}
s\equiv\sin\theta_0=\tanh\frac{\rho_0}{\ell},\qquad
q_\theta\equiv\frac{1-s}{1+s}=e^{-2\rho_0/\ell}.
\label{qtheta-def}
\end{align}
At $\Phi_r=0$, the endpoint-dependent parts of the DES and QES functionals differ exactly by
\begin{align}
\Delta S_0(a)&\equiv S_{\rm DES}^{(0)}(a)-S_{\rm QES}^{(0)}(a)\notag\\
&=\frac{c}{6}\log\!\left[\frac{2\big(a^2+L^2+2aLs\big)}{(1+s)(a+L)^2}\right]
=\frac{c}{6}\log\!\left[1+q_\theta\left(\frac{a-L}{a+L}\right)^2\right].
\label{finite-angle-offshell}
\end{align}
Thus $\Delta S_0(L)=\Delta S_0'(L)=0$: the descriptions have the same rigid saddle, value and first variation. Their off-shell curvatures nevertheless differ at finite angle. The term proportional to $q_\theta$ retains transverse information about a finite-angle bulk chord that is absent from the minimal intrinsic JT matter functional.

Expanding the QES condition and the full first-order DES normal-response condition about the common rigid saddle gives the raw endpoint labels
\begin{align}
a_{\rm QES}^{(y)}=L+2\alpha+\mathcal O(\alpha^2),\qquad
a_{\rm DES}^{(y)}=L+(1+s)\alpha+\mathcal O(\alpha^2),
\qquad \alpha\equiv\frac{\rho_0\Phi_r}{\ell}=L\mu.
\label{vacuum-endpoint-mismatch}
\end{align}
Their difference, $(1-s)\alpha$, is a statement about raw slicing labels. After the brane bends, that label need not coincide with the coordinate inherited by the fixed background slice used in the reduced description.

A natural linearized identification follows from the background projection. From the Poincar\'e/AdS$_2$ slicing map in \cref{AdS2-slicing-map-Poincare}, define
\begin{align}
Y\equiv -x+e^{-\rho_0/\ell}z.
\label{vacuum-projection-Y}
\end{align}
On the rigid brane $\rho=\rho_0$, $Y=y$. For a bent brane $\rho=\rho_0+\delta\rho(y)$,
\begin{align}
Y=y+(1-s)\frac{y\,\delta\rho(y)}{\ell}+\mathcal O(\delta\rho^2).
\label{vacuum-projection-shift}
\end{align}
With the vacuum profile $\delta\rho(y)=\rho_0\Phi_r/y$, one obtains $Y=y+(1-s)\alpha+\mathcal O(\Phi_r^2)$ and hence
\begin{align}
a_{\rm DES}^{(Y)}=a_{\rm DES}^{(y)}+(1-s)\alpha
=L+2\alpha+\mathcal O(\Phi_r^2)
=a_{\rm QES}+\mathcal O(\Phi_r^2).
\label{projected-endpoint-match}
\end{align}
This is naturally interpreted as a tangential/St\"uckelberg compensation associated with describing a normally displaced brane in a fixed reduced coordinate system. The same universal shift applies to every vacuum brane endpoint, including the disjoint configurations below. We use \cref{projected-endpoint-match} as a linearized projection-covariant prescription; a derivation of a unique nonlinear completion directly from the reduced action remains an open problem.

The projection does not remove the genuine finite-angle difference between the two minimal off-shell functionals. Expanding \cref{finite-angle-offshell} about $a=L+\delta a$ gives
\begin{align}
\Delta S_0(a)=\frac{c}{24}q_\theta\frac{(\delta a)^2}{L^2}+\mathcal O(\delta a^3).
\label{finite-angle-quadratic}
\end{align}
Since $\delta a=\mathcal O(\Phi_r)$, this changes the extremized entropy only at $\mathcal O(q_\theta\Phi_r^2)$, the same order as terms omitted by the linear JT reduction. A large brane angle therefore suppresses genuine higher-order finite-angle corrections and improves off-shell agreement. It is not demonstrably necessary for the $O(\Phi_r)$ endpoint location once the comparison is made in the projection-covariant coordinate. This distinction between raw slicing coordinates, projected endpoint data and off-shell functional matching will also be useful in the thermal geometry.

\subsubsection*{A perturbative extremization identity}\label{sec:extremization-lemma}
The warm-up suggests a general simplification used throughout the rest of the paper.  Suppose
\begin{align}
\mathcal{S}(a,\lambda)=\mathcal{S}_0(a)+\lambda\,\mathcal{S}_1(a)+\mathcal{O}(\lambda^2),
\qquad
a_\star=a_0+\lambda a_1+\mathcal{O}(\lambda^2),
\end{align}
with $\mathcal{S}'_0(a_0)=0$.  Expanding the extremality equation gives
\begin{align}
a_1=-\frac{\mathcal{S}_1'(a_0)}{\mathcal{S}_0''(a_0)},
\end{align}
but substituting the shifted saddle back into the functional removes the explicit $a_1$ contribution at first order,
\begin{align}
\mathcal{S}(a_\star,\lambda)=\mathcal{S}_0(a_0)+\lambda\,\mathcal{S}_1(a_0)+\mathcal{O}(\lambda^2).\label{first-order-extremization}
\end{align}
Thus the first-order motion of an extremal endpoint affects the on-shell observable only at second order.  Together with the finite-angle identity \cref{finite-angle-offshell}, this separates two logically distinct sources of approximation: the small-fluctuation expansion controls the order in $\Phi_r$, whereas $q_\theta=e^{-2\rho_0/\ell}$ controls how faithfully the reduced theory reproduces the higher-dimensional functional away from its rigid saddle.  The identity also allows us to evaluate explicit brane-fluctuation corrections on the rigid cross section whenever we work consistently to first order \cite{Deng:2022yll}.

\subsection{Reflected entropy, EWCS and islands}\label{sec:RE-review}
The preceding discussion concerns a single bipartition.  Reflected entropy adds a second layer of structure: for a mixed state $\rho_{AB}$ it asks how the purification itself is partitioned between $A$ and $B$.  The canonical purification regards $\sqrt{\rho_{AB}}$ as a state in $\mathcal H_A\otimes\mathcal H_B\otimes\mathcal H_{A^*}\otimes\mathcal H_{B^*}$, and the reflected entropy is \cite{Dutta:2019gen}
\begin{align}
S_R(A:B)=S(AA^*)_{|\sqrt{\rho_{AB}}\rangle}=S(BB^*)_{|\sqrt{\rho_{AB}}\rangle}.\label{RE-definition}
\end{align}
In two-dimensional CFT this quantity is computed by a two-step replica construction. An even index $m$ prepares the canonical purification, while an index $n$ computes the R\'enyi entropy across $AA^*$. The ordered continuation $m\to1$ followed by $n\to1$ produces higher-point twist correlators. In the semiclassical large-$c$ limit, candidate dominant OPE/BOE conformal-block channels are associated with different bulk cross-section topologies \cite{Dutta:2019gen,Jeong:2019xdr,Fitzpatrick:2014vua,Banerjee:2016qca,Kusuki:2019evw}. Throughout the paper, every factorization of a higher-point replica correlator is understood as its leading asymptotic form in the stated semiclassically dominant channel, not as an exact operator identity. A channel formula is used as the physical reflected entropy only where that channel and its associated outer entanglement wedge dominate.

For a connected holographic entanglement wedge, the leading geometric result is
\begin{align}
S_R(A:B)=2E_W(A:B)=\frac{\operatorname{Area}(\Sigma_{AB})}{2G_N},\label{RE-EWCS}
\end{align}
where $\Sigma_{AB}$ is the minimal entanglement-wedge cross section (EWCS) separating the portions homologous to $A$ and $B$ \cite{Dutta:2019gen,Jeong:2019xdr,Tamaoka:2018ned,Caputa:2018xuf}.  Islands enlarge the homology region, so the separator may now end on an island surface or may divide a reflected island itself.  The generalized reflected-entropy prescription is accordingly \cite{Li:2020ceg,Chandrasekaran:2020qtn,Akers:2022max}
\begin{align}
S_R(A:B)=\min\,\underset{\partial I_A\cap\partial I_B}{\operatorname{ext}}
\left[\frac{\operatorname{Area}(\partial I_A\cap\partial I_B)}{2G_N}
+S_R^{\mathrm{eff}}(A\cup I_A:B\cup I_B)\right].\label{SR-island-intro}
\end{align}
The factor of two relative to the ordinary QES area coefficient reflects the canonical-purification replica geometry.  The gravitational replica derivation and subsequent applications show that these reflected islands probe multipartite entanglement structures that are invisible to the ordinary entropies \cite{Chandrasekaran:2020qtn,Li:2020ceg,KumarBasak:2020ams,Dong:2021oad,Bhattacharya:2020ymw}.

Partial reduction turns this generalized island formula into a second geometric dictionary.  From the defect viewpoint one extremizes a cross section $\Gamma$ together with the effective reflected entropy,
\begin{equation}\label{SR-bdy}
S_R^{\rm bdy}(A:B)=\min\,\operatorname{Ext}_{\Gamma}\left\{\frac{\operatorname{Area}[\Gamma]}{2G_N}+S_R^{\rm eff}\big(A\cup I_{S_R}(A):B\cup I_{S_R}(B)\big)\right\},
\end{equation}
while in the three-dimensional bulk one may equivalently write
\begin{equation}\label{SR-bulk}
S_R^{\rm bulk}(A:B)=\min\,\operatorname{Ext}_{\Sigma}\left\{S_R^{\rm eff}(A:B)+\frac{\operatorname{Area}[\Sigma_{AB}]}{2G_N}\right\}.
\end{equation}
The first organization exposes the island assignment and the CFT replica channel; the second makes the geometry of the EWCS manifest.  Their equivalence is the mixed-state extension of the QES/DES correspondence reviewed above \cite{Li:2021dmf,Ling:2021vxe,Basak:2023bnc,Basu:2023jtf,Dey:2025dlj}.

\begin{figure}[t]
\centering
\includegraphics[width=0.68\linewidth]{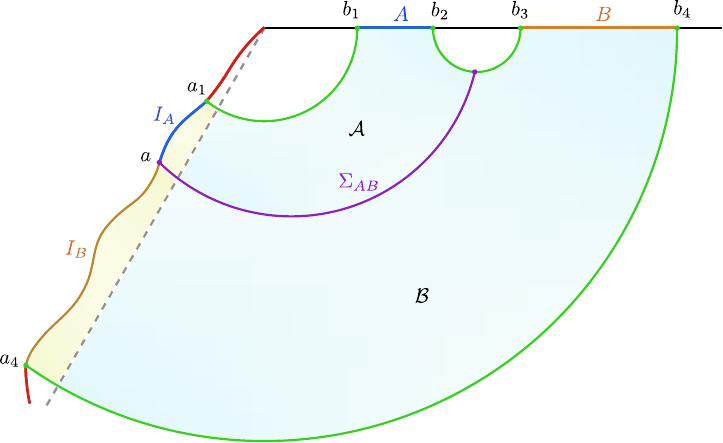}
\caption{Reflected-entropy dictionary in the presence of islands.  In the two-dimensional description, the canonical purification partitions the reflected island into pieces associated with $A$ and $B$; their interface contributes a generalized-area term to $S_R$.  In double holography, the same interface is represented by an EWCS or DES in AdS$_3$.  Because the cross section can terminate on an ordinary RT surface, an island surface or the EOW brane, reflected entropy resolves changes in the internal purification that need not coincide with a transition of the union entropy.}
\label{fig:reflected-entropy-review}
\end{figure}
Figure~\ref{fig:reflected-entropy-review} summarizes this second dictionary.  The important difference from \cref{fig:partial-reduction-review} is that the new extremal object is not the outer boundary of the entanglement wedge but an internal separator of that wedge.  Consequently, the outer entropy saddle can remain unchanged while the reflected-entropy saddle changes topology.  This is precisely the phenomenon responsible for the additional Page-like transitions found below.

The Markov gap provides a compact way to diagnose this internal reorganization.  We define
\begin{align}
\mathcal M(A:B)&=S_R(A:B)-I(A:B),\notag\\
I(A:B)&=S(A)+S(B)-S(A\cup B).\label{markov-gap-def}
\end{align}
The inequality $S_R\ge I$ guarantees $\mathcal M\ge0$.  More importantly, $\mathcal M$ controls a Markov-recovery problem: a nonzero value means that the correlations between $A$ and $B$ cannot be reconstructed from purely local conditional data as they could in an exact quantum Markov chain \cite{Hayden:2021gno}.  In time-symmetric AdS$_3$ geometries it is bounded below by a universal contribution for each endpoint of the EWCS, and the island analysis of \cite{Lu:2022cgq} shows that the same endpoint-counting intuition survives when the cross section ends on island surfaces.  This gives the gap a useful multipartite interpretation.  For a global pure state divided into $A$, $B$ and the purifying remainder $C$, an order-$c$ Markov gap diagnoses an obstruction to organizing the correlations solely as independent pairwise EPR links.  We stress that $\mathcal M$ is not, in general, a tripartite entanglement monotone; here we use it as a geometric diagnostic of non-Markovian and multipartite organization involving the radiation regions, their complement and the gravitating/island degrees of freedom \cite{Hayden:2021gno,Lu:2022cgq,Chandrasekaran:2020qtn}.  This interpretation will be particularly useful for the Page-curve figures in section~\ref{sec:thermal-reflected}.

With these two dictionaries in place, the logic of the rest of the paper is simple.  In the vacuum we classify the possible endpoints of the EWCS and compare each CFT replica channel with its bulk common perpendicular.  At finite temperature the same endpoint topologies compete with time-growing wormhole-crossing surfaces.  The ordinary entropy selects the outer entanglement wedge, reflected entropy selects its optimal internal partition, and the Markov gap diagnoses non-Markovian organization of the correlation structure that is not captured by mutual information alone.

\section{Reflected entropy in the vacuum state}\label{sec:vacuum-reflected}
In this section, we investigate the reflected entropy for various bipartite mixed-state configurations defined on a constant-time slice of the AdS$_3$/BCFT$_2$ setup reviewed in \cref{sec:partial-reduction-review}.  The CFT calculation uses the vacuum twist correlators and their large-$c$ OPE channels, whereas the bulk calculation uses geodesic EWCSs with endpoints allowed on RT surfaces, island surfaces and the EOW brane \cite{Dutta:2019gen,Jeong:2019xdr,Li:2021dmf,Basu:2023jtf,Basak:2023bnc}.  We compute every phase using both the island and defect-extremal-surface prescriptions and compare the two descriptions at the order to which the brane fluctuation is retained.

\subsection{Adjacent subsystems}
Following the standard large-$c$ replica-channel analysis of adjacent intervals \cite{Dutta:2019gen,Jeong:2019xdr,Basu:2023jtf}, we consider $A=(b_1,b_2)$ and $B=(b_2,b_3)$ on a constant-time slice, with $0<b_1<b_2<b_3$.  The first question is whether the RT surface of $A\cup B$ remains a same-boundary dome or instead ends on the EOW brane and supports an island.  Once that outer entanglement saddle is fixed, reflected entropy performs a second minimization inside the resulting wedge.  This nested structure is physically important: the union can already be in an island phase while the optimal cross section still chooses among several inequivalent ways of partitioning that island.  We therefore organize the analysis first by the entanglement saddle and then by the endpoint topology of the cross section.
    
\subsubsection{Connected dome saddle for adjacent intervals}
\begin{figure}[ht]
    \centering
    \includegraphics[width=0.8\linewidth]{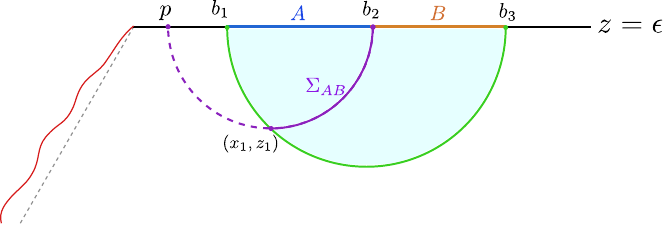}
    \caption{Vacuum adjacent intervals in the connected no-island entanglement saddle.  The solid semicircle is the RT surface homologous to $A\cup B$.  The reflected-entropy cross section starts at the common endpoint $b_2$ and ends orthogonally on the RT surface; because the brane does not participate, this saddle is identical to the standard CFT vacuum result.}
    \label{fig:vac-adj-connected}
\end{figure}

When the two adjacent intervals are sufficiently far from the EOW brane, the brane does not enter the homology region.  The entropy of $A\cup B$ is then carried by the single same-boundary RT semicircle shown in \cref{fig:vac-adj-connected}.  Evaluating its regulated length gives
\begin{align}
	S_1= \frac{{\ell}}{2 G_N} \log\frac{b_3-b_1}{\epsilon}.
\end{align}
Once this no-island entanglement saddle is fixed there is only one admissible connected cross section, shown in \cref{fig:vac-adj-connected}.  Because no brane segment belongs to the reflected island, the boundary calculation reduces to the ordinary effective reflected entropy.  The adjacent replica geometry is therefore governed by the standard three-point twist correlator,
\begin{align}\label{Adj-connected-bdy}
	S_R^{\text{bdy}}(A:B)&= \lim_{{m,n} \to 1}\frac{1}{1-n}\log \frac{      \langle \sigma_{g^{}_A}(b_1)\sigma_{g^{}_B g_A^{-1}}(b_2)\sigma_{g^{-1}_A}(b_3)\rangle}{ \langle\sigma_{g^{}_m}(b_1)\sigma_{g^{-1}_m}(b_3)\rangle^n}, \notag\\      
    &=\frac{c}{3}  \log \frac{2 (b_2-b_1) (b_3-b_2)}{\epsilon (b_3-b_1)}.
\end{align}
In the bulk description, the EWCS for this saddle is the geodesic $\Sigma_{AB}$ shown in \cref{fig:vac-adj-connected}.  It runs from the boundary point $b_2$ to the RT semicircle anchored on $(b_1,b_3)$ and meets that semicircle orthogonally.  A direct minimization gives the orthogonal common perpendicular without introducing an auxiliary boundary extension.  Let
\begin{align}
 c_0=\frac{b_1+b_3}{2},\qquad R=\frac{b_3-b_1}{2},\qquad d=b_2-c_0.
\end{align}
The RT surface is $(x-c_0)^2+z^2=R^2$.  Parametrizing it as $x=c_0+R\cos\varphi$, $z=R\sin\varphi$, the regulated distance from $(b_2,\epsilon)$ obtained from \cref{Poincare-length} is extremized at
\begin{align}
 \cos\varphi_\star=\frac{2Rd}{R^2+d^2},\qquad
 \sin\varphi_\star=\frac{R^2-d^2}{R^2+d^2}.
\end{align}
Using the extremizing angle in the semicircle parametrization fixes the bulk intersection point to
\begin{align}
 x_\star=c_0+\frac{2R^2d}{R^2+d^2},\qquad
 z_\star=\frac{R(R^2-d^2)}{R^2+d^2}.
\end{align}
This also makes explicit that the radius of the outer RT semicircle is $(b_3-b_1)/2$.  Substitution into \cref{Poincare-length} gives
\begin{align}
 S_R^{\mathrm{bulk}}(A:B)
 =\frac{\mathrm{Area}(\Sigma_{AB})}{2G_N}
 =\frac{\ell}{2G_N}\log\frac{2(b_2-b_1)(b_3-b_2)}{\epsilon(b_3-b_1)}.\label{Adj-connected-bulk}
\end{align}
The Brown--Henneaux relation $c=3\ell/(2G_N)$ then gives exact agreement with \cref{Adj-connected-bdy}.  Physically, this saddle is completely insensitive to the gravitational brane sector: the entire purification is encoded in the ordinary connected vacuum wedge.  The answer depends only on the three boundary separations, and the factor $(b_2-b_1)(b_3-b_2)/(b_3-b_1)$ measures how deeply the common endpoint $b_2$ sits inside the outer RT semicircle.  As either adjacent interval shrinks, the cross section collapses; moving the pair toward the EOW brane without changing these separations does not affect this no-island result until the competing island saddle takes over.

\subsubsection{Island saddle for adjacent intervals}
\begin{figure}[ht]
    \centering
    \includegraphics[width=0.65\linewidth]{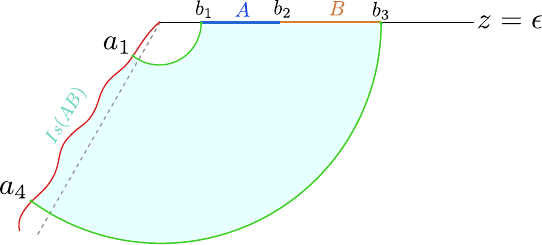}
    \caption{Vacuum adjacent intervals in the island-supported entanglement saddle.  The brane-ending surfaces end on the EOW brane and define the island of $A\cup B$.  Depending on the relative sizes of $A$ and $B$, the reflected-entropy cross section can terminate on either island surface or reach the EOW brane and split the island; the three possibilities are resolved in \cref{fig:vac-adj-A-island,fig:vac-adj-brane,fig:vac-adj-B-island}.}
    \label{fig:vac-adj-island-overview}
\end{figure}
When the outer endpoints approach the EOW brane, the competing homology class consists of two brane-ending island surfaces, as shown in \cref{fig:vac-adj-island-overview}.  Applying the single-interval result \cref{EE_island} to the two outer endpoints $b_1$ and $b_3$ gives the entropy of this island-supported saddle,
\begin{align}
        S_2=&\frac{c}{6}\log\left(\frac{4 b_1 b_3}{\epsilon^2}\right)+\frac{c}{3}\log\left(\frac{2\ell}{\epsilon_y\cos\theta_0}\right)+\frac{c}{3}\log\left(\sec\theta_0+\tan\theta_0\right)\notag\\
        &+\frac{c}{6}\frac{\rho_0\Phi_r}{\ell}\left(\frac{1}{b_1}+\frac{1}{b_3}\right)+\mathcal{O}(\Phi_r^2)\,.
\end{align}
The new terms relative to the connected-dome saddle have a direct interpretation.  The two logarithmic brane contributions are the rigid-brane boundary entropy/defect contribution, while the term proportional to $\Phi_r$ is the leading dynamical JT correction evaluated at the two island endpoints.  Thus the onset of the island phase changes not only the homology class of the entropy surface but also introduces a local gravitational weighting that depends separately on $b_1$ and $b_3$.

Within this island-supported entanglement wedge, the reflected-entropy problem has three inequivalent endpoint topologies.  The common perpendicular may end on the island surface associated with $A$, terminate directly on the EOW brane and thereby partition the island, or end on the island surface associated with $B$.  Their competition resolves how the same total island is apportioned between the two parties---information that is invisible in the union entropy $S_2$.  We analyze these three possibilities in turn.

\subsubsection*{Cross section ending on the $A$-island surface} \label{sec:vac-adj-A-island-cross-section}
\begin{figure}[ht]
    \centering
    \includegraphics[width=0.8\linewidth]{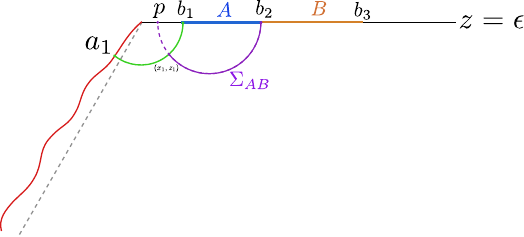}
    \caption{Adjacent-interval island saddle when subsystem $A$ is the smaller component.  The cross section terminates on the island surface associated with $A$.  No independent island cross section lies on the brane, so the reflected entropy reduces to an effective matter contribution plus the bulk geodesic shown in the figure.}
    \label{fig:vac-adj-A-island}
\end{figure}
\paragraph{Boundary description.} Suppose $A$ is the smaller adjacent component.  The minimizing cross section then ends on the island surface emanating from $b_1$, as shown in \cref{fig:vac-adj-A-island}; it never reaches the EOW brane itself.  Consequently there is no independent island-cross-section area term.  The reflected entropy is entirely the effective matter contribution for $A$ against the enlarged region $B\cup I_{S_R}(B)$.  The relevant replica channel is a three-point function involving the endpoints $b_1,b_2$ and the island endpoint $a_1$,
\begin{equation}
	S_R^{\rm bdy}(A:B)= S_R^{\rm eff}(A:B\cup I_{S_{R}}(B))= \lim_{{m,n} \to 1}\frac{1}{1-n}\log \frac{      \langle \sigma_{g^{}_A}(b_1)\sigma_{g^{}_B g_A^{-1}}(b_2)\sigma_{g^{-1}_A}( a_1 )\rangle}{ \langle\sigma_{g^{}_m}(b_1)\sigma_{g^{-1}_m}(a_1)\rangle^n},
\end{equation}  
where $a_1$ denotes the intersection of the $A$-island surface with the EOW brane.  It is useful to introduce an endpoint-specific expansion parameter
\begin{align}
\mu_1\equiv\frac{\rho_0\Phi_r}{\ell b_1}.
\end{align}
Evaluating the three-point function in the semiclassically dominant large-$c$ channel corresponding to this geometry, including the brane Weyl factor, gives
\begin{align}\label{sr-vac-adj-A-island-bdy}
        S_R^{\rm bdy}(A:B)&=\frac{c}{3}\log\frac{b_2^2-b_1^2}{b_1\epsilon}-\frac{c}{3} \left(\frac{b_2-b_1}{b_2+b_1} \right)\mu_1+\mathcal{O}(\mu_1^2),\notag\\
        &=\frac{c}{3}\log\frac{b_2^2-b_1^2}{b_1\epsilon}-\frac{c}{3}\frac{\rho_0\Phi_r}{\ell b_1}\left(\frac{b_2-b_1}{b_2+b_1}\right)+\mathcal{O}(\Phi_r^2),
\end{align}
The island endpoint appearing here is fixed by the entropy saddle rather than by a second reflected-entropy extremization.  In terms of $\mu_1$ it is
\begin{align}
a_1=\frac{b_1}{2}\left(1+\mu_1+\sqrt{1+6\mu_1+\mu_1^2}\right).
\end{align}
The first line of \cref{sr-vac-adj-A-island-bdy} displays the perturbative structure, whereas the second makes the local dilaton dependence explicit.  Notice already that the leading correction is negative: the brane fluctuation moves the supporting island surface in a direction that shortens the internal cross section.

\paragraph{Bulk description.} The same configuration is the common perpendicular from the boundary point $b_2$ to the $A$-island geodesic, shown in \cref{fig:vac-adj-A-island}.  Since the reflected island is assigned entirely to $B$, there is no independent lower-dimensional reflected-island generalized-area term. In the higher-dimensional description the ordinary EWCS area remains the geometric contribution, namely the length of this common perpendicular.  On the $t=0$ slice every Poincar\'e geodesic is a semicircle.  Let $(x_c,r)$ be the center and radius of the island geodesic joining $(b_1,\epsilon)$ to the brane point specified by $(a_1,\theta_1)$.  Its two endpoint conditions are
\begin{align}
    (b_1-x_c)^2=r^2, \quad \left(a_1 \sin \theta_1 +x_c \right)^2+ a_1^2 \cos^2 \theta_1 =r^2,
\end{align}
Solving these conditions fixes the supporting semicircle,
    \begin{align}\label{radius_center}
        x_c=\frac{b_1^2-a_1^2}{2(b_1+a_1\sin\theta_1)}~~,~~r=\frac{b_1^2+a_1^2+2b_1 a_1 \sin\theta_1}{2(b_1+a_1\sin\theta_1)},
    \end{align}
    where $\theta_1$ is the local brane angle at the endpoint.  To find the common perpendicular, parametrize an arbitrary point on this semicircle by $(x_1,z_1)$.  Its regulated distance from the adjacent endpoint $b_2$ can be written as
    \begin{align}
       \ell \log\left[\frac{(b_2-x_1)^2+z_1^2}{z_1 \epsilon}\right]=\ell \log\left[\frac{b_2^2-2 b_2 x_1+r^2+x_c(2 x_1-x_c)}{\epsilon  \sqrt{r^2-(x_1-x_c)^2}}\right],
    \end{align}
    where the circle equation $(x_1-x_c)^2+z_1^2=r^2$ has been used to eliminate $z_1$.  Orthogonality of the minimizing geodesic to the island semicircle is equivalent to extremizing this distance with respect to $x_1$, which yields
    \begin{align}
        x_1=x_c+\frac{2 r^2 (b_2-x_c)}{(b_2-x_c)^2+r^2},
    \end{align}
    Evaluating the regulated distance at the extremal point gives the geometric contribution to the reflected entropy,
    \begin{align}
    S_R^{\rm bulk}(A:B)= \frac{\operatorname{Area}[\Sigma_{AB}]}{2 G_N}&=\frac{\ell}{2G_N}\log\left[\frac{(b_2-x_c)^2-r^2}{r \,\epsilon }\right]\notag\\&=\frac{\ell}{2G_N}\log\left[\frac{2 (b_2-b_1) \left(a_1^2+a_1 (b_1+b_2) \sin \theta_1+b_1 b_2\right)}{\epsilon\left(a_1^2+2b_1a_1\sin\theta_1+b_1^2\right)}\right].
    \end{align}
    For a rigid brane, $a_1=b_1$ and $\theta_1=\theta_0$; the common-perpendicular length then reduces to
    \begin{align}
        S_R^{\rm bulk}(A:B)=\frac{\ell}{2G_N}\log\frac{b_2^2-b_1^2}{b_1\epsilon}.
    \end{align}
    In the presence of small brane fluctuations, we have from the extremization of the entanglement entropy of $A\cup B$,
    \begin{align}\label{island_location}
        a_1=b_1+\mu_1(1+\sin\theta_0)b_1+\mathcal{O}(\mu_1^2),
    \end{align}
    The fluctuation also changes the local brane angle.  To the same order it is convenient to keep the exact geometric relation
    \begin{align}\label{brane_fluctuation}
        \theta_1 = \cos ^{-1}\left[\operatorname{sech}\left(\frac{\rho_0+\Delta\rho(b_1)}{\ell}\right)\right]~~,~~\Delta\rho(b_1)=\rho_0\frac{\Phi_r}{a_1}.
    \end{align}
    Substituting the perturbed endpoint and angle into the common-perpendicular length gives
    \begin{align}
        S_R^{\rm bulk}(A:B)=&\frac{\ell}{2G_N}\log\frac{b_2^2-b_1^2}{b_1\epsilon}-\frac{\ell\mu_1}{2G_N}\frac{b_2-b_1}{b_2+b_1}+\mathcal{O}(\mu_1^2)\notag\\
        =&\frac{\ell}{2G_N}\log\frac{b_2^2-b_1^2}{b_1\epsilon}-\frac{\rho_0\Phi_r}{2G_N b_1}\frac{b_2-b_1}{b_2+b_1}+\mathcal{O}(\Phi_r^2).
    \end{align}
Using the Brown--Henneaux relation, the boundary and bulk answers agree through first order in $\Phi_r$.  The sign of the correction follows directly from the common-perpendicular projection: increasing the DES endpoint $a_1$ moves the island geodesic toward the boundary insertion at $b_2$, thereby decreasing the cross-section length.  Its zero-temperature behavior is consistent with the thermal result below.  The important physical point is that a positive brane fluctuation does not simply add a universal positive ``area'' correction to every mixed-state observable.  Here it displaces the supporting island geodesic in a direction that shortens the internal cross section, so the reflected entropy decreases at first order.  The sign therefore probes the local geometry of the purification rather than merely the generalized entropy of the island endpoint.

\subsubsection*{Cross section ending on the EOW brane}
\begin{figure}[ht]
    \centering
    \includegraphics[width=0.65\linewidth]{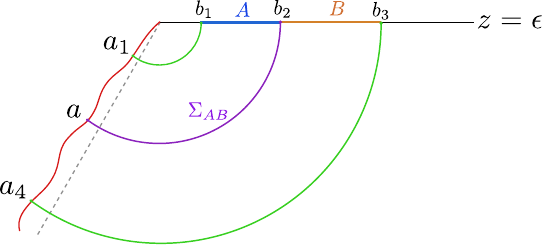}
    \caption{Adjacent-interval island saddle in the intermediate regime where the optimal cross section reaches the EOW brane.  Its endpoint partitions the island between the two reflected subsystems, so the generalized reflected entropy contains both the brane-area/dilaton term and the effective matter reflected entropy.}
    \label{fig:vac-adj-brane}
\end{figure}

\paragraph{Boundary description.} For intermediate relative sizes of $A$ and $B$, neither island surface provides the shortest separator.  The minimizing cross section instead reaches the EOW brane, as in \cref{fig:vac-adj-brane}, and its endpoint partitions the total island into the pieces assigned to the two reflected subsystems.  This is the first vacuum configuration in which the generalized reflected entropy contains a genuine island-cross-section term.  The matter contribution is the reflected entropy of the two enlarged regions and is fixed by a two-point twist correlator,
\begin{align}
    S_R^{\rm eff}(A\cup I_{S_{R}}(A):B \cup I_{S_{R}}(B))= &\lim_{{m,n} \to 1}\frac{1}{1-n}\log       \langle \sigma_{g^{}_A g_B^{-1}}(b_2) \sigma_{g^{}_A g_B^{-1}}( a_2 )\rangle,\notag\\
    &=\frac{c}{3}  \log \left[\frac{\ell (a_2+b_2)^2}{a_2 ~\epsilon  ~ \epsilon_y   \operatorname{sech} \frac{\rho_0}{\ell} }\right].
\end{align}
Here $a_2$ is the brane location at which the reflected island is divided, and $\epsilon_y$ is the UV cutoff of the brane CFT.  The gravitational contribution associated with this internal boundary follows from the same partial-reduction dictionary as the ordinary QES area term \cite{Deng:2022yll},
\begin{align}\label{Area_IS}
    \operatorname{Area}(\Gamma)= \frac{\rho_0}{\ell}\left(1+\frac{\Phi_r}{a_2}\right).
\end{align}
Adding the matter and gravitational pieces gives the generalized reflected-entropy functional that must be extremized over $a_2$,
\begin{equation}\label{adj-Brane-EW}
	S_R^{\text{bdy}}(A:B)=	\frac{c}{3}  \log \left[\frac{\ell (a_2+b_2)^2}{a_2 ~\epsilon ~ \epsilon_y   \operatorname{sech} \frac{\rho_0}{\ell} }\right]+\frac{c}{3}\frac{\rho_0}{\ell}\left(1+\frac{\Phi_r}{a_2}\right).
\end{equation}
The stationarity condition has the same algebraic structure as the single-interval QES problem.  With $\mu_2\equiv\rho_0\Phi_r/(\ell b_2)$, its physical solution is
\begin{align}
        a_2=\frac{b_2}{2} \left(1+\mu_2 +\sqrt{1+6\mu_2+\mu_2 ^2}\right),
\end{align}
Expanding this saddle for $\mu_2\ll1$ and substituting it back into \cref{adj-Brane-EW} yields
\begin{align}
        S_R^{\text{bdy}}(A:B)=\frac{c}{3}\log\left(\frac{2 b_2}{\epsilon}\right)+\frac{c}{3}\log\left(\frac{2\ell \operatorname{sech} \frac{\rho_0}{\ell}}{\epsilon_y}\right)+\frac{c}{3}\frac{\rho_0}{\ell}+\frac{c\mu_2}{3}+\mathcal{O}(\mu_2^2)\,.
\end{align}

\paragraph{Bulk description.} In the three-dimensional description, $\Sigma_{AB}$ is the geodesic from the common boundary endpoint $(b_2,\epsilon)$ to the EOW brane.  Because the cross section ends on the brane, the higher-dimensional functional contains both this geodesic length and the effective reflected entropy of the brane degrees of freedom.  The latter is represented by the BCFT correlator
{\small\begin{equation}\label{SR-eff-adj-def}
	S_R^{\rm eff}(A:B)=S_R^{\rm eff}(I_{A}:I_{B})= \lim_{{m,n} \to 1}\frac{1}{1-n}\log \frac{   \Pi_{i} ~ \Omega_{i}^{2 h_{i}}  \langle \sigma_{g^{}_A}(a_1)\sigma_{g^{}_B g_A^{-1}}(a)\sigma_{g^{-1}_B}(a_3)\rangle_{\mathrm{BCFT}^{\bigotimes mn}}}{ \Omega(a_1)^{2 n h_{mn}} \Omega(a_3)^{2 n h_{mn}} \langle\sigma_{g^{}_m}(a_1)\sigma_{g^{-1}_m}(a_3)\rangle^n_{\mathrm{BCFT}^{\bigotimes m}}},
\end{equation}}
The correlator admits both BOE and OPE decompositions.  In the semiclassical channel appropriate to the brane-ending saddle, the BOE contribution dominates \cite{Li:2021dmf}; the three-point function is, in the dominant large-$c$ channel, approximated by one-point functions.  Including the brane Weyl factors leaves a position-independent defect contribution,
\begin{equation}\label{SR-eff-adj-value}
	S_R^{\rm eff}(I_{A}:I_{B})= \frac{c}{3} \log \frac{2 \ell}{\epsilon_y  \cos \theta_0}.
\end{equation}
The geometric part of \cref{SR-bulk} is the regulated distance from $(b_2,\epsilon)$ to a variable brane point $(-a\sin\theta,a\cos\theta)$.  Using the Poincar\'e distance formula gives
  \begin{align}
        \frac{\operatorname{Area}(\Sigma_{AB})}{2G_N}=\frac{\ell}{2G_N}\operatorname{arccosh}\left[\frac{\left(b_2+a\sin\theta\right)^2+a^2\cos^2\theta}{2a\epsilon\cos\theta}\right].
    \end{align}
Writing the local angle as $\theta( a)=\arccos[\operatorname{sech}((\rho_0+\Delta\rho(a))/\ell)]$ and expanding to first order in the transverse displacement gives   
\begin{align}
        \frac{\operatorname{Area}(\Sigma_{AB})}{2G_N}=\frac{\ell}{2G_N}\left(\log\left[\frac{a^2+b_2^2+2 a b_2\sin\theta_0}{a\epsilon\cos\theta_0}\right]+\frac{\Delta\rho}{\ell}\frac{(a^2+b_2^2)\sin\theta_0+2a b_2}{a^2+b_2^2+2 a b_2\sin\theta_0}\right).
    \end{align}
By the perturbative extremization identity \cref{first-order-extremization}, the explicit $\Delta\rho$ correction can be evaluated at the rigid-brane saddle $a=b_2$.  Adding \cref{SR-eff-adj-value} then gives the bulk generalized functional
 \begin{align}
        S_R^{\rm bulk}(A:B)=\frac{\ell}{2G_N} \log\left[\frac{a^2+b_2^2+2 a b_2\sin\theta_0}{a\epsilon\cos\theta_0}\right]+\frac{\rho_0}{2G_N}\frac{\Phi_r}{a}+\frac{c}{3}\log\frac{2\ell}{\epsilon_y\cos\theta_0}.\label{sr-vac-adj-brane-bulk}
    \end{align}
Its stationarity condition is
  \begin{align}
     \frac{a (a^2-b_2^2)}{a^2+2 a b_2 \sin\theta_0+b_2^2}- \mu_2 b_2=0.
    \end{align}
The solution continuously connected to $a=b_2$ has the expansion
\begin{align}
        a=b_2+b_2(1+\sin\theta_0)\mu_2+\mathcal{O}(\mu_2^2),
    \end{align}
Evaluating the functional at this saddle yields
\begin{align}
        S_R^{\text{bulk}}(A:B)=\frac{\ell}{2 G_N}\left[\log\frac{2 b_2}{\epsilon}+\log\frac{2\ell \operatorname{sech} \frac{\rho_0}{\ell}}{\epsilon_y}+\frac{\rho_0}{\ell}+\mu_2\right]+\mathcal{O}(\mu_2^2)\,,
\end{align}
where we have used $\theta_0=\arccos\!\left[\operatorname{sech}\!\left(\frac{\rho_0}{\ell}\right)\right]$. The Brown--Henneaux relation again makes the two descriptions identical through first order.  In contrast with the island-surface saddle, the correction is now positive: the cross section itself ends on the gravitating brane, so the JT dilaton enters as a genuine generalized-area cost rather than only through a displacement of a supporting geodesic.  This change of sign is a useful physical distinction between the two endpoint topologies.

\subsubsection*{Cross section ending on the $B$-island surface}
\begin{figure}[ht]
    \centering
    \includegraphics[width=0.65\linewidth]{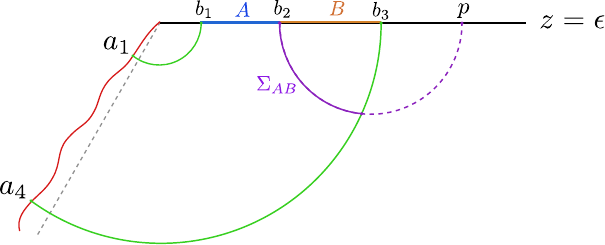}
    \caption{Mirror adjacent-interval island saddle in which subsystem $B$ is the smaller component.  The cross section terminates on the $B$-island surface and is related to \cref{fig:vac-adj-A-island} by exchanging the two parties.}
    \label{fig:vac-adj-B-island}
\end{figure}

The third possibility is the mirror configuration in which $B$ is the smaller component, shown in \cref{fig:vac-adj-B-island}.  The optimal separator now ends on the island surface anchored at $b_3$.  Defining $\mu_3\equiv\rho_0\Phi_r/(\ell b_3)$, the boundary result follows from the $A$-controlled saddle by the exchange $b_1\leftrightarrow b_3$ together with the corresponding reversal of orientation,
\begin{align}\label{sr-vac-adj-A-island-bdyii}
        S_R^{\rm bdy}(A:B)&=\frac{c}{3}\log\frac{b_3^2-b_2^2}{b_3 \epsilon}+\frac{c}{3}\frac{b_3-b_2}{b_3+b_2}\mu_3+\mathcal{O}(\mu_3^2),\notag\\
        &=\frac{c}{3}\log\frac{b_3^2-b_2^2}{b_3\epsilon}+\frac{c}{3}\frac{\rho_0\Phi_r}{\ell b_3}\frac{b_3-b_2}{b_3+b_2}+\mathcal{O}(\Phi_r^2).
\end{align}
The bulk common-perpendicular problem is related by the same reflection.  The first-order correction now has the opposite orientation dependence because the supporting island geodesic is displaced on the other side of the common endpoint $b_2$.  After the endpoint-specific parameter $\mu_3$ is used, the Brown--Henneaux relation gives agreement with \cref{sr-vac-adj-A-island-bdyii}.  Together with \cref{fig:vac-adj-A-island}, this mirror saddle makes clear that the smaller adjacent party determines which island surface carries the optimal cross-section endpoint.


    \subsection{Disjoint subsystems}
    We next separate the two parties and take $A=(b_1,b_2)$ and $B=(b_3,b_4)$ with $0<b_1<b_2<b_3<b_4$. The central gap $b_3-b_2$ makes connectivity an additional saddle-selection problem. A crucial competitor is the disconnected entanglement wedge. Define
    \begin{align}
    S_{\rm dome}(u,v)&=\frac{c}{3}\log\frac{v-u}{\epsilon},\notag\\
    S_{\rm isl}(u,v)={}&\frac{c}{6}\log\frac{4uv}{\epsilon^2}
    +\frac{c}{3}\log\frac{2\ell}{\epsilon_y\cos\theta_0}
    +\frac{c}{3}\log(\sec\theta_0+\tan\theta_0)\notag\\
    &+\frac{c\rho_0\Phi_r}{6\ell}\left(\frac1u+\frac1v\right)+\mathcal O(\Phi_r^2).
    \label{vac-single-interval-saddles}
    \end{align}
    The leading disconnected union entropy is
    \begin{align}
    S_{\rm disc}(A\cup B)={}&\min\{S_{\rm dome}(b_1,b_2),S_{\rm isl}(b_1,b_2)\}\notag\\
    &+\min\{S_{\rm dome}(b_3,b_4),S_{\rm isl}(b_3,b_4)\}.
    \label{vac-disconnected-union}
    \end{align}
    This includes dome--dome, island--dome, dome--island and island--island disconnected configurations. The connected reflected-entropy saddles below are physical only when their outer entropy saddle beats \cref{vac-disconnected-union}. If the disconnected saddle dominates, the classical entanglement wedge factorizes, no connected EWCS exists, and
    \begin{align}
    I(A:B)=0+O(c^0),\qquad S_R(A:B)=0+O(c^0).
    \label{vac-disconnected-SR-zero}
    \end{align}
    Thus existence of a common perpendicular for a candidate pair of geodesics is distinct from dominance of the connected wedge. Subject to this condition, we compare the dominant large-$c$ replica channels with their bulk common perpendiculars.

    \subsubsection{Connected dome saddle for disjoint intervals}
\begin{figure}[ht]
    \centering
    \includegraphics[width=0.8\linewidth]{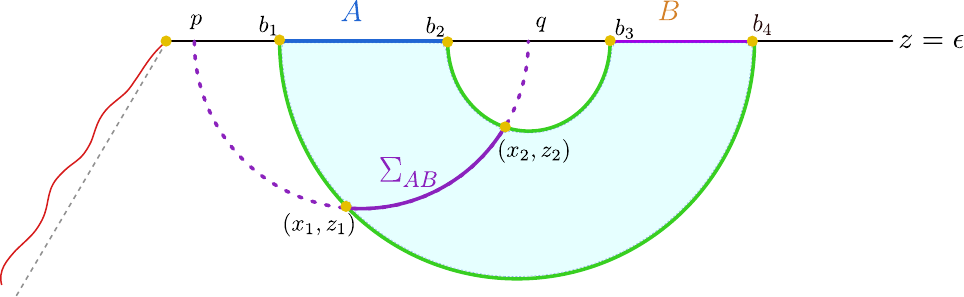}
    \caption{Vacuum disjoint intervals in the connected no-island entanglement saddle.  The two dome geodesics bound a connected entanglement wedge, and the reflected entropy is twice the length of the common perpendicular between them.  Increasing the separation of the intervals lengthens this cross section until the connected saddle ceases to dominate.}
    \label{fig:vac-dis-connected}
\end{figure}
When both intervals lie far enough from the EOW brane, the connected no-island homology class is bounded by the two same-boundary dome geodesics shown in \cref{fig:vac-dis-connected}: an outer dome connecting $(b_1,b_4)$ and an inner dome connecting $(b_2,b_3)$.  Their regulated lengths give
\begin{equation}
	S_1= \frac{\ell}{2 G_N} \log\left[\frac{b_3-b_2}{\epsilon}\right]+\frac{\ell}{2 G_N} \log\left[\frac{b_4-b_1}{\epsilon}\right].
\end{equation}
For this connected entanglement wedge there is a unique minimal separator, the common perpendicular in \cref{fig:vac-dis-connected}.  On the CFT side the corresponding large-$c$ channel is encoded in the four-point reflected-entropy correlator
\begin{equation}
	S_R^{\text{bdy}}(A:B)= \lim_{{m,n} \to 1}\frac{1}{1-n}\log\frac{\langle \sigma_{g_A}(b_1)\sigma_{g_A^{-1}}(b_2)\sigma_{g_B}(b_3)\sigma_{g^{-1}_B}(b_4)\rangle}{\langle \sigma_{g_m}(b_1)\sigma_{g_m^{-1}}(b_2)\sigma_{g_m}(b_3)\sigma_{g^{-1}_m}(b_4)\rangle^n}.
\end{equation}
The semiclassical four-point block depends only on the conformal cross ratio.  In the connected channel it gives \cite{Dutta:2019gen,Fitzpatrick:2014vua}
\begin{equation}\label{SR-Dis-1}
	S_R^{\text{bdy}}(A:B)=\frac{c}{3}  \log \left[\frac{1+\sqrt{X}}{1-\sqrt{X}}\right],
\end{equation} 
where all dependence on the four endpoints is captured by the standard conformal cross ratio
\begin{equation}\label{Cross-ratio}
	X= \frac{(b_2-b_1)(b_4-b_3)}{(b_3-b_1) (b_4-b_2)}.
\end{equation}
In the bulk description, the EWCS is the unique common perpendicular between the RT semicircles anchored on $(b_1,b_4)$ and $(b_2,b_3)$.  The result can be obtained without carrying the individual intersection coordinates through the algebra: an $SL(2,\mathbb R)$ transformation maps the two geodesics to concentric geodesics, so their separation is fixed entirely by the conformal cross ratio $X$ in \cref{Cross-ratio} \cite{Dutta:2019gen,Tamaoka:2018ned}.  The common-perpendicular length satisfies
\begin{align}
 \cosh\frac{\mathcal L_{\rm EW}}{\ell}=\frac{1+X}{1-X},
\end{align}
Using this invariant common-perpendicular length in the holographic relation $S_R=\mathrm{Area}(\Sigma_{AB})/(2G_N)$ gives
\begin{align}
 S_R^{\mathrm{bulk}}(A:B)
 &=\frac{\ell}{2G_N}\,\operatorname{arccosh}\!\left(\frac{1+X}{1-X}\right)\notag\\
 &=\frac{\ell}{2G_N}\log\!\left(\frac{1+\sqrt X}{1-\sqrt X}\right).\label{SR-Dis-1-bulk}
\end{align}
Equivalently, this common-perpendicular invariant can be written directly in terms of the four endpoints as
\begin{align}
 \frac{1+X}{1-X}
 =\frac{b_1b_2+b_1b_3-2b_1b_4-2b_2b_3+b_2b_4+b_3b_4}
 {(b_1-b_4)(b_2-b_3)}.
\end{align}
For ordered endpoints $0<X<1$, so the argument of the $\operatorname{arccosh}$ is larger than one and the common perpendicular is real throughout this kinematic range. This is only an existence statement. The physical connected RT channel must also beat the disconnected pairing $(b_1,b_2)\cup(b_3,b_4)$. In the vacuum no-island sector,
\begin{align}
S_{\rm conn}<S_{\rm disc}^{\rm dome}\quad\Longleftrightarrow\quad X>\frac12,
\label{vac-cross-ratio-transition}
\end{align}
for the standard large-$c$ RT competition. Hence \cref{SR-Dis-1,SR-Dis-1-bulk} are the leading reflected entropy only in the dominant connected block/channel; for $X\leq1/2$ the leading wedge is disconnected and \cref{vac-disconnected-SR-zero} applies. Using $c=3\ell/(2G_N)$ reproduces the connected-channel CFT answer \cref{SR-Dis-1}.

\subsubsection{Mixed island--dome saddle for disjoint intervals}
\begin{figure}[ht]
    \centering
    \includegraphics[width=0.65\linewidth]{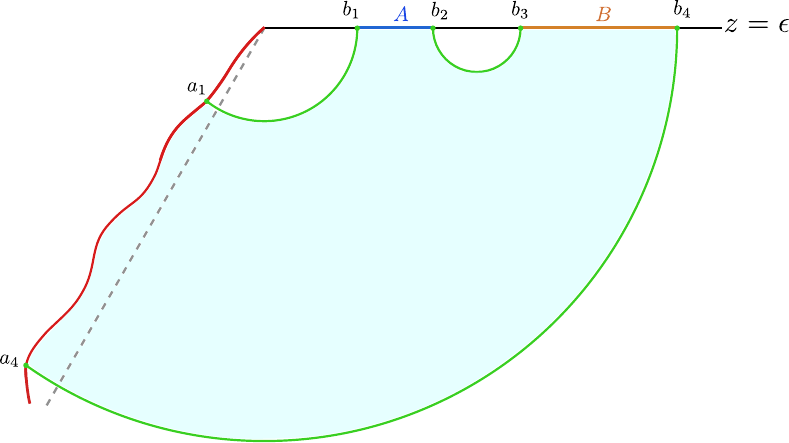}
    \caption{Vacuum disjoint intervals in the mixed island--dome entanglement saddle.  The outer endpoints are connected to the EOW brane by island surfaces while the inner endpoints support a dome.  The reflected-entropy cross section may end on an island surface, on the brane, or on the mirror island surface.}
    \label{fig:vac-dis-mixed-overview}
\end{figure}

Closer to the EOW brane, a competing homology class attaches the two outer endpoints $b_1$ and $b_4$ to the brane while the inner endpoints remain connected by the central dome, shown in \cref{fig:vac-dis-mixed-overview}.  The entropy is therefore the sum of two single-endpoint island contributions and the ordinary dome contribution,
\begin{align}
        S_2={}&\frac{c}{6}\log\frac{4 b_1 b_4}{\epsilon^2}
        +\frac{c}{3}\log\frac{2\ell}{\epsilon_y\cos\theta_0}
        +\frac{c}{3}\log\left(\sec\theta_0+\tan\theta_0\right)
        +\frac{c}{3}\log\left(\frac{b_3-b_2}{\epsilon}\right)\notag\\
        &+\frac{c\rho_0\Phi_r}{6\ell}\left(\frac{1}{b_1}+\frac{1}{b_4}\right)+\mathcal{O}(\Phi_r^2)\,.
\end{align}
The resulting entanglement wedge again admits three inequivalent ways to separate $A$ from $B$: the cross section can end on the $A$-island surface, terminate on the EOW brane, or end on the mirror $B$-island surface.  Unlike the adjacent problem, one endpoint of the separator now begins on the central dome, so the finite gap $b_3-b_2$ enters the reflected entropy directly.
    
\subsubsection*{Cross section ending on the $A$-island surface}
\begin{figure}[ht]
    \centering
    \includegraphics[width=0.65\linewidth]{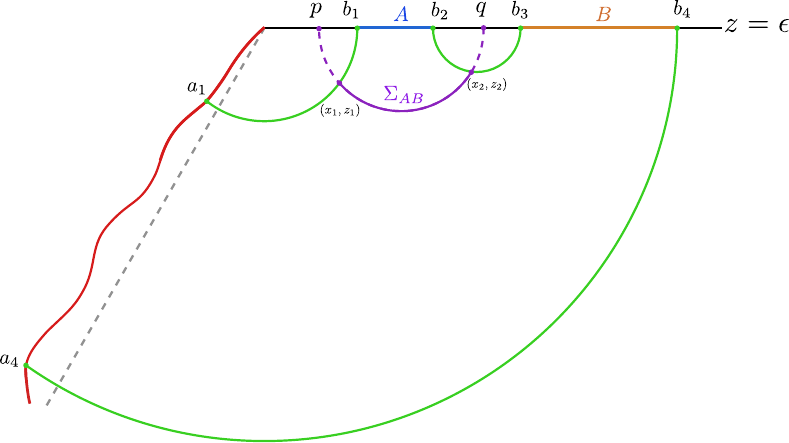}
    \caption{Disjoint-interval mixed saddle when the cross section terminates on the island surface associated with subsystem $A$.  The relevant bulk object is the common perpendicular between the central dome and that island surface.}
    \label{fig:vac-dis-A-island}
\end{figure}

\paragraph{Boundary description.} When $A$ is the smaller disjoint component, the minimal separator leaves the central dome and terminates on the island surface anchored at $b_1$, as shown in \cref{fig:vac-dis-A-island}.  The cross section does not reach the EOW brane, so no independent island-cross-section area term appears.  The boundary problem is instead the effective reflected entropy between $A$ and the enlarged region $B\cup I_{S_R}(B)$, represented by the four-point correlator
{\small\begin{equation}
S_R^{\rm bdy}(A:B)= S_R^{\rm eff}(A:B\cup I_{S_{R}}(B))= \lim_{{m,n} \to 1}\frac{1}{1-n}\log\frac{\langle \sigma_{g^{-1}_B}(a_1) \sigma_{g_A}(b_1)\sigma_{g_A^{-1}}(b_2)\sigma_{g_B}(b_3)\rangle_{mn}}{\langle \sigma_{g^{-1}_m}(a_1) \sigma_{g_m}(b_1)\sigma_{g_m^{-1}}(b_2)\sigma_{g_m}(b_3)\rangle^n_{ m}},
\end{equation}}
where the island endpoint is fixed by the outer entropy saddle.  With $\mu_1=\rho_0\Phi_r/(\ell b_1)$ it is $a_1=b_1[1+2\mu_1+\mathcal O(\mu_1^2)]$.  Evaluating the dominant large-$c$ four-point block whose monodromy/endpoint pairing is dual to the dome--island common perpendicular, and expanding to first order, gives
\begin{align}
 S_R^{\rm bdy}(A:B)=&\frac{c}{3} \log \left[\frac{\sqrt{(b_1+b_2)(b_3-b_1)}+\sqrt{(b_2-b_1)(b_3+b_1)}}{\sqrt{(b_1+b_2)(b_3-b_1)}- \sqrt{(b_2-b_1)(b_3+b_1)}}\right]\notag\\
 &-\frac{c \mu_1}{3}\sqrt{\frac{(b_2-b_1) (b_3-b_1)}{\left(b_2+b_1\right) \left(b_3+b_1\right)}}+\mathcal{O}(\mu_1^2).
\end{align}

\paragraph{Bulk description.} Geometrically, $\Sigma_{AB}$ is the common perpendicular between the central dome and the $A$-island geodesic.  Because the reflected island is assigned entirely to $B$, there is no independent lower-dimensional generalized-area contribution from a reflected-island separator.  The higher-dimensional geometric contribution does not vanish: it is the ordinary EWCS area, namely the length of this common perpendicular.  For arbitrary endpoints $(x_1,z_1)$ and $(x_2,z_2)$ on the two supporting geodesics, \cref{Poincare-length} gives
  \begin{align}\label{SR_dis_2_island}
        \frac{\operatorname{Area}\left[\Sigma_{AB}\right]}{2 G_N}&=\frac{\ell}{2G_N}\operatorname{arccosh}\left[\frac{(x_1-x_2)^2+z_1^2+z_2^2}{2z_1z_2}\right].
    \end{align}
The two supporting RT/DES geodesics are semicircles.  Denoting their centers and radii by $(x_{c_1},r_1)$ and $(x_{c_2},r_2)$, the candidate endpoints obey
\begin{align}
    (x_1-x_{c_{1}})^2+z_1^2=r_1^2 ~~,   (x_2- x_{c_{2}})^2+z_2^2= r_2^2.  
\end{align}
Eliminating $z_1$ and $z_2$ with these circle equations rewrites the distance as
 \begin{align}
        \frac{\operatorname{Area}\left[\Sigma_{AB}\right]}{2 G_N}&=\frac{\ell}{2G_N}\operatorname{arccosh}\left[\frac{r_1^2+r_2^2-2 x_1x_2+2 x_1 x_{c_1}+2 x_2x_{c_2}-x_{c_1}^2-x_{c_2}^2}{2 \sqrt{r_1^2-(x_1-x_{c_1})^2} \sqrt{r_2^2-(x_2-x_{c_2})^2}}\right].
    \end{align}
The physical common perpendicular is obtained by varying both endpoint coordinates.  The stationarity equations have two algebraic branches,
    \begin{align}
        x_1&=x_2=\frac{r_2^2-r_1^2+x_{c_1}^2-x_{c_2}^2}{2(x_{c_1}-x_{c_2})}\,,\notag\\
        x_1&=x_{c_1}-\frac{2 r_1^2 \left(x_{c_2}-x_{c_1}\right)}{r_2^2-r_1^2-\left(x_{c_2}-x_{c_1}\right)^2}~~,~~x_2=x_{c_2}-\frac{2 r_2^2 \left(x_{c_2}-x_{c_1}\right)}{r_2^2-r_1^2+\left(x_{c_2}-x_{c_1}\right)^2}.
    \end{align}
The coincident-coordinate branch does not represent the separator between the two distinct geodesics.  The nontrivial branch is the common perpendicular and gives the invariant distance
    {\footnotesize\begin{align}
        \frac{\operatorname{Area}\left[\Sigma_{AB}\right]}{2 G_N}&=\frac{\ell}{2G_N}\operatorname{arccosh}\left[\frac{(x_{c_1}-x_{c_2})^2-r_1^2-r_2^2}{2r_1 r_2}\right],\notag\\
        &=\frac{\ell}{2G_N}\operatorname{arccosh}\left[\frac{2 a_1 \left(b_1^2-b_2 b_3\right) \sin\theta_1+a_1^2 (2 b_1-b_2-b_3)+b_1^2( b_2+b_3)-2 b_1 b_2b_3}{(b_2-b_3) \left(2 b_1 a_1 \sin\theta_1+a_1^2+b_1^2\right)}\right],
    \end{align}}
where we have substituted the value of $x_{c_{1}}, r_1$ from \cref{radius_center} and $x_{c_2}= \frac{b_2+b_3}{2}$, $r_2= \frac{b_3-b_2}{2}$. In the above expression we may introduce the brane fluctuation, given in \cref{brane_fluctuation}, which leads to the following
    {\scriptsize\begin{align}
        \operatorname{Area}\left[\Sigma_{AB}\right]=\ell\,\operatorname{arccosh}\left[\frac{2 a_1 \left(b_1^2-b_2 b_3\right) \tanh \frac{\rho_0+ \Delta \rho(b_1)}{\ell}+a_1^2 (2 b_1-b_2-b_3)+b_1^2 (b_2+b_3)-2 b_1 b_2b_3)}{(b_2-b_3) \left(2 b_1 a_1 \tanh \frac{\rho_0+ \Delta \rho(b_1)}{\ell}+a_1^2+b_1^2\right)}\right].
    \end{align}}
The supporting island surface is then displaced according to \cref{island_location}.  Substituting that displacement and expanding the common-perpendicular length to first order in $\mu_1$ yields    
    \begin{align}
        S_R^{\text{bulk}}(A:B)=&\frac{\ell}{2 G_N}
        \operatorname{arccosh}\!\left[1+\frac{(b_1-b_2)(b_1+b_3)}{b_1(b_2-b_3)}\right]\notag\\
        &-\frac{\ell\mu_1}{2G_N}
        \frac{(b_1-b_2)(b_1-b_3)}{\sqrt{(b_1^2-b_2^2)(b_1^2-b_3^2)}}
        +\mathcal O(\mu_1^2),\notag\\
        =&\frac{\ell}{2G_N}
        \operatorname{arccosh}\!\left[1+\frac{(b_1-b_2)(b_1+b_3)}{b_1(b_2-b_3)}\right]\notag\\
        &-\frac{\rho_0\Phi_r}{2G_N b_1}
        \sqrt{\frac{(b_1-b_2)(b_1-b_3)}{(b_1+b_2)(b_1+b_3)}}
        +\mathcal O(\Phi_r^2).
    \end{align}
The Brown--Henneaux relation gives agreement between the two descriptions through first order.  The rigid part of this answer measures the separation between the central dome and the island surface rather than the separation between two ordinary domes.  Its dependence on $b_1,b_2,b_3$ therefore resolves how the central gap competes with the proximity of the outer endpoint to the brane.  The negative $\Phi_r$ correction has the same geometric origin as in the adjacent $A$-controlled saddle: the fluctuation displaces the supporting island surface so as to shorten the common perpendicular.

    \subsubsection*{Cross section ending on the EOW brane}
\begin{figure}[ht]
    \centering
    \includegraphics[width=0.65\linewidth]{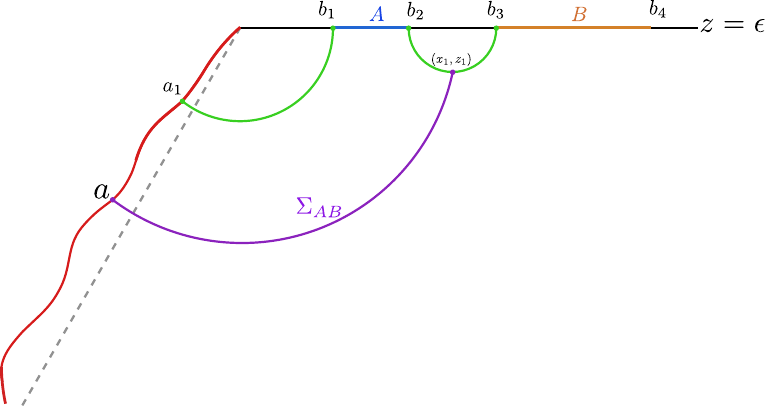}
    \caption{Disjoint-interval mixed saddle in which the central-dome cross section reaches the EOW brane.  The brane endpoint divides the reflected island and therefore carries the leading JT/dilaton correction to the reflected entropy.}
    \label{fig:vac-dis-brane}
\end{figure}

\paragraph{Boundary description.} In the intermediate regime the common perpendicular from the central dome reaches the EOW brane rather than either island surface, as illustrated in \cref{fig:vac-dis-brane}.  Its brane endpoint $a$ separates the reflected island into the portions assigned to $A$ and $B$, so the generalized reflected entropy contains both an internal gravitational term and a matter correlator.  In the semiclassically dominant brane-ending channel, the effective matter contribution is fixed by the three-point twist correlator $\langle \sigma_{g_A^{-1}}(b_2)\sigma_{g_B}(b_3)\sigma_{g_Bg_A^{-1}}(a)\rangle$ and is
\begin{align}\label{SR_eff_brane}
	S_R^{\rm eff}(A\cup I_{S_{R}}(A):B \cup I_{S_{R}}(B))= \frac{c}{3}  \log \left[\frac{2 \ell (a+b_2)(a+b_3)\cosh \frac{\rho_0}{\ell}}{\epsilon_y  (b_3-b_2) a }\right],
\end{align}
The first term in \cref{SR-bdy} is the island-cross-section area \cref{Area_IS}.  Adding it to the effective matter contribution gives the functional to be extremized over the brane location $a$,
\begin{equation}\label{SR-Dis-Brane}
	S_R^{\text{bdy}}(A:B)=	\frac{c}{3}  \log \left[\frac{2 \ell (a+b_2)(a+b_3)}{\epsilon_y  (b_3-b_2) a \operatorname{sech}\!\left(\frac{\rho_0}{\ell}\right) }\right] + \frac{c}{3}\frac{\rho_0}{\ell}\left(1+\frac{\Phi_r}{a}\right).
\end{equation}
At zero fluctuation the saddle lies at the geometric mean $a_0=\sqrt{b_2b_3}$, reflecting the symmetry of the two inner endpoints in logarithmic coordinates.  Perturbing the stationarity condition in $\Phi_r$ moves it to
\begin{align}
        a=\sqrt{b_2b_3}+\frac{\rho_0\Phi_r}{\ell}\left(\frac{b_2+b_3+2\sqrt{b_2b_3}}{2\sqrt{b_2b_3}}\right)+\mathcal{O}(\Phi_r^2).
\end{align}
Evaluating the generalized functional at this saddle yields the boundary reflected entropy
\begin{align}
    S_R^{\text{bdy}}(A:B)= \frac{c}{3} \log \left[ \frac{ (b_2+b_3+ 2 \sqrt{b_2 b_3})}{(b_3-b_2)}\right]+ \frac{c}{3} \log \frac{2 \ell \cosh \frac{\rho_0}{\ell}}{\epsilon_y}+ \frac{c}{3} \frac{\rho_0}{\ell} \frac{ \Phi_r}{\sqrt{b_2 b_3}}+\mathcal{O}(\Phi_r^2).
\end{align}

\paragraph{Bulk description.} The higher-dimensional cross section joins a variable point on the central dome to a variable point on the EOW brane.  The effective brane-matter term is the same constant contribution \cref{SR-eff-adj-value}; the nontrivial problem is to extremize the geodesic connecting the two supporting objects.  Parameterize the point on the central dome by an angle $\phi$ and define its coordinates by
\begin{align}
 x_\phi&\equiv\frac{b_2+b_3}{2}+\frac{b_3-b_2}{2}\sin\phi,
 &z_\phi&\equiv\frac{b_3-b_2}{2}\cos\phi .\label{central-dome-point}
\end{align}
The brane endpoint is $(-a\sin\theta,a\cos\theta)$.  In terms of \cref{central-dome-point}, the Poincar\'e distance to be extremized is
\begin{align}\label{dis_brane_length}
\frac{\operatorname{Area}[\Sigma_{AB}]}{2G_N}
=\frac{\ell}{2G_N}\operatorname{arccosh}\!\left[
\frac{(x_\phi+a\sin\theta)^2+z_\phi^2+a^2\cos^2\theta}
{2z_\phi a\cos\theta}\right].
\end{align}
For fixed $a$, orthogonality to the central dome follows by extremizing with respect to $\phi$, giving
    \begin{align}
        \phi=\arcsin\left[\frac{(b_2-b_3) (b_2+b_3+2 a \sin \theta)}{2 a^2+2 a (b_2+b_3) \sin \theta +b_2^2+b_3^2}\right].
    \end{align}
Eliminating $\phi$ with this stationarity condition expresses the cross-section length entirely in terms of the brane coordinate,
    \begin{align}
        \frac{\operatorname{Area}[\Sigma_{AB}]}{2G_N}=\frac{\ell}{2G_N}\operatorname{arccosh}\left[\frac{\sqrt{\left(a^2+2 a b_2 \sin \theta +b_2^2\right) \left(a^2+2 a b_3 \sin \theta +b_3^2\right)}}{a (b_3-b_2) \cos \theta }\right].
    \end{align}
The local brane angle depends on the transverse fluctuation.  Expanding the invariant length to first order in $\Delta\rho$ gives
    \begin{align}
        \frac{\operatorname{Area}[\Sigma_{AB}]}{2G_N}= &\frac{\ell}{2G_N}\operatorname{arccosh}\left[\frac{\sqrt{\left(a^2+2 a b_2 \sin \theta_0 +b_2^2\right) \left(a^2+2 a b_3 \sin \theta_0 +b_3^2\right)}}{a (b_3-b_2) \cos \theta_0 }\right]\notag\\&\qquad\qquad\quad\quad+\frac{\sin\theta_0 \left(a^2+b_2b_3\right)+a (b_2+b_3)}{\sqrt{\left(a^2+2 a b_2 \sin \theta_0+b_2^2\right) \left(a^2+2 a b_3 \sin \theta_0+b_3^2\right)}}\frac{\Delta\rho(a)}{2G_N}.
    \end{align}
By \cref{first-order-extremization}, the explicit fluctuation term can be evaluated at the rigid-brane saddle $a_0=\sqrt{b_2b_3}$.  Adding the brane-matter term then produces the bulk generalized reflected-entropy functional    
    \begin{align}
        S_R^{\rm bulk}(A:B)=&\frac{\ell}{2G_N}\operatorname{arccosh}\left[\frac{\sqrt{\left(a^2+2 a b_2 \sin \theta_0 +b_2^2\right) \left(a^2+2 a b_3 \sin \theta_0 +b_3^2\right)}}{a (b_3-b_2) \cos \theta_0 }\right]\notag\\&+\frac{\rho_0\Phi_r}{2 G_N a}+\frac{\ell}{2 G_N}\log\frac{2\ell}{\epsilon_y\cos\theta_0},
    \end{align}
where $\Delta\rho(a)=\rho_0\Phi_r/a$.  The stationarity condition shifts the bulk endpoint to
    \begin{align}
        a=\sqrt{b_2b_3}+\frac{\rho_0\Phi_r}{\ell}\frac{b_2+b_3+2\sqrt{b_2b_3}\sin\theta_0}{2\sqrt{b_2b_3}}+\mathcal{O}(\Phi_r^2).
    \end{align}
Evaluating the functional at this shifted endpoint and retaining terms through first order gives    
    {\small\begin{align}
        S_R^{\rm bulk}(A:B)&=\frac{\ell}{2 G_N}\left(\operatorname{arccosh}\left[\frac{b_2+b_3+2\sqrt{b_2 b_3}\sin\theta_0}{(b_3-b_2)\cos\theta_0}\right]+\frac{\rho_0\Phi_r}{\ell\sqrt{b_2b_3}}+\log\frac{2\ell}{\epsilon_y\cos\theta_0}\right)+\mathcal{O}(\Phi_r^2)\notag\\&=\frac{\ell}{2 G_N}\left(\operatorname{arccosh}\left[\frac{b_3+b_2}{b_3-b_2}\right]+\operatorname{arccosh}(\sec\theta_0)+\frac{\rho_0\Phi_r}{\ell\sqrt{b_2b_3}}+\log\frac{2\ell}{\epsilon_y\cos\theta_0}\right)+\mathcal{O}(\Phi_r^2).
    \end{align}}
The Brown--Henneaux relation again gives agreement between the boundary and bulk answers through first order.  Here the central dome and the brane jointly determine the purification: the common perpendicular leaves the dome and ends on the EOW brane, so the JT correction measures the gravitational dressing of an internal mixed-state correlation rather than the entropy of either subsystem separately.  The zeroth-order saddle $a=\sqrt{b_2b_3}$ is the hyperbolic analogue of a geometric mean and reflects the symmetry between the two inner endpoints; the first-order shift records how the fluctuating brane biases that balance.

    \subsubsection*{Cross section ending on the $B$-island surface}
\begin{figure}[ht]
    \centering
    \includegraphics[width=0.65\linewidth]{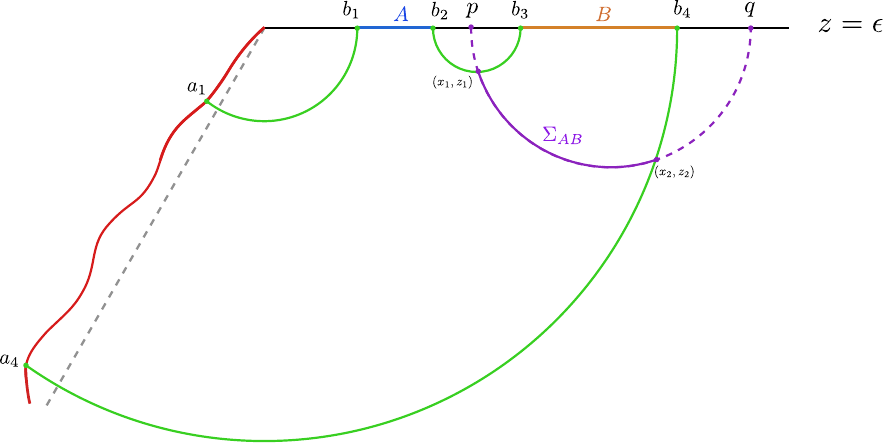}
    \caption{Mirror disjoint-interval saddle in which subsystem $B$ controls the reflected-entropy cross section.  The EWCS terminates on the island surface associated with the outer endpoint $b_4$.  Its value follows from the $A$-controlled configuration by the endpoint reflection $b_1\leftrightarrow b_4$, $b_2\leftrightarrow b_3$.}
    \label{fig:vac-dis-B-island}
\end{figure}
The geometry of this mirror configuration is shown in \cref{fig:vac-dis-B-island}.  In the third saddle subsystem $B$ is the smaller component, and the configuration is obtained from the $A$-island-surface saddle by reflecting the interval ordering and exchanging the roles of the two parties.  Defining the local fluctuation parameter $\mu_4=\rho_0\Phi_r/(\ell b_4)$, it is convenient to introduce
\begin{align}
 X_B=\frac{(b_4-b_3)(b_4+b_2)}{(b_4+b_3)(b_4-b_2)}.
\end{align}
Evaluating the dominant large-$c$ four-point block in this mirror dome--island channel and expanding to first order in $\mu_4$ gives
\begin{align}\label{SR-dis-vac-phaseIII}
 S_R^{\mathrm{bdy}}(A:B)
 =&\frac{c}{3}\log\!\left(\frac{1+\sqrt{X_B}}{1-\sqrt{X_B}}\right)
 -\frac{c\mu_4}{3}\sqrt{\frac{(b_4-b_3)(b_4-b_2)}{(b_4+b_3)(b_4+b_2)}}
 +\mathcal{O}(\mu_4^2).
\end{align}
The bulk calculation is the corresponding distance from the dome geodesic to the island surface ending at $b_4$.  Repeating the extremization of the $A$-island-surface saddle with $b_1\leftrightarrow b_4$ and $b_2\leftrightarrow b_3$ gives precisely \cref{SR-dis-vac-phaseIII} after $c=3\ell/(2G_N)$.  The equality is understood through first order in the brane fluctuation, in accordance with \cref{first-order-extremization}.


    \section{BTZ: eternal black hole}\label{sec:BTZ-review}
We now turn to the thermofield-double state.  This setting introduces the main dynamical ingredient of the paper: surfaces that cross the Einstein--Rosen bridge grow with boundary time, while same-boundary domes and island surfaces are stationary at fixed endpoints.  The Page curves studied in \cref{sec:thermal-reflected} are built from repeated competitions among these elementary pieces, so it is useful to fix the thermal notation and the single-interval saddles before addressing multipartite reflected entropy.

We work with the nonrotating BTZ black brane,
\begin{align}
\d s^2=-\frac{r^2-r_h^2}{\ell^2}\d t^2+\frac{\ell^2}{r^2-r_h^2}\d r^2+\frac{r^2}{\ell^2}\d\phi^2,
\qquad t\sim t+i\beta,
\end{align}
whose horizon radius is related to the inverse temperature by
\begin{align}
r_h=\frac{2\pi\ell^2}{\beta}.
\end{align}
It is convenient to introduce the inverse thermal length
\begin{align}
\kappa\equiv\frac{r_h}{\ell^2}=\frac{2\pi}{\beta},
\end{align}
so that an equal-time thermal chord of coordinate size $\Delta x$ is naturally expressed through $\sinh(\kappa\Delta x/2)$.  We use $\kappa$ selectively in the longer thermal formulas to keep their structure visible, while retaining $r_h$ in expressions where the radial BTZ scale is geometrically important.  For two-sided configurations we use
\begin{align}
L_{ij}\equiv L_i-L_j,
\qquad
R_{ij}\equiv R_i-R_j,
\end{align}
and reserve
\begin{align}
\nu\equiv\frac{2\pi\rho_0\Phi_r}{\beta\ell}
\end{align}
for the dimensionless thermal brane fluctuation.  This notation keeps $\nu$ distinct from the endpoint-dependent vacuum parameters $\mu_i$ used in section~\ref{sec:vacuum-reflected}.  Two combinations that repeatedly appear in the Page-transition conditions are
\begin{align}
\mathcal C_i&\equiv\cosh\!\left[\frac{r_h(L_i-R_i)}{\ell^2}\right],\notag\\
\mathcal Q_i&\equiv\coth\!\left(\frac{r_hL_i}{\ell^2}\right)+\coth\!\left(\frac{r_hR_i}{\ell^2}\right).\label{thermal-CQ}
\end{align}
Here $\mathcal C_i$ is the static left--right separation appearing in a wormhole-crossing geodesic, while $\mathcal Q_i$ is the linear response of a pair of island surfaces to the JT fluctuation.

We also denote the dimensionless background brane position by $\alpha_0\equiv\rho_0/\ell$.  The brane-angle combination that enters the island/Page formulas is
\begin{align}
\mathcal B_0\equiv2\sec\theta_0\left(\sec\theta_0+\tan\theta_0\right).\label{B0-def}
\end{align}
Writing this quantity directly in terms of $\theta_0$ keeps the subsequent expressions manifestly hyperbolic.  A second useful notation packages the BTZ distance from a bath endpoint $L$ to a variable brane point $a$.  With
\begin{align}
u_\pm(a,L)\equiv\frac{\pi(a\pm L)}{\beta},
\end{align}
we define
\begin{align}
\mathcal F_{\alpha}(a,L)&\equiv
\cosh\alpha\left[\sinh^2u_++\sinh^2u_-\right]
+\sinh\alpha\left[\sinh^2u_+-\sinh^2u_-\right],\label{thermal-F}\\
\mathcal G_{\alpha}(a,L)&\equiv
\cosh\alpha\left[\sinh^2u_+-\sinh^2u_-\right]
+\sinh\alpha\left[\sinh^2u_++\sinh^2u_-\right].\label{thermal-G}
\end{align}
The functions $\mathcal F$ and $\mathcal G$ respectively encode the rigid-brane geodesic and its normal response to a transverse displacement.  This form is algebraically equivalent to the usual BTZ expressions but avoids long strings of exponentials and makes the dependence on the brane rapidity transparent.

For later bookkeeping, three elementary entropy contributions are especially useful.  We denote by $D$ a same-boundary dome, by $H$ a wormhole-crossing HM surface, and by $J$ a pair of island/DES surfaces:
\begin{align}
D(x_i,x_j)&\equiv \frac{\ell}{2G_N}\log\!\left[\frac{2\ell^2}{r_h\epsilon}\sinh\!\left(\frac{r_h|x_i-x_j|}{2\ell^2}\right)\right],\label{thermal-D}\\
H(L_i,R_i;t)&\equiv \frac{\ell}{4G_N}\log\!\left[2\left(\frac{\ell^2}{r_h\epsilon}\right)^2
\left(\cosh\frac{r_h(L_i-R_i)}{\ell^2}+\cosh\frac{2r_ht}{\ell^2}\right)\right],\label{thermal-H}\\
J(L_i,R_i)&\equiv \frac{\ell}{4G_N}\log\!\left[\left(\frac{\beta}{\pi\epsilon}\right)^2
\sinh\frac{2\pi L_i}{\beta}\sinh\frac{2\pi R_i}{\beta}\right]
+\frac{\ell}{2G_N}\left[\frac{\rho_0}{\ell}+\log\frac{2\ell\cosh(\rho_0/\ell)}{\epsilon_y}\right]\notag\\
&\hspace{1cm}+\frac{\ell\nu}{4G_N}\left(\coth\frac{2\pi L_i}{\beta}+\coth\frac{2\pi R_i}{\beta}\right)+\mathcal O(\nu^2).\label{thermal-J}
\end{align}
The half-argument in $D$ and the overall factor of $2$ in $H$ follow from the thermal-cylinder geometry.  At fixed endpoints, $D$ and $J$ are time independent, whereas $H$ grows through the $\cosh(2r_ht/\ell^2)$ term.  Every Page transition below can be understood as the point at which the growing cost of one or more $H$ contributions is overtaken by static dome or island contributions \cite{Hartman:2013qma,Deng:2022yll,Basu:2023jtf}.

\subsection{Competing island and HM saddles}
Before treating multipartite regions, consider the elementary two-sided interval
$A=(0_L,L_1)\cup(0_R,R_1)$.  Its entropy has two qualitatively different candidates.  The island saddle attaches each bath endpoint to the nearby EOW brane and is time independent at fixed endpoints.  The HM saddle instead joins the two asymptotic boundaries through the black-hole interior and grows as the wormhole stretches.  The competition between these two objects is the basic Page mechanism that will be embedded repeatedly in the multi-interval configurations of \cref{sec:thermal-reflected}.

\subsubsection*{Island saddle}
The thermal island calculation is most transparent after rewriting BTZ in the same AdS$_2$ slicing that underlies the zero-temperature partial reduction.  Starting from the BTZ coordinates, the required change of variables is
\begin{align}
r&=r_h\cosh\left(\frac{\rho}{\ell}\right)
\sqrt{\coth^2\left(\frac{r_hx}{\ell^2}\right)-\tanh^2\left(\frac{\rho}{\ell}\right)},\notag\\
\phi&=\frac{\ell^2}{r_h}\operatorname{artanh}\left[\tanh\left(\frac{\rho}{\ell}\right)\tanh\left(\frac{r_hx}{\ell^2}\right)\right].\label{AdS2-slicing-map-BTZ}
\end{align}
In these coordinates the three-dimensional metric takes the warped-product form
\begin{align}
\d s^2=\d\rho^2+\frac{r_h^2}{\ell^2}\cosh^2\left(\frac{\rho}{\ell}\right)
\frac{-\d t^2+\d x^2}{\sinh^2\left(\frac{r_hx}{\ell^2}\right)}.\label{BTZ-AdS2-sliced-metric}
\end{align}
Equation~\eqref{BTZ-AdS2-sliced-metric} makes the finite-temperature partial reduction almost identical to the vacuum construction.  Defining
\begin{align}
\d \hat s_2^2\equiv\frac{r_h^2}{\ell^2}
\frac{-\d t^2+\d x^2}{\sinh^2(\kappa x)},
\qquad \hat R=-\frac{2}{\ell^2},
\end{align}
we may write the bulk metric simply as $\d s_3^2=\d\rho^2+\cosh^2(\rho/\ell)\d\hat s_2^2$.  The constant-$\rho_0$ brane therefore carries an AdS$_2$ black-hole metric of radius
\begin{align}
\ell_2=\ell\cosh\left(\frac{\rho_0}{\ell}\right),
\qquad
R^{(2)}=-\frac{2}{\ell_2^2}.
\end{align}
Now reduce the finite strip $0\leq\rho\leq\rho_0+\Delta\rho(x)$.  The zeroth-order integration over the strip produces the topological two-dimensional Einstein term and identifies $G_N^{(2)}=G_N/\rho_0$.  Varying the upper integration limit gives the term linear in the local brane displacement.  After including the EOW-brane boundary term, the result is
\begin{align}
I_{\rm eff}=I_{\rm top}
+\frac{1}{16\pi G_N^{(2)}}\int\d^2x\sqrt{-g^{(2)}}\,
\frac{\Delta\rho(x)}{\rho_0}
\left[R^{(2)}+\frac{2}{\ell_2^2}\right]
+I_{\rm CFT}+\mathcal O(\Delta\rho^2).\label{BTZ-partial-reduction-action}
\end{align}
Using $\ell_2=\ell\cosh(\rho_0/\ell)$, this is precisely the JT action \eqref{2d-action} with $\Phi=\Delta\rho/\rho_0$.  The derivation also clarifies an important point: temperature changes the coordinate patch of the AdS$_2$ slice, but not the local partial-reduction dictionary.  The background equation $R^{(2)}+2/\ell_2^2=0$ is temperature independent, while the state dependence enters through the thermal dilaton profile and the matter correlators.  This is why the same brane fluctuation that generated the vacuum JT mode continues to control the finite-temperature island saddle.

\paragraph{Two-dimensional island calculation.}
Restricting the BTZ slicing to the EOW brane at $\rho=\rho_0$ gives
\begin{align}
\d s^2_\textrm{brane}=\ell^2\cosh^2\left(\frac{\rho_0}{\ell}\right)\d s^2_{\mathrm{AdS}_2}.
\end{align}
The intrinsic AdS$_2$ black-hole metric is conveniently written in light-cone coordinates as
\begin{align}
\d s^2_{\mathrm{AdS}_2}=-\frac{4\pi^2}{\beta^2}
\frac{\d y^+\d y^-}{\sinh^2\!\left[\pi(y^+-y^-)/\beta\right]}
=-\Omega(y)^{-2}\d y^+\d y^-,
\end{align}
where $y^\pm=t\pm y$ and $y>0$ is the intrinsic spatial coordinate on the AdS$_2$ black-hole exterior.  The matter correlator is evaluated in the flat thermal frame, for which the total Weyl factor is
\begin{align}
\hat\Omega\,\Omega(y)=\left|\frac{\beta}{2\pi\ell\cosh(\rho_0/\ell)}
\sinh\frac{2\pi y}{\beta}\right|.
\end{align}
The thermal JT solution carries the dilaton profile \cite{Almheiri:2019yqk}
\begin{align}
\Phi(y)=\frac{2\pi\Phi_r}{\beta}\coth\frac{\pi(y^+-y^-)}{\beta}
=\frac{2\pi\Phi_r}{\beta}\coth\frac{2\pi y}{\beta}.\label{dilaton-thermal}
\end{align}
For the left bath endpoint $L_1$, let $a$ denote a candidate island endpoint on the brane.  The generalized entropy balances the dilaton cost against the thermal matter entropy connecting $L_1$ to $a$,
\begin{align}\label{island-gen-ee}
S_\textrm{gen}(a)=\frac{c}{6}\frac{\rho_0}{\ell}\left(1+\frac{2\pi\Phi_r}{\beta}\coth\frac{2\pi a}{\beta}\right)
+\frac{c}{6}\log\left[\frac{2\beta}{\pi\epsilon}\frac{\ell\cosh(\rho_0/\ell)}{\epsilon_y}
\frac{\sinh^2[\pi(a+L_1)/\beta]}{\sinh(2\pi a/\beta)}\right].
\end{align}
The QES condition therefore reads
\begin{align}
\partial_aS_\textrm{gen}=0
\quad\Longleftrightarrow\quad
\frac{\sinh[\pi(a-L_1)/\beta]}{\sinh[\pi(a+L_1)/\beta]}
=\frac{2\pi\rho_0\Phi_r}{\beta\ell}\frac{1}{\sinh(2\pi a/\beta)}.\label{finite-T-extremal-condition}
\end{align}
In the perturbative regime $\nu\ll1$, the solution connected continuously to the rigid-brane endpoint $a=L_1$ is
\begin{align}\label{location-island-BTZ}
a=L_1+\frac{\beta}{\pi}\nu+\mathcal O(\nu^2).
\end{align}
Evaluating the generalized entropy at this saddle and adding the right-copy contribution gives
\begin{align}
S_\textrm{QES}={}&\frac{c}{6}\log\left[\left(\frac{\beta}{\pi\epsilon}\right)^2
\sinh\frac{2\pi L_1}{\beta}\sinh\frac{2\pi R_1}{\beta}\right]
+\frac{c}{3}\left[\frac{\rho_0}{\ell}+\log\frac{2\ell\cosh(\rho_0/\ell)}{\epsilon_y}\right]\notag\\
&+\frac{c\nu}{6}\left(\coth\frac{2\pi L_1}{\beta}+\coth\frac{2\pi R_1}{\beta}\right)+\mathcal O(\nu^2).
\end{align}
At fixed endpoints this saddle is time independent and therefore supplies the Page plateau.  The linear JT correction is largest when an endpoint is close to the thermal origin, where the dilaton gradient is strongest, and becomes less important as the interval endpoints are moved farther out.

\paragraph{Defect-extremal-surface calculation.}
In the three-dimensional description the same saddle is a DES ending on the EOW brane.  Restricting \cref{AdS2-slicing-map-BTZ} to a brane point labelled by the intrinsic coordinate $a$ gives
\begin{align}\label{EOW-point}
(r,\phi)=\left(
 r_h\cosh\frac{\rho}{\ell}\sqrt{\coth^2\frac{r_ha}{\ell^2}-\tanh^2\frac{\rho}{\ell}},
 -\frac{\ell^2}{r_h}\operatorname{artanh}\left[\tanh\frac{\rho}{\ell}\tanh\frac{r_ha}{\ell^2}\right]
\right).
\end{align}
The regulated geodesic from this brane point to the asymptotic endpoint $(L_1,\ell^2/\epsilon)$ is compactly expressed using \cref{thermal-F},
\begin{align}
\frac{\operatorname{Area}(\Gamma_A)}{4G_N}
=\frac{\ell}{4G_N}\log\left[\frac{\beta}{\pi\epsilon}
\frac{\mathcal F_{\rho/\ell}(a,L_1)}{\sinh(2\pi a/\beta)}\right].\label{thermal-DES-length}
\end{align}
A transverse displacement $\rho=\rho_0+\Delta\rho$ changes this length through the companion function $\mathcal G$.  To first order,
\begin{align}
\frac{\operatorname{Area}(\Gamma_A)}{4G_N}={}&\frac{\ell}{4G_N}\log\left[\frac{\beta}{\pi\epsilon}
\frac{\mathcal F_{\alpha_0}(a,L_1)}{\sinh(2\pi a/\beta)}\right]
+\frac{\Delta\rho}{4G_N}\frac{\mathcal G_{\alpha_0}(a,L_1)}{\mathcal F_{\alpha_0}(a,L_1)}
+\mathcal O(\Delta\rho^2).\label{thermal-DES-expand}
\end{align}
The ratio $\mathcal G/\mathcal F$ is precisely the normal response of the geodesic to the brane displacement, giving a direct three-dimensional interpretation of the JT correction.  At the rigid saddle $a=L_1$ one has $\mathcal G/\mathcal F=1$ and $\partial_a(\mathcal G/\mathcal F)=0$.  Consequently the endpoint shift and the on-shell entropy through $O(\nu)$ can be obtained from the local near-saddle representative
\begin{align}\label{EE-des-gen}
S_\textrm{DES}^{\rm loc}(a)={}&\frac{\ell}{4G_N}\log\left[\frac{\beta}{\pi\epsilon}
\frac{\mathcal F_{\alpha_0}(a,L_1)}{\sinh(2\pi a/\beta)}\right]
+\frac{\rho_0}{4G_N}\frac{2\pi\Phi_r}{\beta}\coth\frac{2\pi a}{\beta}
+\frac{c}{6}\log\frac{2\ell\cosh\alpha_0}{\epsilon_y},
\end{align}
which is not meant to replace the complete off-shell $O(\Delta\rho)$ expression in \cref{thermal-DES-expand} away from the rigid saddle.  Its stationarity condition, sufficient through first order, is
\begin{align}
\frac{\beta}{2\pi}\partial_a\log\mathcal F_{\alpha_0}(a,L_1)
=\coth\frac{2\pi a}{\beta}+\nu\operatorname{csch}^2\frac{2\pi a}{\beta}+\mathcal O(\nu^2).\label{thermal-DES-extremality}
\end{align}
For a brane far into the AdS$_2$ wedge this reduces to the QES equation \cref{finite-T-extremal-condition}.  At finite $\rho_0$, the endpoint has the perturbative location
\begin{align}\label{BTZ-island-location}
a=L_1+\frac{\beta}{2\pi}\left(1+\tanh\frac{\rho_0}{\ell}\right)\nu+\mathcal O(\nu^2).
\end{align}
Evaluating the DES functional at this endpoint and adding the right-copy contribution yields
\begin{align}\label{island_EE}
S_\textrm{DES}={}&\frac{\ell}{4G_N}\log\left[\left(\frac{\beta}{\pi\epsilon}\right)^2
\sinh\frac{2\pi L_1}{\beta}\sinh\frac{2\pi R_1}{\beta}\right]
+\frac{\ell}{2G_N}\left[\frac{\rho_0}{\ell}+\log\frac{2\ell\cosh(\rho_0/\ell)}{\epsilon_y}\right]\notag\\
&+\frac{\ell\nu}{4G_N}\left(\coth\frac{2\pi L_1}{\beta}+\coth\frac{2\pi R_1}{\beta}\right)+\mathcal O(\nu^2).
\end{align}
The Brown--Henneaux relation makes this identical to the QES result through first order. The finite-temperature geometry exhibits the same distinction between raw slicing labels, projected endpoint data and the off-shell functional. Using \cref{thermal-F}, the rigid-brane geodesic factor may be rewritten as $\mathcal F_{\alpha_0}=e^{\alpha_0}\sinh^2u_++e^{-\alpha_0}\sinh^2u_-$. Comparing the zero-dilaton parts of the thermal DES and QES functionals gives
\begin{align}
S_{\rm DES}^{(0)}(a)-S_{\rm QES}^{(0)}(a)
=\frac{c}{6}\log\!\left[1+e^{-2\alpha_0}\frac{\sinh^2u_-(a,L_1)}{\sinh^2u_+(a,L_1)}\right].
\label{thermal-finite-angle-offshell}
\end{align}
This reduces to \cref{finite-angle-offshell} as $\beta\to\infty$. At $a=L_1$ the numerator vanishes quadratically, so the value and first derivative agree while the second derivative retains finite-angle information. In raw slicing coordinates,
\begin{align}
a_{\rm QES}^{(y)}-a_{\rm DES}^{(y)}
=\frac{\beta}{2\pi}(1-\tanh\alpha_0)\nu+\mathcal O(\nu^2).
\label{thermal-endpoint-mismatch}
\end{align}
The corresponding linear projection is conveniently expressed in the uniformizing coordinate
\begin{align}
F(y)\equiv\sinh(\kappa y),\qquad \kappa\equiv\frac{2\pi}{\beta}.
\end{align}
To first order in the brane displacement we take
\begin{align}
F(Y)=F(y)\left[1+(1-s)\frac{\delta\rho(y)}{\ell}\right]+\mathcal O(\delta\rho^2),
\qquad s=\tanh\alpha_0.
\label{thermal-projection-F}
\end{align}
For $\delta\rho/\ell=\nu\coth(\kappa y)$ this gives
\begin{align}
F(Y)=\sinh(\kappa y)+(1-s)\nu\cosh(\kappa y)+\mathcal O(\nu^2),
\end{align}
so that
\begin{align}
Y=y+\frac{1-s}{\kappa}\nu+\mathcal O(\nu^2).
\label{thermal-projection-shift}
\end{align}
This shift precisely compensates \cref{thermal-endpoint-mismatch}; the projected QES and DES endpoint locations agree through first order. As in the vacuum, this is a natural linearized projection prescription rather than a complete derivation of the finite-angle reduced theory. The nonzero off-shell difference \cref{thermal-finite-angle-offshell} begins quadratically about the rigid saddle and can affect the on-shell entropy at $O(\nu^2)$. The large-angle limit suppresses these genuine higher-order finite-angle terms, while the leading projected endpoint matching does not require $s\to1$.

\subsubsection*{HM saddle}
\paragraph{Field-theory description.}
The competing two-sided saddle is encoded by a thermal twist two-point function with one insertion on each TFD copy.  With the right insertion analytically continued by half of the Euclidean thermal circle, the entropy is
\begin{align}\label{Def-EE}
S_A=\lim_{n\to1}\frac{1}{1-n}
\log\left\langle\Phi_n(L_1,t_0)\Phi_n\!\left(R_1,-t_0+\frac{i\beta}{2}\right)\right\rangle.
\end{align}
This correlator grows with Lorentzian time because the two insertions separate along opposite time directions on the two copies.

\paragraph{Bulk description.}
The dual extremal surface is the HM geodesic joining $(L_1,t_0,\ell^2/\epsilon)$ to $(R_1,-t_0+i\beta/2,\ell^2/\epsilon)$.  Substituting these endpoints into the BTZ embedding-space distance gives, before taking the asymptotic cutoff limit,
\begin{align}
\mathcal L=\ell\operatorname{arccosh}\left[
\frac{\ell^4\cosh\frac{r_h(L_1-R_1)}{\ell^2}-(r_h^2\epsilon^2-\ell^4)
\cosh\frac{r_h(t_L-t_R)}{\ell^2}}{r_h^2\epsilon^2}\right].
\end{align}
Using $t_L=t_0$ and $t_R=-t_0+i\beta/2$, the regulated entropy becomes
\begin{align}
S_{\rm worm}(t_0)=\frac{\ell}{4G_N}\log\left[
2\left(\frac{\ell^2}{r_h\epsilon}\right)^2
\left(\cosh\frac{r_h(L_1-R_1)}{\ell^2}+\cosh\frac{2r_ht_0}{\ell^2}\right)\right].
\end{align}
The second hyperbolic cosine is the origin of the linear late-time growth: for $r_ht_0/\ell^2\gg1$, the logarithm grows as $2r_ht_0/\ell^2$.  This is the geometric manifestation of the stretching wormhole and is the only source of time dependence in the elementary Page competition.

\subsection*{Single-interval Page transition}
The Page time is obtained by equating the time-growing wormhole entropy with the static island entropy.  Using \cref{thermal-CQ,B0-def,island_EE}, the solution through first order in the brane fluctuation is
\begin{align}
t_e=\frac{\ell^2}{2r_h}\operatorname{arccosh}\left[
\frac{2\ell^2\mathcal B_0^2}{\epsilon_y^2}
\sinh\frac{r_hL_1}{\ell^2}\sinh\frac{r_hR_1}{\ell^2}
\left(1+\nu\mathcal Q_1\right)-\mathcal C_1\right]+\mathcal O(\nu^2).
\end{align}
The sign of the $\nu$ correction is physically transparent: a positive fluctuation increases the generalized cost of the island saddle, so the growing wormhole must evolve for longer before the island becomes favorable.  Thus, in this perturbative regime the JT mode delays the Page transition.

\begin{figure}[ht]
\centering
\includegraphics[width=0.56\linewidth]{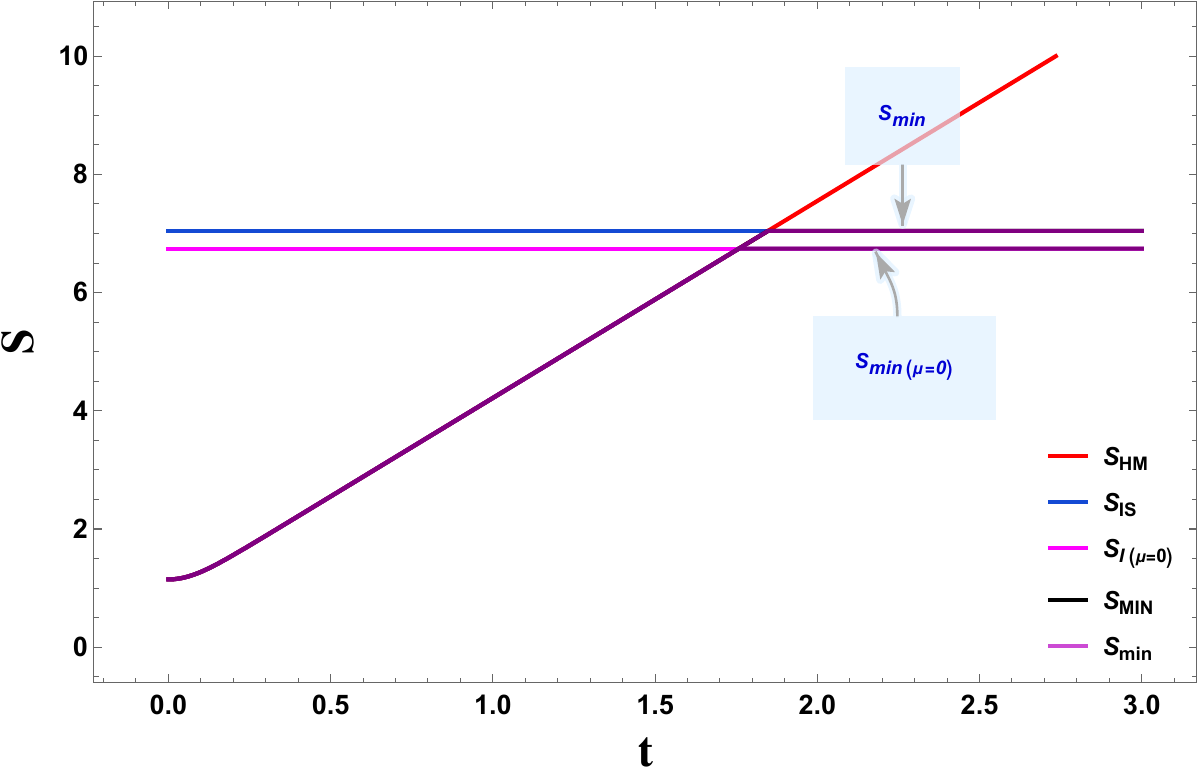}
\caption{Elementary Page competition for a two-sided interval in the eternal BTZ geometry.  The wormhole-crossing HM entropy grows with boundary time, while the island/DES entropy is stationary at fixed endpoints.  The physical entropy is the lower envelope of the two branches, and their intersection gives the Page time.  The same competition between time-growing and stationary building blocks reappears in every multi-interval curve in section~\ref{sec:thermal-reflected}.}
\label{fig:single-page-btz}
\end{figure}
Figure~\ref{fig:single-page-btz} illustrates the mechanism directly.  Before the crossing, the entanglement wedge connects the two asymptotic boundaries through the black-hole interior.  After the crossing, it is more economical for the homology surface to terminate on the brane, and the island reconstructs the degrees of freedom that would otherwise have been encoded deep in the wormhole.  Reflected entropy will refine this story by asking how the wedge is partitioned internally before and after the outer entropy transition.

\section{Thermal reflected entropy in the eternal black-hole geometry}\label{sec:thermal-reflected}
We now turn from the elementary two-sided Page transition to the internal correlation structure of multipartite radiation regions.  The setup is the thermofield-double state of the BTZ/BCFT system reviewed in \cref{sec:BTZ-review}.  For both adjacent and disjoint pairs of subregions, we first determine the dominant entanglement wedge of $A\cup B$ and then extremize the reflected-entropy cross section inside that wedge.  This second minimization is the main source of new physics: a change in the cross-section topology can occur within a fixed entanglement saddle and therefore need not coincide with the Page transition of the union.  The analysis builds on thermal BCFT and double-holographic reflected-entropy constructions \cite{Li:2021dmf,Ling:2021vxe,Basu:2023jtf,Basu:2025sqk,Dey:2025dlj}, but keeps the leading dynamical JT correction generated by the fluctuating brane.

\subsection{Adjacent subsystems}
We take adjacent regions $A=(L_1,L_2)\cup(R_1,R_2)$ and $B=(L_2,L_3)\cup(R_2,R_3)$, with the same ordering on the two TFD boundaries.  Four outer entanglement saddles can compete: two wormhole-crossing surfaces, a mixed island/wormhole configuration, two same-boundary domes, and a fully island-supported configuration.  These saddles determine the allowed homology region, but they do not by themselves determine $S_R$.  Inside each wedge the reflected-entropy cross section can still choose among inequivalent endpoints, and the dominant choice depends on the relative sizes of $A$ and $B$.

On the CFT side all of these possibilities are encoded in different large-$c$ channels of the same six-point replica correlation function.  We denote the bath insertions on the left copy by $P_i$ and those on the right copy by $Q_i$.  Before selecting a channel, the effective reflected entropy is represented by
\begin{align}\label{adj-SR-def}
    S_R^{\rm eff}(A:B)= \lim_{m,n \to 1}\frac{1}{1-n}\log \frac{\langle \sigma_{g^{}_A}({P_1})\sigma_{g^{}_B g_A^{-1}}({P_{2}})\sigma_{g_B^{-1}}({P_{3}})\sigma_{g^{}_B}({Q_{3}})\sigma_{g^{}_A g_B^{-1}}({Q_{2}})\sigma_{g_A^{-1}}({Q_{1}})\rangle }{ \langle \sigma_{g^{}_m}({P_1}) \sigma_{g_m^{-1}}({P_{3}})\sigma_{g^{}_m}({Q_{3}}) \sigma_{g_m^{-1}}({Q_{1}}) \rangle^n },
\end{align}
where $P_i= (L_i, t_0,\frac{\ell^2}{\epsilon})$ and $Q_i= (R_i, -t_0+\frac{i \beta}{2},\frac{\ell^2}{\epsilon})$.

\subsubsection{Two wormhole-crossing HM surfaces}\label{sec:thermal-adj-two-wormholes}
\begin{figure}[ht]
    \centering
    \includegraphics[width=0.55\linewidth]{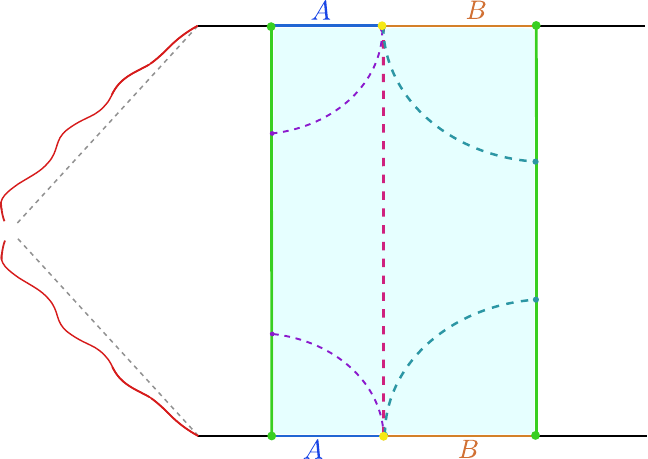}
    \caption{Schematic diagram of the distinct reflected-entropy, or equivalently bulk-EWCS, saddles between subsystems $A$ and $B$ when the entropy of $A\cup B$ is supported by two wormhole-crossing HM surfaces.  The candidate cross sections are indicated by dashed curves.}
    \label{fig:-FT-adj-2HM}
\end{figure}
When both adjacent regions extend sufficiently far from the brane, the extremal surface for the union contains two independent wormhole-crossing HM geodesics, one associated with the outer endpoint pair $(L_1,R_1)$ and one with $(L_3,R_3)$. These HM geodesics are shown as solid curves in \cref{fig:-FT-adj-2HM}. Each contribution grows with $t_0$, so this saddle is naturally favored at early times and eventually becomes expensive.  Using the regulated HM length \cref{HM-length}, the union entropy is
\begin{align}\label{HM_EE}
    S_1=&\frac{c}{6}~  \log \left[2\left(\frac{\ell^2}{ r_h \epsilon }\right)^2 \left(\cosh \frac{r_h \left(L_1-R_1\right)}{\ell^2}+\cosh \frac{2 r_h t_0}{\ell^2}\right)\right]\notag\\
    &+ \frac{c}{6} ~ \log \left[2\left(\frac{\ell^2}{r_h \epsilon }\right)^2 \left(\cosh \frac{r_h \left(L_3-R_3\right)}{\ell^2}+\cosh \frac{2 r_h t_0}{\ell^2}\right)\right].
\end{align}
The two logarithms have the same time dependence but retain the independent left--right offsets $L_i-R_i$.  Consequently an asymmetric choice of endpoints can make one wormhole-crossing component substantially more expensive than the other even before any Page transition occurs.  Reflected entropy can resolve this asymmetry because its cross section may couple preferentially to one of the two components.

The outer entanglement surface does not touch the brane in this saddle, so there is no lower-dimensional reflected-island separator and hence no reflected-island generalized-area contribution in \cref{SR-bdy}.  The higher-dimensional formula \cref{SR-bulk} still contains the ordinary EWCS area, which is the geometric contribution evaluated below.  Nevertheless, the entanglement wedge admits three inequivalent internal separations.  The cross section can run directly across the wormhole, or it can terminate on the HM surface associated with whichever adjacent party is smaller.  These alternatives have the same outer homology but probe different parts of the canonical purification, which is why a reflected-entropy transition may occur without any change of the entropy saddle of $A\cup B$.  We now evaluate the three possibilities in turn from the replica and bulk viewpoints.

\subsubsection*{Cross section traversing the wormhole}
\paragraph{Boundary description.} For the genuinely wormhole-crossing cross section, the large-$c$ channel pairs the middle twists directly across the two TFD copies.  The six-point numerator is, in the dominant large-$c$ channel, approximated by three two-point functions, while the denominator has the corresponding dominant-channel normalization factorization.  After the cancellations, only the two-point function connecting the middle endpoints remains:
\begin{align}
    S_R^{\text{bdy}}(A:B)= \lim_{m,n \to 1}\frac{1}{1-n}\log \langle \sigma_{g^{}_B g_A^{-1}}({P_{2}}) \sigma_{g^{}_A g_B^{-1}}({Q_{2}}) \rangle.
\end{align}
The thermal two-point function depends only on the invariant separation between the middle endpoints $P_2$ and $Q_2$.  After the replica continuation, the wormhole-crossing branch is
\begin{align}\label{sr-thermal-adj-wormhole}
    S_R^{\text{bdy}}({A:B)}=\frac{c}{3}  \log \left[2\left(\frac{\ell^2}{ r_h \epsilon }\right)^2 \left(\cosh \frac{r_h \left(L_2-R_2\right)}{\ell^2}+\cosh \frac{2 r_h t_0}{\ell^2}\right)\right].
\end{align}
The expression contains exactly the same thermal invariant as an ordinary HM geodesic, but with the reflected-entropy normalization $c/3$.  Hence this branch grows linearly at late times and directly tracks correlations whose canonical purification continues to extend through the black-hole interior.

\paragraph{Bulk description.} The bulk realization is the HM cross section anchored at the middle endpoints $L_2$ and $R_2$, shown as a dashed curve in \cref{fig:-FT-adj-2HM}.  It is the unique geodesic separator that traverses the wormhole and is compatible with the present outer entanglement wedge.  Evaluating the regulated BTZ length on these two middle endpoints therefore gives twice the HM entropy contribution,
\begin{align}
    S_R^{\text{bulk}}(A:B)=\frac{\ell}{2 G_N}  \log \left[2\left(\frac{\ell^2}{ r_h \epsilon }\right)^2 \left(\cosh \frac{r_h \left(L_2-R_2\right)}{\ell^2}+\cosh \frac{2 r_h t_0}{\ell^2}\right)\right].
\end{align}
Using the Brown--Henneaux relation $c=3\ell/(2G_N)$, the boundary and bulk expressions coincide exactly.  This result grows at the same late-time rate as a wormhole-crossing geodesic, so it directly measures the part of the canonical purification that continues to probe the black-hole interior.  In the Page-curve language of \cite{Basu:2023jtf}, this is the most nonlocal reflected-entropy branch: once a same-boundary or island-supported cross section becomes cheaper, the physical $S_R$ can depart from this growth even if part of the union wedge still crosses the wormhole.
 
\subsubsection*{Cross section controlled by the smaller subsystem $A$}

\paragraph{Boundary description.}
If $A$ is the smaller adjacent component, the optimal cross section no longer follows the middle wormhole bridge.  The corresponding replica channel isolates the degrees of freedom associated with $B$ in a two-point factor and leaves a four-point block that encodes the cross section controlled by $A$.  In the dominant large-$c$ channel, the numerator of \cref{adj-SR-def} has the leading factorized form
\begin{align}
 & \langle \sigma_{g^{}_A}({P_1})\sigma_{g^{}_B g_A^{-1}}({P_{2}})\sigma_{g_B^{-1}}({P_{3}})\sigma_{g^{}_B}({Q_{3}})\sigma_{g^{}_A g_B^{-1}}({Q_{2}})\sigma_{g_A^{-1}}({Q_{1}})\rangle \notag\\&\quad\quad \simeq \langle \sigma_{g^{}_A}({P_1})\sigma_{g^{}_B g_A^{-1}}({P_{2}})\sigma_{g^{}_A g_B^{-1}}({Q_{2}})\sigma_{g_A^{-1}}({Q_{1}})\rangle \langle \sigma_{g_B^{-1}}({P_{3}})\sigma_{g^{}_B}({Q_{3}}) \rangle.
\end{align}
The semiclassical channel identified in \cite{Basu:2023jtf} resolves this four-point block into two three-point blocks sharing the outer twist insertion.  This dominant-channel block decomposition is the boundary counterpart of allowing the common perpendicular to end on the HM support rather than crossing the whole wormhole:
\begin{align}
    &\langle \sigma_{g^{}_A}({P_1})\sigma_{g^{}_B g_A^{-1}}({P_{2}})\sigma_{g^{}_A g_B^{-1}}({Q_{2}})\sigma_{g_A^{-1}}({Q_{1}})\rangle \notag\\ &\quad \quad \simeq \langle \sigma_{g^{}_A}({P_1})\sigma_{g^{}_B g_A^{-1}}({P_{2}})\sigma_{g_A^{-1}}({Q_{1}})  \rangle \langle \sigma_{g^{}_A}({P_1}) \sigma_{g^{}_A g_B^{-1}}({Q_{2}})\sigma_{g_A^{-1}}({Q_{1}})\rangle.
\end{align}
The denominator has the corresponding dominant-channel normalization decomposition.  Cancelling the spectator factors leaves two thermal three-point blocks,
\begin{align}
    S_R^{\rm bdy}(A:B)= &\lim_{m,n \to 1}\frac{1}{1-n}\log \frac{\langle \sigma_{g^{}_A}({P_1})\sigma_{g^{}_B g_A^{-1}}({P_{2}})\sigma_{g_A^{-1}}({Q_{1}})  \rangle}{\langle \sigma_{g^{}_m}({P_1}) \sigma_{g_m^{-1}}({Q_{1}})\rangle^n} \notag\\
    &+ \lim_{m,n \to 1}\frac{1}{1-n}\log \langle \sigma_{g^{}_A}({P_1}) \sigma_{g^{}_A g_B^{-1}}({Q_{2}})\sigma_{g_A^{-1}}({Q_{1}})\rangle
\end{align}
Using the appropriate conformal factors and the thermal three-point twist correlator gives a compact expression.  We use $\kappa=r_h/\ell^2=2\pi/\beta$ to display the thermal separations:
\begin{align}\label{sr-thermal-adj-A-controlled}
S_R^{\mathrm{bdy}}(A:B)=&\frac{c}{3}\log\!\left[\frac{4\ell^2}{r_h\epsilon}
\sinh\!\left(\frac{\kappa L_{21}}{2}\right)
\sqrt{\frac{\cosh[\kappa(L_2-R_1)]+\cosh(2\kappa t_0)}
{\cosh[\kappa(L_1-R_1)]+\cosh(2\kappa t_0)}}\right]\notag\\
&+\frac{c}{3}\log\!\left[\frac{4\ell^2}{r_h\epsilon}
\sinh\!\left(\frac{\kappa R_{21}}{2}\right)
\sqrt{\frac{\cosh[\kappa(R_2-L_1)]+\cosh(2\kappa t_0)}
{\cosh[\kappa(R_1-L_1)]+\cosh(2\kappa t_0)}}\right].
\end{align}
The half-arguments in the same-boundary $\sinh$ functions are fixed by the thermal cylinder map, $\sinh(\pi\Delta x/\beta)=\sinh(\kappa\Delta x/2)$.  They are essential for the zero-temperature limit and for agreement with the bulk result \cite{Basu:2023jtf,Dey:2025dlj}.  Physically, each logarithm is a boundary-to-HM common-perpendicular length: the factor $\sinh(\kappa L_{21}/2)$ measures the size of the smaller adjacent interval, while the square-root ratio records how the common perpendicular is displaced by the asymmetric location of the supporting wormhole surface.  This makes the branch sensitive to left--right asymmetry even though the outer saddle contains two HM components.

\paragraph{Bulk description.}

In the bulk description, the cross section connects the common endpoint of the subsystems to the HM surface anchored at $L_1$ and $R_1$, as shown by the dashed curves in \cref{fig:-FT-adj-2HM}.  When the two anchoring positions differ, the wormhole-crossing geodesic is not a constant-$x$ curve; the full embedding-space parametrization is therefore required before the endpoint of the reflected-entropy cross section is extremized.

Let $\gamma(P_1,Q_1)$ denote the HM geodesic.  For this invariant formula the endpoints $A,B$ are represented by boundary null rays $p_A,p_B$, whereas $P$ is kept at the regulated bulk point $X_P$.  With the scale-invariant combinations $\zeta_{PA}=-X_P\!\cdot p_A/\ell$, $\zeta_{PB}=-X_P\!\cdot p_B/\ell$ and $\zeta_{AB}=-p_A\!\cdot p_B$, the minimal distance is
\begin{align}
 \cosh\frac{d(P,\gamma_{AB})}{\ell}
 =\sqrt{\frac{2\zeta_{PA}\zeta_{PB}}{\zeta_{AB}}}.\label{point-to-HM-invariant}
\end{align}
Rescaling either boundary null representative cancels between numerator and denominator; the UV regulator enters only through $X_P$.
For the left and right copies this yields
\begin{align}
\frac{\mathcal L_L}{\ell}=&\log\!\left[\frac{4\ell^2}{r_h\epsilon}
\sinh\!\left(\frac{\kappa L_{21}}{2}\right)
\sqrt{\frac{\cosh[\kappa(L_2-R_1)]+\cosh(2\kappa t_0)}
{\cosh[\kappa(L_1-R_1)]+\cosh(2\kappa t_0)}}\right],\label{generic-HM-left}\\
\frac{\mathcal L_R}{\ell}=&\log\!\left[\frac{4\ell^2}{r_h\epsilon}
\sinh\!\left(\frac{\kappa R_{21}}{2}\right)
\sqrt{\frac{\cosh[\kappa(R_2-L_1)]+\cosh(2\kappa t_0)}
{\cosh[\kappa(R_1-L_1)]+\cosh(2\kappa t_0)}}\right].\label{generic-HM-right}
\end{align}
After evaluating the two common-perpendicular lengths and adding the effective contribution, the reflected entropy takes the form
\begin{align}
 S_R^{\mathrm{bulk}}(A:B)=\frac{\mathcal L_L+\mathcal L_R}{2G_N},
\end{align}
which agrees exactly with \cref{sr-thermal-adj-A-controlled}.  The special constant-$x$ HM profile at the end of the appendix applies only for $x_L=x_R$; using it in a generic asymmetric configuration would discard the spatial tilt of the supporting wormhole geodesic and therefore change the common-perpendicular invariant.  In the symmetric limit the result reduces to the corresponding braneworld expression \cite{Basu:2025sqk}.

\subsubsection*{Cross section controlled by the smaller subsystem $B$}

\paragraph{Boundary description.}
When subsystem $B$ is smaller than $A$, the relevant OPE channel is the mirror of the previous one.  The outer $A$ sector becomes a spectator two-point function, while the nontrivial four-point block is built from the twists surrounding $B$.  This is the boundary signature that the common perpendicular is controlled by the right-hand adjacent component rather than by the full wormhole.  Explicitly, in the dominant large-$c$ mirror channel the numerator of \cref{adj-SR-def} has the leading factorized form
\begin{align}
 & \langle \sigma_{g^{}_A}({P_1})\sigma_{g^{}_B g_A^{-1}}({P_{2}})\sigma_{g_B^{-1}}({P_{3}})\sigma_{g^{}_B}({Q_{3}})\sigma_{g^{}_A g_B^{-1}}({Q_{2}})\sigma_{g_A^{-1}}({Q_{1}})\rangle \notag\\&\quad\quad \simeq \langle \sigma_{g_A^{}}({P_{1}})\sigma_{g^{-1}_A}({Q_{1}})\rangle\,
\langle \sigma_{g^{}_B g_A^{-1}}({P_{2}}) \sigma_{g^{-1}_B}({P_3})\sigma_{g^{}_A g_B^{-1}}({Q_{2}})\sigma_{g_B^{}}({Q_{3}})\rangle.
\end{align}
The large-$c$ block then resolves into two three-point functions in exactly the channel obtained by exchanging $A\leftrightarrow B$ in the smaller-$A$ calculation \cite{Basu:2023jtf}.  The normalization correlator has the same dominant-channel factorization, so the spectator factors cancel.  To display the surviving channel without obscuring its structure, define the normalized three-point blocks
\begin{align}
\mathcal R_L^{(B)}&\equiv
\frac{\langle \sigma_{g_B}(Q_3)\sigma_{g_Bg_A^{-1}}(P_2)\sigma_{g_B^{-1}}(P_3)\rangle}
{\langle \sigma_{g_m^{-1}}(P_3)\sigma_{g_m}(Q_3)\rangle^n},\notag\\
\mathcal R_R^{(B)}&\equiv
\langle \sigma_{g_B}(Q_3)\sigma_{g_Ag_B^{-1}}(Q_2)\sigma_{g_B^{-1}}(P_3)\rangle .
\end{align}
The boundary reflected entropy in this channel is therefore
\begin{align}
S_R^{\rm bdy}(A:B)=\lim_{m,n\to1}\frac{1}{1-n}
\left[\log\mathcal R_L^{(B)}+\log\mathcal R_R^{(B)}\right].
\end{align}
Evaluating the two thermal three-point functions with their cylinder Weyl factors gives the explicit mirror branch
\begin{align}\label{sr-thermal-adj-B-controlled}
     S_R^{\rm bdy}(A:B)=& \frac{c}{3} \log \left[ \frac{4 \ell^2}{r_h \epsilon} \sinh \left(\frac{r_h L_{32}}{2\ell^2} \right)\sqrt{\frac{\cosh \frac{r_h(L_2-R_3)}{\ell^2}+ \cosh \frac{2 r_h t_0}{\ell^2}}{\cosh\frac{r_h (L_3-R_3)}{\ell^2}+ \cosh \frac{2 r_h t_0 }{\ell^2}}} \right]\notag\\
    &+\frac{c}{3} \log \left[ \frac{4 \ell^2}{r_h \epsilon} \sinh \left( \frac{r_h R_{32}}{2\ell^2} \right)\sqrt{\frac{\cosh \frac{r_h(R_2-L_3)}{\ell^2}+ \cosh \frac{2 r_h t_0}{\ell^2}}{\cosh\frac{r_h (R_3-L_3)}{\ell^2}+ \cosh \frac{2 r_h t_0 }{\ell^2}}} \right].
\end{align}
The two logarithms have the same geometric interpretation as in the smaller-$A$ branch, but with the supporting HM surface anchored at $(L_3,R_3)$.  In particular, the same-boundary chord factors now measure the size of $B$, while the square-root ratios record the asymmetric displacement of the common perpendicular relative to that outer wormhole-crossing geodesic.  The result is therefore related to the smaller-$A$ answer by the physical exchange of the two adjacent parties, not by imposing any left--right symmetry.

\paragraph{Bulk description.}
The bulk cross section is the mirror of the smaller-$A$ saddle: it joins the endpoint $P_2$ (and $Q_2$ on the second copy) to the general HM surface anchored at $(P_3,Q_3)$, depicted as dashed curves in \cref{fig:-FT-adj-2HM}.  It is important not to use the constant-$x$ specialization of the HM profile unless $L_3=R_3$. Using the general embedding-space invariant in \cref{point-to-HM-invariant}, with $(P_1,Q_1)$ replaced by $(P_3,Q_3)$, gives
\begin{align}\label{sr-thermal-adj-B-controlled-bulk}
S_R^{\rm bulk}(A:B)=&\frac{\ell}{2G_N}\log\!\left[\frac{4\ell^2}{r_h\epsilon}\sinh\!\frac{r_hL_{32}}{2\ell^2}
\sqrt{\frac{\cosh\frac{r_h(L_2-R_3)}{\ell^2}+\cosh\frac{2r_ht_0}{\ell^2}}{\cosh\frac{r_h(L_3-R_3)}{\ell^2}+\cosh\frac{2r_ht_0}{\ell^2}}}\right]\notag\\
&+\frac{\ell}{2G_N}\log\!\left[\frac{4\ell^2}{r_h\epsilon}\sinh\!\frac{r_hR_{32}}{2\ell^2}
\sqrt{\frac{\cosh\frac{r_h(R_2-L_3)}{\ell^2}+\cosh\frac{2r_ht_0}{\ell^2}}{\cosh\frac{r_h(R_3-L_3)}{\ell^2}+\cosh\frac{2r_ht_0}{\ell^2}}}\right].
\end{align}
After $c=3\ell/(2G_N)$ this agrees term by term with the boundary result.  The half-arguments of the thermal $\sinh$ factors follow from $r_h=2\pi\ell^2/\beta$ and are essential for this matching.  Exchanging the two adjacent parties maps the two common-perpendicular geometries into one another while leaving the full HM invariant intact, as required by the $A\leftrightarrow B$ symmetry.

\subsubsection{Island-supported inner pair with a remaining HM bridge}\label{sec:thermal-adj-mixed-island-wormhole}
\begin{figure}[ht]
    \centering
    \includegraphics[width=0.55\linewidth]{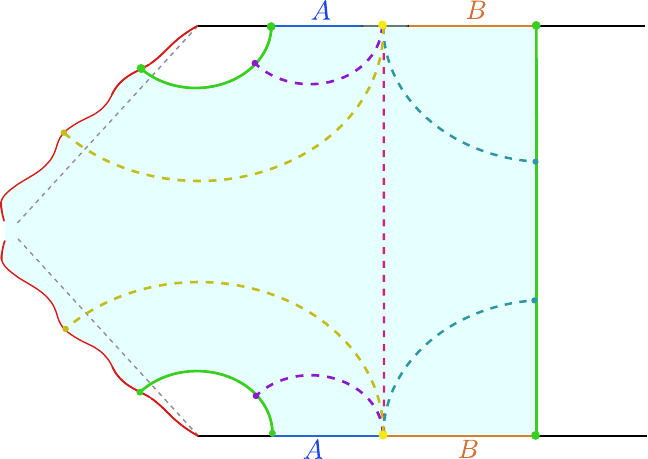}
    \caption{Schematic illustration of the distinct reflected-entropy, or equivalently bulk-EWCS, saddles between $A$ and $B$ when the entropy of $A\cup B$ is supported by an island pair together with one remaining HM surface.  Candidate cross sections are shown by dashed curves.}
    \label{fig:-FT-adj-1HM}
\end{figure}
In this mixed saddle subsystem $A$ lies close enough to the gravitating interface to have already acquired an island, whereas subsystem $B$ still extends far enough into the bath for its outer surface to cross the wormhole.  The union therefore combines a stationary island contribution with a time-growing HM contribution, depicted by the outer geodesics in \cref{fig:-FT-adj-1HM}.  In the building-block notation of \cref{thermal-H,thermal-J}, this decomposition is
\begin{align}
  S_2=J(L_1,R_1)+H(L_3,R_3;t_0),\label{adj-EE-S2-block}
\end{align}
Expanding the compact expression into its island and HM pieces makes the cutoff dependence and the first-order brane correction manifest,
\begin{align}
S_2=&\frac{\ell}{4G_N}\log\!\left[
\left(\frac{\beta}{\pi\epsilon}\right)^2
\sinh\frac{2\pi L_1}{\beta}\sinh\frac{2\pi R_1}{\beta}\right]\notag\\
&+\frac{\ell}{2G_N}\left(\frac{\rho_0}{\ell}+\log\frac{2\ell\cosh(\rho_0/\ell)}{\epsilon_y}\right)
+\frac{\ell\nu}{4G_N}\left(\coth\frac{2\pi L_1}{\beta}+\coth\frac{2\pi R_1}{\beta}\right)\notag\\
&+\frac{\ell}{4G_N}\log\!\left[
\frac{2}{(\kappa\epsilon)^2}
\left(\cosh\!\big[\kappa(L_3-R_3)\big]+\cosh(2\kappa t_0)\right)\right]
+\mathcal O(\nu^2).
\end{align}
The decomposition \cref{adj-EE-S2-block} is also useful below because it makes the Page-transition equations transparent: the transition from the mixed island--wormhole saddle to the fully island-supported saddle is simply the elementary replacement $H(L_3,R_3;t)\to J(L_3,R_3)$.  More importantly, the mixed outer geometry allows four inequivalent internal separators.  Depending on the relative interval sizes, the reflected-entropy cross section can remain wormhole-supported, end on the $A$-island surface, reach the EOW brane and split the reflected island, or be controlled by the opposite party.  These alternatives distinguish whether the mixed-state purification is predominantly stored in the black-hole interior, in the island geometry, or directly in the brane degrees of freedom.  We evaluate them below.

\subsubsection*{Cross section traversing the remaining wormhole}  
The first internal saddle, shown by dashed curves in \cref{fig:-FT-adj-1HM}, preserves the same wormhole-crossing separator that was available before the outer $A$-pair underwent its island transition.  Although the entanglement wedge of $A\cup B$ has changed globally, the local separator still joins the middle endpoints across the two TFD copies.  Since the local extremization problem and the surviving replica block are unchanged, its reflected entropy is still given by \cref{sr-thermal-adj-wormhole}, provided the same wormhole-pairing block remains the dominant large-$c$ channel inside the mixed outer wedge.  This provides a useful example in which the union has already developed an island while the optimal reflected-entropy cross section continues to probe the black-hole interior.

\subsubsection*{Cross section ending on the EOW brane}  
\paragraph{Boundary description.} 

When the adjacent regions are sufficiently balanced, the shortest separator can reach the EOW brane instead of ending on one of the supporting island surfaces.  This is the genuinely gravitational reflected-entropy saddle of the mixed wedge: the separator itself partitions the island of $A\cup B$, so the generalized reflected entropy contains both an explicit brane-area contribution and a matter reflected entropy.  Because the two TFD copies contribute additively, it is sufficient to derive the left-copy term and then add its right-copy image.  In the brane-ending semiclassical channel, the spectator twists have already been fused into the outer island/HM blocks, and the nontrivial matter contribution is the thermal two-point twist correlator between the bath junction $P_2$ and the brane endpoint $a_{P_2}$.  This is the leading block associated with the brane-ending EWCS topology and is used only in the region where that channel is dominant,
\begin{align}
    S_R^{\rm eff}(A\cup I_{S_{R}}(A):B \cup I_{S_{R}}(B))= &\lim_{{m,n} \to 1}\frac{1}{1-n} \log       \langle \sigma_{g^{}_A g_B^{-1}}(P_2) \sigma_{g^{}_A g_B^{-1}}( a_{P_{2}} )\rangle,\notag\\
    &=\frac{c}{3} \log\left[\frac{2\beta}{\pi\epsilon}\frac{\ell\cosh\left(\frac{\rho_0}{\ell}\right)}{\epsilon_y}\frac{\sinh ^2\frac{\pi  (L_2+a_2)}{\beta }}{\sinh\frac{2 \pi  a_2}{\beta }}\right].
\end{align}
The geometric part is the generalized area of the reflected-island separator on the brane.  To first order in the transverse fluctuation it is
\begin{align}
    \operatorname{Area}(\Gamma)= \frac{\rho_0}{\ell}\left(1+\frac{ \nu \ell}{ \rho_0}\coth\frac{2\pi a_2}{\beta}\right)
\end{align}
Combining the matter and geometric pieces gives a one-variable generalized reflected-entropy functional,
\begin{align}\label{sr-thermal-adj-brane-bdy}
    S_{R}^{\text{bdy }}(A:B)=\frac{c}{3}\frac{\rho_0}{\ell}\left(1+\frac{ \nu \ell}{ \rho_0}\coth\frac{2\pi a_2}{\beta}\right)+ \frac{c}{3} \log\left[\frac{2\beta}{\pi\epsilon}\frac{\ell\cosh\left(\frac{\rho_0}{\ell}\right)}{\epsilon_y}\frac{\sinh ^2\frac{\pi  (L_2+a_2)}{\beta }}{\sinh\frac{2 \pi  a_2}{\beta }}\right].
\end{align}
The stationarity condition fixes where the separator lands on the fluctuating brane.  Expanding the physical solution for small $\nu$ gives
\begin{align}
    a_2= L_2+ \frac{\beta}{\pi} \nu + \mathcal{O}(\nu^2).
\end{align}
Evaluating the functional on this saddle and adding the right-copy contribution yields
\begin{align}
    S_{R}^{\text{bdy }}(A:B)= &\frac{c}{3}\log\left[\left(\frac{\beta}{\pi\epsilon}\right)^2  \sinh\frac{2\pi L_2}{\beta} \sinh\frac{2\pi R_2}{\beta}  \right]+ \frac{2c}{3} \left(\frac{\rho_0}{\ell}+\log\frac{2\ell\cosh(\rho_0/\ell)}{\epsilon_y}\right)\notag\\
        &+ \frac{c \nu }{3} \left(\coth\frac{2\pi L_2}{\beta}+\coth\frac{2\pi R_2}{\beta}\right) +\mathcal{O}(\nu^2).
\end{align}
The first-order correction is proportional to the sum of two $\coth$ factors and is positive for positive $\nu$.  This sign has a direct geometric interpretation: unlike the island-surface saddle, where the fluctuation merely moves a supporting geodesic, here the cross section itself acquires a brane-area cost.  The result therefore cleanly isolates the dynamical JT contribution to the reflected-island separator.

\paragraph{Bulk description.}
In the bulk description, the extremal curve joins point $P_2$ to a point $a_{P_{2}}$ on the EOW brane in the left TFD copy and also point $Q_2$ to a point $a_{Q_{2}}$ on the EOW brane in the right TFD copy and thereby splits the reflected island into two parts. These extremal curves are shown as dashed curves in \cref{fig:-FT-adj-1HM}.  This is the reflected-entropy analogue of the defect-extremal-surface problem reviewed earlier.  The local DES problem is identical to the single-interval bath-to-brane extremization reviewed in section~\ref{sec:BTZ-review}; only the bath endpoint changes from $(L_1,R_1)$ to $(L_2,R_2)$.  The bulk reflected entropy is therefore
 \begin{align}\label{sr-thermal-adj-brane-bulk}
         S_R^{\rm bulk}(A:B)=&\frac{\ell}{2 G_N}\log\left[\left(\frac{\beta}{\pi\epsilon}\right)^2 \sinh \frac{2\pi L_2}{\beta} \sinh \frac{2\pi R_2}{\beta}\right]+ \frac{\ell}{G_N}  \left(\frac{ \rho_0}{\ell }+\log\frac{2\ell\cosh(\rho_0/\ell)}{\epsilon_y}\right)\notag\\
         &+\frac{ \ell \nu}{2 G_N} \left(\coth\frac{2\pi L_2}{\beta}+\coth\frac{2\pi R_2}{\beta}\right)  +\mathcal{O}(\nu^2).
\end{align}
Using $c=3\ell/(2G_N)$, the bulk result agrees with the boundary answer through $\mathcal O(\nu)$.  The equality is nontrivial because the two descriptions organize the correction differently: the two-dimensional calculation packages it as a dilaton area plus a thermal matter correlator, while the three-dimensional calculation interprets the same term as the response of a defect extremal surface to a transverse brane displacement.

\subsubsection*{Cross section ending on the $A$-island surface}

\paragraph{Boundary description.}
In this case we assume that the subsystem $A$ is the smaller. Hence the cross section now terminates on the $A$-island surface rather than on the EOW brane.  There is consequently no independent brane segment in the reflected-island partition. The reflected entropy for this saddle reduces to the effective reflected entropy $S_{R}^\text{eff} (A: B \cup I_{S_{R}}(B ))$ in the boundary description, which is computed by using the following three-point twist-field correlator
\begin{align}
	S_R^{\rm bdy}(A:B)= S_R^{\rm eff}(A:B\cup I_{S_{R}}(B))=& \lim_{{m,n} \to 1}\frac{1}{1-n}\log \frac{      \langle \sigma_{g^{}_A}(P_1)\sigma_{g^{}_B g_A^{-1}}(P_2)\sigma_{g^{-1}_A}( a_{P_{1}} )\rangle}{ \langle\sigma_{g^{}_m}(P_1)\sigma_{g^{-1}_m}( a_{P_{1}} )\rangle^n}\notag\\
    & +\lim_{{m,n} \to 1}\frac{1}{1-n}\log \frac{      \langle \sigma_{g^{}_A}(Q_1)\sigma_{g^{}_B g_A^{-1}}(Q_2)\sigma_{g^{-1}_A}( a_{Q_{1}} )\rangle}{ \langle\sigma_{g^{}_m}(Q_1)\sigma_{g^{-1}_m}( a_{Q_{1}} )\rangle^n},
\end{align}  
where $a_{P_{1}}$ and $a_{Q_{1}}$ are the endpoints of the $A$-island surfaces on the left and right EOW branes.  These three-point functions are the surviving blocks in the dominant large-$c$ channel obtained after the spectator sector of the six-point correlator is fused according to the island-supported outer topology; this is the same semiclassical channel identified in the corresponding rigid-braneworld calculation \cite{Basu:2023jtf}.  The thermal blocks carry the appropriate bath and brane Weyl factors and are not being used as an exact factorization of the full six-point operator product.  Substituting the perturbative island locations from \cref{location-island-BTZ} and expanding consistently to first order in $\nu$ gives
{\footnotesize\begin{align}\label{sr-thermal-adj-A-island-bdy}
S_R^{\rm bdy}(A:B)=&\frac{c}{3}\log\!\left[\frac{4}{\kappa\epsilon}\,
\frac{\sinh\frac{\kappa L_{21}}{2}\,\sinh\frac{\kappa(L_2+a_L)}{2}}
{\sinh\frac{\kappa(L_1+a_L)}{2}}\right]+(L\rightarrow R),\qquad
 a_L=L_1+\frac{2\nu}{\kappa}+\mathcal O(\nu^2)\notag\\
=&\frac{c}{3}\log\!\left[\frac{4}{\kappa\epsilon}\,
\frac{\sinh\frac{\kappa L_{21}}{2}\,\sinh\frac{\kappa(L_1+L_2)}{2}}
{\sinh(\kappa L_1)}\right]
+\frac{c\nu}{3}\left[\coth\frac{\kappa(L_1+L_2)}{2}-\coth(\kappa L_1)\right]\notag\\
&+\frac{c}{3}\log\!\left[\frac{4}{\kappa\epsilon}\,
\frac{\sinh\frac{\kappa R_{21}}{2}\,\sinh\frac{\kappa(R_1+R_2)}{2}}
{\sinh(\kappa R_1)}\right]
+\frac{c\nu}{3}\left[\coth\frac{\kappa(R_1+R_2)}{2}-\coth(\kappa R_1)\right]+\mathcal O(\nu^2).
\end{align}}

\paragraph{Bulk description.}  In the bulk description, the extremal curves $\Sigma_{AB}$ connect point $L_2$ to a point $(x_1,r_1)$ on the island surface anchored between points $L_1$ and $a_{L_{1}}$ in the left TFD copy and also point $R_2$ to a point $(x_2,r_2)$ on the island surface that joins $R_1$ and $a_{R_{1}}$ in the right TFD copy,  as illustrated by the dashed curves in \cref{fig:-FT-adj-1HM}. Since the entire island is assigned to subsystem $B$, there is no separate leading effective reflected-entropy contribution localized on a reflected-island separator in this channel.  The higher-dimensional contribution is therefore the ordinary EWCS area, i.e. the extremal length of the curves $\Sigma_{AB}$. In the following we explain the computation of the length of $\Sigma_{AB}$ in the left TFD copy and subsequently include the analogous contribution from the right TFD copy to obtain the reflected entropy.
To determine the contact point explicitly, take an arbitrary point $(x_1,t_0,r_1)$ on the supporting island geodesic and connect it to the regulated bath endpoint $(L_2,t_0,\ell^2/\epsilon)$.  The BTZ invariant \cref{BTZ-length} gives the corresponding trial length
\begin{align}\label{btz-adj-boundary-island-length}
\operatorname{Area}[\Sigma_{AB}]={}&\ell\,\operatorname{arccosh}\!\left[
\frac{\ell^2 r_1\cosh\frac{r_h(L_2-x_1)}{\ell^2}
-\sqrt{(\ell^4-r_h^2\epsilon^2)(r_1^2-r_h^2)}}{r_h^2\epsilon}\right]\notag\\
={}&\ell\log\!\left[\frac{2\ell^2}{r_h\epsilon}
\frac{\cosh\frac{r_h(L_2-x_1)}{\ell^2}-m_1\cosh\frac{r_h(x_1-x_{c_1})}{\ell^2}}
{\sqrt{1-m_1^2\cosh^2\frac{r_h(x_1-x_{c_1})}{\ell^2}}}\right].
\end{align}
where, in obtaining the second line, we employed the island-geodesic profile $r_1=r_h/\sqrt{1-m_1^2\cosh^2 y}$ with $y\equiv r_h(x_1-x_{c_1})/\ell^2$.  It is more transparent to keep the extremization in hyperbolic form than to solve it through exponentials.  Writing $\delta\equiv r_h(L_2-x_{c_1})/\ell^2$, stationarity of \cref{btz-adj-boundary-island-length} is equivalent to
\begin{align}
 -\frac{\sinh(\delta-y)+m_1\sinh y}{\cosh(\delta-y)-m_1\cosh y}
 +\frac{m_1^2\cosh y\sinh y}{1-m_1^2\cosh^2y}=0.\label{adj-island-stationarity}
\end{align}
This equation has a unique solution on the physical branch of the island geodesic.  It is precisely the condition that the reflected-entropy geodesic meet the island surface orthogonally.
At the orthogonal intersection, the auxiliary point $x_1$ drops out in favor of the island-geodesic data.  The extremal cross-section length reduces to    
\begin{align}
      \operatorname{Area}[\Sigma_{AB}]=  \ell\,\log\left[\frac{2\ell^2}{r_h\epsilon}  \left(\frac{1-m_1 
        \cosh\frac{r_h(L_2-x_{c_{1}})}{\ell^2}}{\sqrt{1-m_1^2}}\right)\right].
\end{align}
where $(m_1,x_{c_1})$ is fixed by the endpoint conditions collected in \cref{center_radius_island}.  Eliminating these auxiliary parameters produces a form involving only hyperbolic functions.  Defining
\begin{align}
s_i^-\equiv\sinh\frac{r_h(a-L_i)}{2\ell^2},\qquad s_i^+\equiv\sinh\frac{r_h(a+L_i)}{2\ell^2},
\end{align}
the extremized length is
\begin{align}
\operatorname{Area}[\Sigma_{AB}]=\ell\log\left[\frac{4}{\kappa\epsilon}\,
\frac{\cosh(\rho/\ell)(s_1^-s_2^-+s_1^+s_2^+)+\sinh(\rho/\ell)(s_1^+s_2^+-s_1^-s_2^-)}
{\operatorname{csch}\frac{r_hL_{21}}{2\ell^2}\left\{\cosh(\rho/\ell)[(s_1^-)^2+(s_1^+)^2]+\sinh(\rho/\ell)[(s_1^+)^2-(s_1^-)^2]\right\}}\right].\label{adj-island-length-hyperbolic}
\end{align}
This representation makes the reflection-even and reflection-odd dependence on the brane position explicit and is numerically better behaved for large $|\rho|$ than the equivalent exponential form.
Finally, substituting the DES saddle \cref{BTZ-island-location} and the brane fluctuation into the extremized length, and then using the first-order extremization identity \cref{first-order-extremization}, gives a compact left-copy contribution
\begin{align}
\mathcal K_L\equiv{}&\log\!\left[\frac{4}{\kappa\epsilon}\,
\frac{\sinh(\kappa L_{21}/2)\sinh[\kappa(L_1+L_2)/2]}{\sinh(\kappa L_1)}\right]\notag\\
&+\nu\left[\coth\frac{\kappa(L_1+L_2)}{2}-\coth(\kappa L_1)\right],
\qquad \mathcal K_R\equiv \mathcal K_L\big|_{L_i\to R_i}.
\end{align}
The full two-sided reflected entropy is therefore
\begin{align}
S_R^{\rm bulk}(A:B)=\frac{\ell}{2G_N}\left(\mathcal K_L+\mathcal K_R\right)
+\mathcal O(\nu^2).\label{SR-adj2-iii-bulk}
\end{align}
This is identical to \cref{sr-thermal-adj-A-island-bdy} after the Brown--Henneaux relation.  The $\nu$ correction is negative for $L_2>L_1$, in agreement with the zero-temperature limit of the corresponding island-surface saddle in section~\ref{sec:vacuum-reflected}.  The explicit $x_1$ extremization above displays the same invariant geometry; only the final first-order expansion has been simplified.

\subsubsection*{Mirror cross section controlled by subsystem $B$}  
When $B$ is the smaller adjacent component the geometry is obtained by exchanging the two parties.  The optimal cross section ends on the wormhole-crossing surface associated with the outer endpoint of $B$, and the corresponding replica channel is the mirror of the preceding calculation.  The resulting reflected entropy is given by \cref{sr-thermal-adj-B-controlled}.

\subsubsection{Late-time same-boundary dome saddle}
\begin{figure}[ht]
    \centering
    \includegraphics[width=0.55\linewidth]{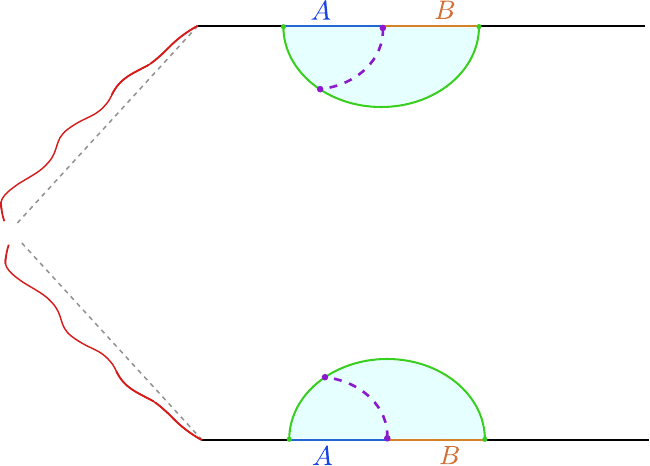}
    \caption{Schematic illustration of the reflected-entropy, or equivalently bulk-EWCS, saddle between $A$ and $B$ when the entropy of $A\cup B$ is supported by same-boundary domes.  The cross section is shown by a dashed curve.}
    \label{fig:-FT-adj-dome}
\end{figure}
For sufficiently small adjacent regions the union is reconstructed independently on each boundary and the wormhole no longer participates.  The entanglement surface consists of one same-boundary dome on the left and its counterpart on the right.  These contributions are time independent, so this saddle is a natural late-time competitor to the growing wormhole-crossing configuration.  In terms of the elementary chord function $D$, the entropy is
\begin{align}\label{adj-EE-S3}
S_3=D(L_1,L_3)+D(R_1,R_3)
=\frac{\ell}{2G_N}\log\!\left[\frac{2\ell^2}{r_h\epsilon}\sinh\frac{r_hL_{31}}{2\ell^2}\right]
+\frac{\ell}{2G_N}\log\!\left[\frac{2\ell^2}{r_h\epsilon}\sinh\frac{r_hR_{31}}{2\ell^2}\right].
\end{align}
This late-time outer saddle contains no brane segment and admits a unique connected separator between the two same-boundary domes.  Hence there is no reflected-island area term: $S_R$ is entirely the matter/geodesic contribution associated with the thermal three-point block on each boundary.  Because all supporting surfaces are time independent, this branch provides the natural saturation value of the reflected entropy after wormhole-supported saddles have become too costly. 

\paragraph{Boundary description.}
In the boundary description, the reflected entropy reduces to $S_R^{\rm eff}(A:B)$ as there is no island cross section. Hence the dominant same-boundary large-$c$ channel approximates the six-point numerator of \cref{adj-SR-def} by two three-point blocks:
\begin{align}
    & \langle \sigma_{g^{}_A}({P_1})\sigma_{g^{}_B g_A^{-1}}({P_{2}})\sigma_{g_B^{-1}}({P_{3}})\sigma_{g^{}_B}({Q_{3}})\sigma_{g^{}_A g_B^{-1}}({Q_{2}}) \sigma_{g_A^{-1}}({Q_{1}})\rangle \notag\\&\quad\quad \simeq \langle \sigma_{g^{}_A}({P_1})\sigma_{g^{}_B g_A^{-1}}({P_{2}})\sigma_{g_B^{-1}}({P_{3}}) \rangle \langle \sigma_{g_A^{-1}}({Q_{1}}) \sigma_{g^{}_A g_B^{-1}}({Q_{2}})  \sigma_{g^{}_B}({Q_{3}}) \rangle.
\end{align}
The normalization correlator has the same leading factorization in that channel.  Dividing by these spectator two-point functions isolates one normalized thermal three-point block on each TFD copy, so the reflected entropy is
\begin{align}
    	S_R^{\rm bdy}(A:B)=& \lim_{{m,n} \to 1}\frac{1}{1-n}\log \frac{      \langle \sigma_{g^{}_A}(P_1)\sigma_{g^{}_B g_A^{-1}}(P_2)\sigma_{g^{-1}_A}( P_3 )\rangle}{ \langle\sigma_{g^{}_m}(P_1)\sigma_{g^{-1}_m}( P_3 )\rangle^n} \notag\\
        &+\lim_{{m,n} \to 1}\frac{1}{1-n}\log \frac{      \langle \sigma_{g^{}_A}(Q_1)\sigma_{g^{}_B g_A^{-1}}(Q_2)\sigma_{g^{-1}_A}( Q_3 )\rangle}{ \langle\sigma_{g^{}_m}(Q_1)\sigma_{g^{-1}_m}( Q_3 )\rangle^n}.
\end{align}
Each normalized three-point block reduces to the logarithm of a ratio of thermal chord distances.  Adding the left- and right-copy contributions gives the time-independent boundary result
\begin{align}
    S_R^{\rm bdy} (A:B)= \frac{c}{3} \log \left[ \frac{4 \ell^2}{r_h \epsilon} \frac{\sinh \frac{r_h L_{21}}{2\ell^2} \sinh \frac{r_h L_{32}}{2\ell^2} }{\sinh \frac{r_h L_{31}}{2\ell^2}} \right]+\frac{c}{3} \log \left[ \frac{4 \ell^2}{r_h \epsilon} \frac{\sinh \frac{r_h R_{21}}{2\ell^2} \sinh \frac{r_h R_{32}}{2\ell^2} }{\sinh \frac{r_h R_{31}}{2\ell^2}} \right].
\end{align}

\paragraph{Bulk description.} The bulk cross section is a boundary-to-dome common perpendicular on each TFD copy.  On the left it joins the regulated junction point $(L_2,t_0,\ell^2/\epsilon)$ to a variable point $(x_1,t_0,r_1)$ on the dome anchored at $(L_1,L_3)$; the right-copy construction is obtained by the half-thermal-circle continuation and the replacement $L_i\to R_i$.  Treating the endpoint on the left dome as variational, \cref{BTZ-length} gives the trial functional
\begin{align}\label{dome-length}
\mathcal{L}=\ell\,\operatorname{arccosh}\!\left[
\frac{\ell^2 r_1\cosh\frac{r_h(L_2-x_1)}{\ell^2}
-\sqrt{(\ell^4-r_h^2\epsilon^2)(r_1^2-r_h^2)}}{r_h^2\epsilon}\right].
\end{align}
Instead of introducing an exponential coordinate for the point on the dome, we parametrize that dome by the signed proper distance $\sigma$ used in \cref{dome-RT-para}.  The extremality condition $\partial_\sigma\mathcal L=0$ is the geometric orthogonality condition between the reflected-entropy cross section and the dome RT surface.  Solving it and substituting back into \cref{dome-length} gives directly
\begin{align}
\mathcal L_{\rm ext}=\ell\log\left[\frac{4}{\kappa\epsilon}\,
\frac{\sinh\frac{r_hL_{21}}{2\ell^2}\sinh\frac{r_hL_{32}}{2\ell^2}}
{\sinh\frac{r_hL_{31}}{2\ell^2}}\right].\label{boundary-to-dome-length}
\end{align}
The three thermal chord factors have a simple interpretation: the numerator measures the distances from the junction point $L_2$ to the two endpoints of the dome, while the denominator removes the overall size of the RT geodesic.  This is the thermal-cylinder analogue of the familiar boundary-to-semicircle distance in Poincar\'e AdS$_3$.
Dividing the extremal common-perpendicular length by $2G_N$ and adding its right-copy image gives the bulk reflected entropy
\begin{align}
    S_R^{\text{bulk}}(A:B)=\frac{\ell}{2G_N}\log\!\left[\frac{4\ell^2}{r_h\epsilon}
\frac{\sinh\frac{r_h L_{21}}{2\ell^2}\sinh\frac{r_h L_{32}}{2\ell^2}}
{\sinh\frac{r_h L_{31}}{2\ell^2}}\right]
+\frac{\ell}{2G_N}\log\!\left[\frac{4\ell^2}{r_h\epsilon}
\frac{\sinh\frac{r_h R_{21}}{2\ell^2}\sinh\frac{r_h R_{32}}{2\ell^2}}
{\sinh\frac{r_h R_{31}}{2\ell^2}}\right].
\end{align}
Using the Brown--Henneaux relation, this is identical to the boundary result.  The half-arguments are required by the thermal cylinder map and also ensure the correct $r_h\to0$ limit.  Since the answer depends only on the three same-boundary chord lengths, it is insensitive to the wormhole interior and to the JT brane fluctuation.  The transition onto this branch therefore has a clean physical meaning: the mixed-state purification has become localized entirely in the exterior thermal geometry.

\subsubsection{Fully island-supported saddle}
\begin{figure}[ht]
    \centering
    \includegraphics[width=0.55\linewidth]{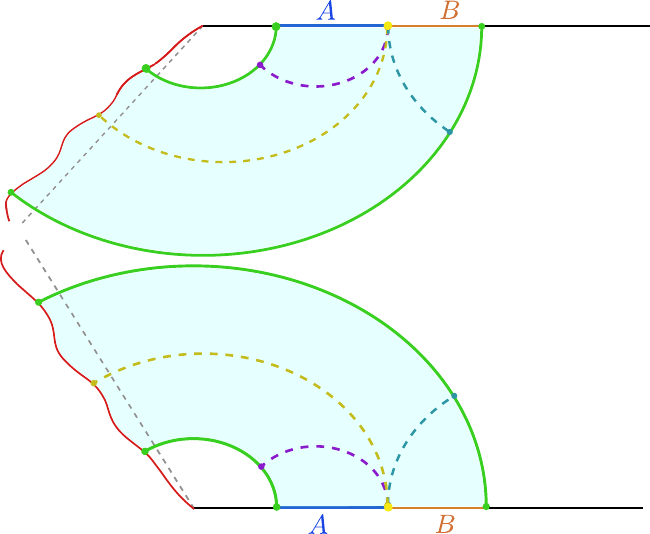}
    \caption{Schematic illustration of the distinct reflected-entropy, or equivalently bulk-EWCS, saddles between $A$ and $B$ when the entropy of $A\cup B$ is supported by a fully island-supported saddle.  Candidate cross sections are shown by dashed curves.}
    \label{fig:-FT-adj-Island}
\end{figure}
When both outer endpoint pairs lie sufficiently close to the brane, the union is supported entirely by island surfaces.  There are four bath-to-brane geodesics, and no part of the entanglement surface crosses the wormhole. These geodesics are shown as solid curves in \cref{fig:-FT-adj-Island}. The compact building-block form of this fully island-supported saddle is
\begin{align}\label{adj-EE-S4}
S_4=J(L_1,R_1)+J(L_3,R_3).
\end{align}
Expanding the two $J$ blocks separates the ordinary thermal entropy, the constant boundary-entropy contribution, and the leading JT correction,
\begin{align}
S_4=&\frac{\ell}{4G_N}\log\!\left[\left(\frac{\beta}{\pi\epsilon}\right)^4
\prod_{x\in\{L_1,L_3,R_1,R_3\}}\sinh\frac{2\pi x}{\beta}\right]
+\frac{\ell}{G_N}\left(\frac{\rho_0}{\ell}+\log\frac{2\ell\cosh(\rho_0/\ell)}{\epsilon_y}\right)\notag\\
&+\frac{\ell\nu}{4G_N}\sum_{x\in\{L_1,L_3,R_1,R_3\}}\coth\frac{2\pi x}{\beta}+\mathcal O(\nu^2).
\end{align}
Every term is time independent, so the fully island-supported outer saddle is the other natural late-time endpoint of the Page evolution.  The $\nu$ correction is nevertheless nontrivial: it records how the two island endpoints respond to the thermal dilaton profile.  Inside this wedge three separators can compete: a cross section that reaches the EOW brane, one ending on the $A$-island surface, and its mirror ending on the $B$-island surface.  The associated boundary formulas are the leading large-$c$ blocks obtained by the same local OPE/BOE pairings as in the mixed island--wormhole wedge; replacing a spectator HM surface by an island surface changes the outer homology class but not the local separator.  They are therefore used only where the corresponding island-supported block is semiclassically dominant.

\subsubsection*{Cross section ending directly on the EOW brane}
The first internal saddle is the brane-ending separator already encountered in the mixed island--wormhole wedge.  The second outer island pair modifies the global homology but leaves the local bath-to-brane extremization at the common junction unchanged.  The reflected entropy is therefore again given by the generalized brane result \cref{sr-thermal-adj-brane-bulk}.  Physically, this is the branch in which the canonical purification is divided directly by a gravitational separator on the EOW brane, even though the entire outer entanglement wedge is already island supported.

\subsubsection*{Cross section ending on the $B$-island surface}

When $B$ is the smaller party, the optimal separator ends on the $B$-island surface.  Since it never reaches the brane, there is no independent lower-dimensional generalized-area term from a reflected-island separator; the boundary answer is the effective reflected entropy of the enlarged regions, while the dual higher-dimensional contribution remains the ordinary EWCS area.  The geometry is obtained from the $A$-controlled island-surface saddle by exchanging the two outer endpoint pairs, so the boundary result is
{\small\begin{align}\label{sr-thermal-adj-B-island-bdy}
      S_R^{\text{bdy}}(A:B)&=  \frac{c}{3} \log \left[ \frac{4 \ell^2}{r_h \epsilon} ~  \frac{\sinh \frac{r_h L_{32}}{2 \ell^2} ~ \sinh \frac{r_h (L_3+L_2)}{2 \ell^2}}{\sinh \frac{ r_h L_3}{\ell^2}}\right]+ \frac{c}{3} \left(\coth \frac{r_h (L_3+L_2)}{2\ell^2} - \coth \frac{r_h L_3}{\ell^2} \right)\nu \notag\\
    &+ \frac{c}{3} \log \left[ \frac{4 \ell^2}{r_h \epsilon} ~  \frac{\sinh \frac{r_h R_{32}}{2 \ell^2} ~ \sinh \frac{r_h (R_3+R_2)}{2 \ell^2}}{\sinh \frac{ r_h R_3}{\ell^2}}\right]+ \frac{c}{3} \left(\coth \frac{r_h (R_3+R_2)}{2\ell^2} - \coth \frac{r_h R_3}{\ell^2} \right)\nu.
\end{align}}
The bulk calculation is the mirror island-surface common perpendicular.  Exchanging $(L_1,R_1)$ with $(L_3,R_3)$ maps the earlier $B$-controlled saddle into the present geometry, and the resulting DES length agrees with \cref{sr-thermal-adj-B-island-bdy} through first order in the brane fluctuation after using the Brown--Henneaux relation.  The fact that the local expression is unchanged while the global entanglement wedge is different is a useful illustration of why the reflected-entropy and union-entropy saddle diagrams contain different information.

\subsubsection*{Cross section ending on the $A$-island surface}
The remaining island-surface saddle is obtained by exchanging the roles of $A$ and $B$ in the previous configuration.  Its cross section terminates on the $A$-island surface and therefore has the same local invariant as the corresponding island-surface saddle of the mixed wedge; the result is \cref{sr-thermal-adj-A-island-bdy}.  This pair of mirror branches makes explicit that the smaller party, rather than the topology of the outer union alone, selects which supporting island surface carries the optimal endpoint.
  
\subsubsection{Page-like transitions of reflected entropy}
In this subsection we organize the time dependence by the topology of the dominant entanglement saddle rather than by a numerical phase label.  The entanglement entropy of $A\cup B$ is always the lower envelope of the four saddles derived above, whereas the reflected entropy is obtained by minimizing over the cross-section topologies compatible with the dominant entanglement wedge.  Each transition therefore has a simple geometric interpretation: a wormhole-crossing HM bridge can be replaced by a same-boundary dome, by an island pair, or sequentially by a combination of the two.  We display both the entanglement-entropy and reflected-entropy competition, together with the Markov gap, so that the geometric reorganization of the purification can be read directly from the curves.

\subsubsection*{From one HM bridge to a dome saddle}

\begin{figure}[ht]
\centering
\begin{subfigure}[b]{0.49\textwidth}
\centering
\includegraphics[width=\linewidth]{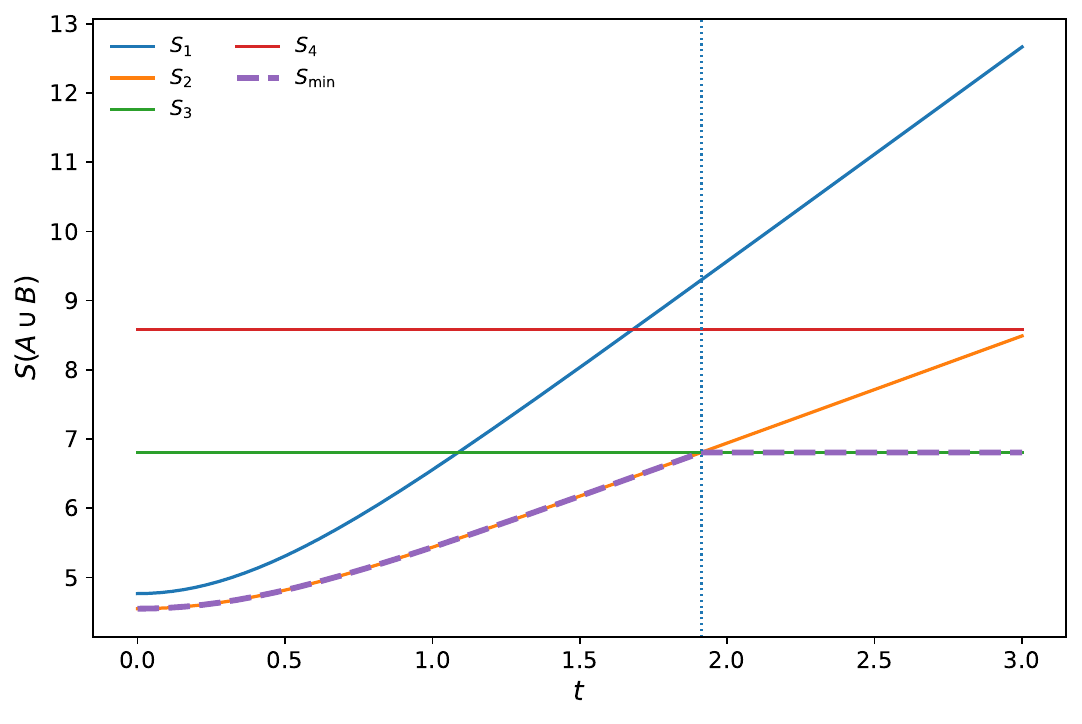}
\caption{Competing union-entropy saddles and their lower envelope.}
\end{subfigure}
\hfill
\begin{subfigure}[b]{0.49\textwidth}
\centering
\includegraphics[width=\linewidth]{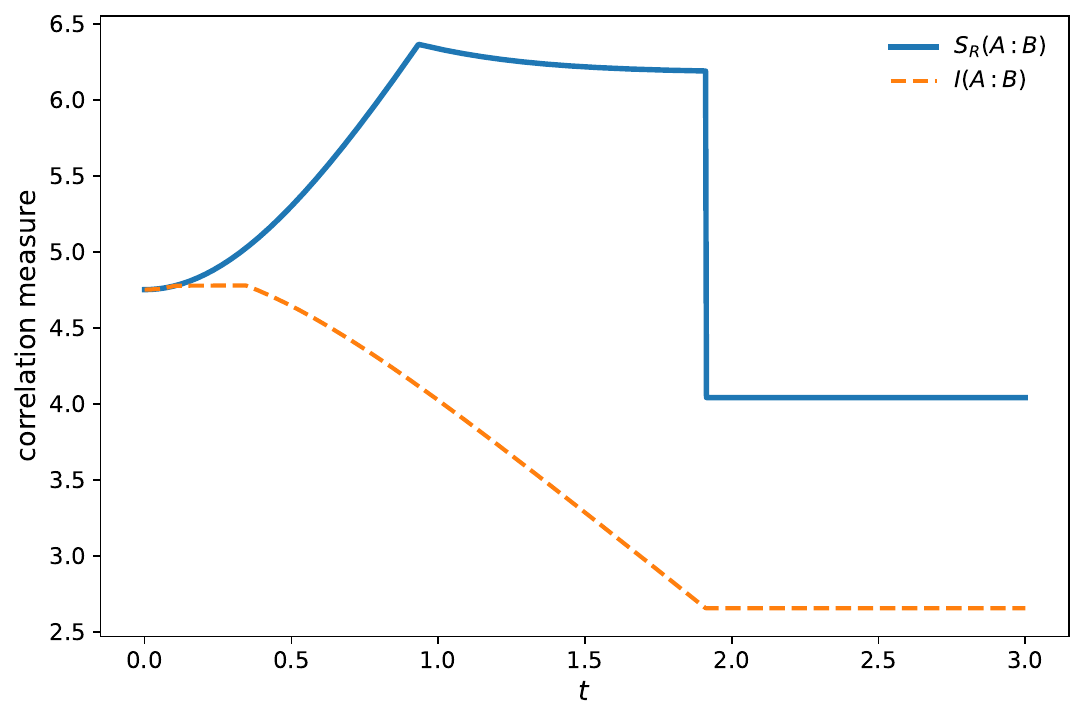}
\caption{Physical reflected entropy $S_R$ and mutual information $I$.}
\end{subfigure}
\par\medskip
\begin{subfigure}[b]{0.62\textwidth}
\centering
\includegraphics[width=\linewidth]{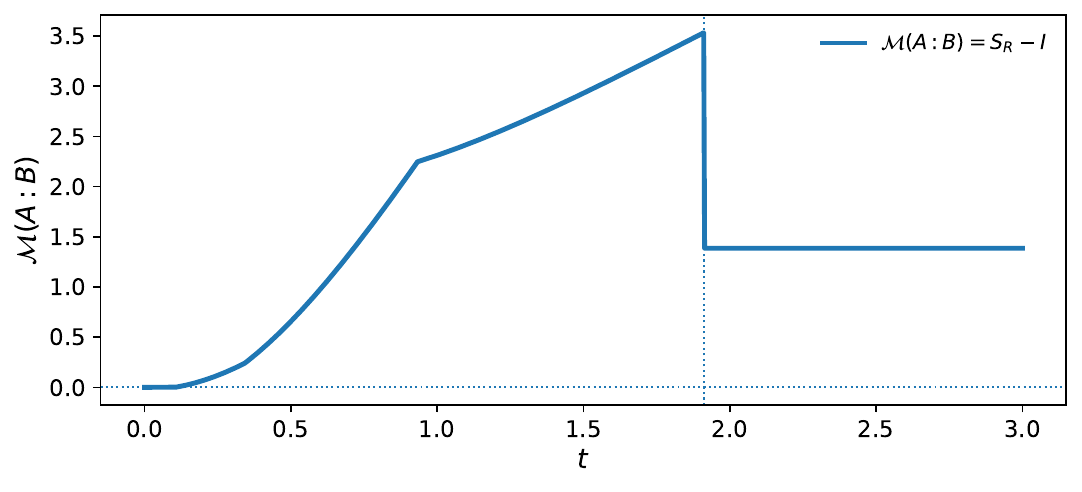}
\caption{Markov gap $\mathcal M=S_R-I$, shown on a separate axis to avoid obscuring the reflected-entropy curve.}
\end{subfigure}
\caption{Adjacent intervals: transition from a single surviving HM bridge to a same-boundary dome.  Panel (a) shows the four candidate entropies of $A\cup B$ and their dashed lower envelope; the dotted vertical line marks the outer-saddle transition.  Panel (b) shows the physical reflected entropy and mutual information on unobstructed axes.  Panel (c) displays the Markov gap separately.  The fact that $\mathcal M$ can change nonanalytically even when the union entropy follows a smooth branch is a direct diagnostic of a reorganization of the internal purification.}
\label{fig:adj-page-bridge-dome}
\end{figure}
The competition displayed in \cref{fig:adj-page-bridge-dome} shows that the entanglement and reflected-entropy transitions need not be identical: the outer homology class changes at the Page crossing, while the internal cross section may reorganize within the same wedge.  In the tripartite viewpoint $(A,B,C)$, with $C$ the purifying complement, the growth of panel (c) before the transition reflects an increasing obstruction to recovering the $A$--$B$ correlations from conditional data on $C$.  The drop when the dome takes over signals that part of this wormhole-mediated tripartite structure has been converted into a more local same-boundary purification, although the remaining positive gap shows that the state is not reduced to pairwise correlations alone \cite{Hayden:2021gno,Lu:2022cgq}.

The wormhole-to-same-boundary transition occurs when the mixed saddle $S_2=J(L_1,R_1)+H(L_3,R_3;t)$ becomes degenerate with the two-dome saddle $S_3=D(L_1,L_3)+D(R_1,R_3)$.  Solving $S_2(t)=S_3$ for the boundary time gives
\begin{align}\label{t-adj-2-3}
 t_{E_{2\to3}}=&\frac{\ell^2}{2r_h}\operatorname{arccosh}\!\left[
 \frac{2\epsilon_y^2\,\sinh^2\!\frac{r_hL_{31}}{2\ell^2}\,\sinh^2\!\frac{r_hR_{31}}{2\ell^2}}
 {\ell^2\mathcal B_0^2\sinh\frac{r_hL_1}{\ell^2}\sinh\frac{r_hR_1}{\ell^2}}
 \left(1-\nu \mathcal Q_1\right)-\mathcal C_3\right].
\end{align}
The crossing in \cref{t-adj-2-3} is simply the competition between the growing HM block and the static dome block.  The separate reflected-entropy kinks in \cref{fig:adj-page-bridge-dome} arise from the internal cross-section minimization and need not coincide with this outer-saddle crossing.

\subsubsection*{From one HM bridge to an island saddle}
\begin{figure}[ht]
\centering
\begin{subfigure}[b]{0.49\textwidth}
\centering
\includegraphics[width=\linewidth]{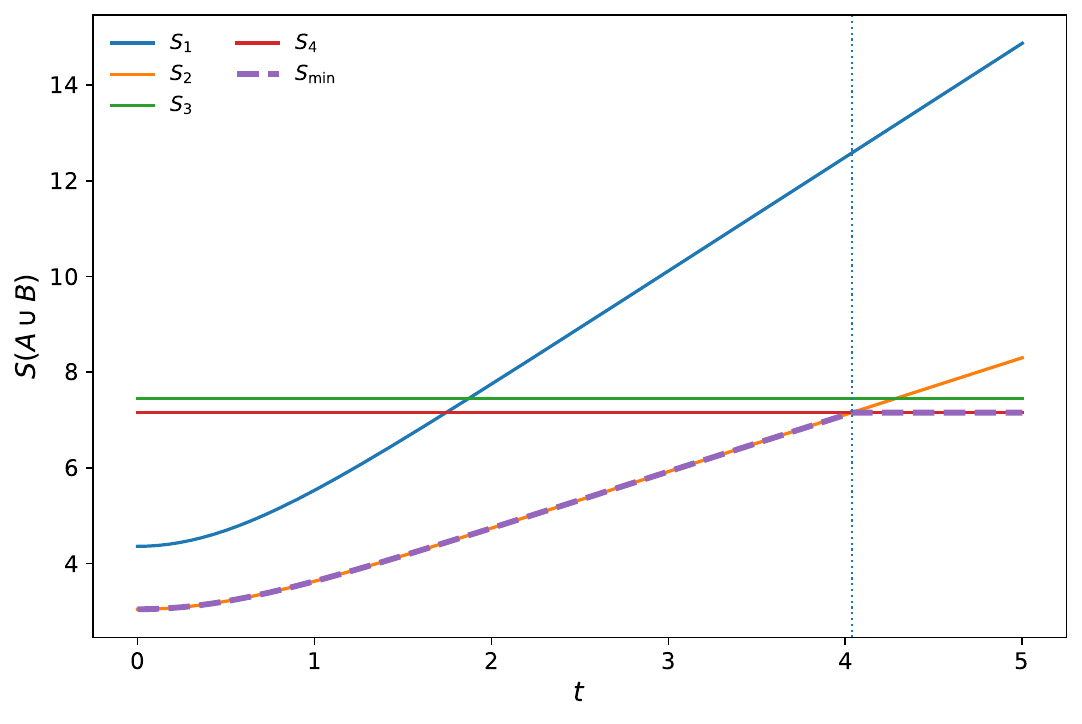}
\caption{Competing union-entropy saddles and their lower envelope.}
\end{subfigure}
\hfill
\begin{subfigure}[b]{0.49\textwidth}
\centering
\includegraphics[width=\linewidth]{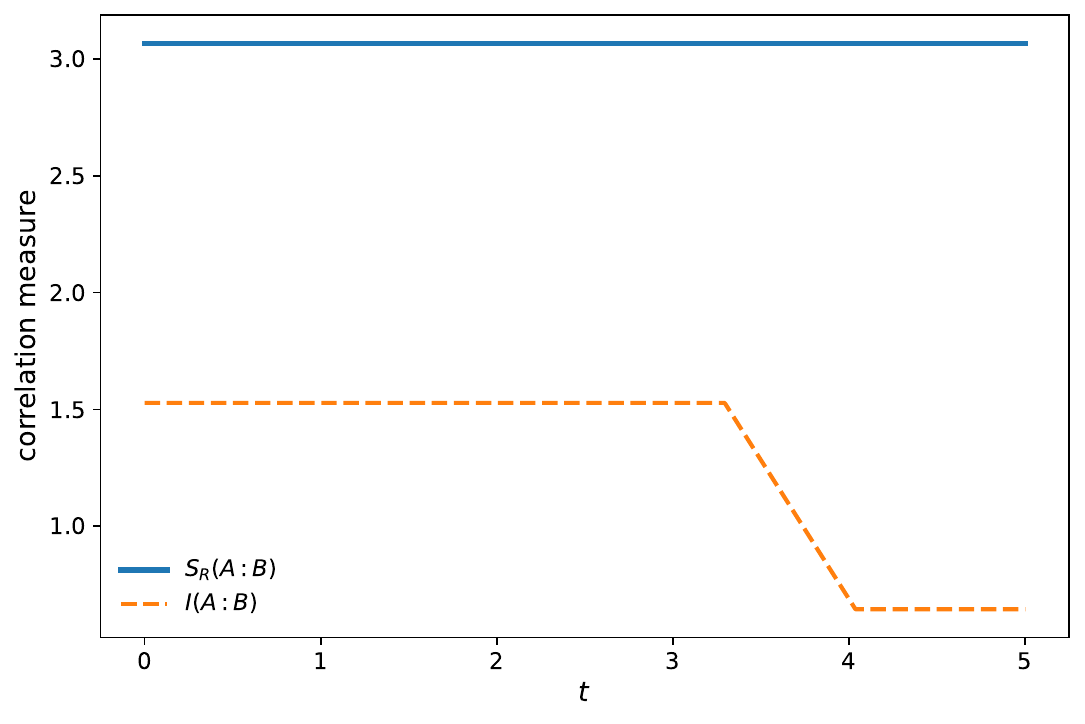}
\caption{Physical reflected entropy $S_R$ and mutual information $I$.}
\end{subfigure}
\par\medskip
\begin{subfigure}[b]{0.62\textwidth}
\centering
\includegraphics[width=\linewidth]{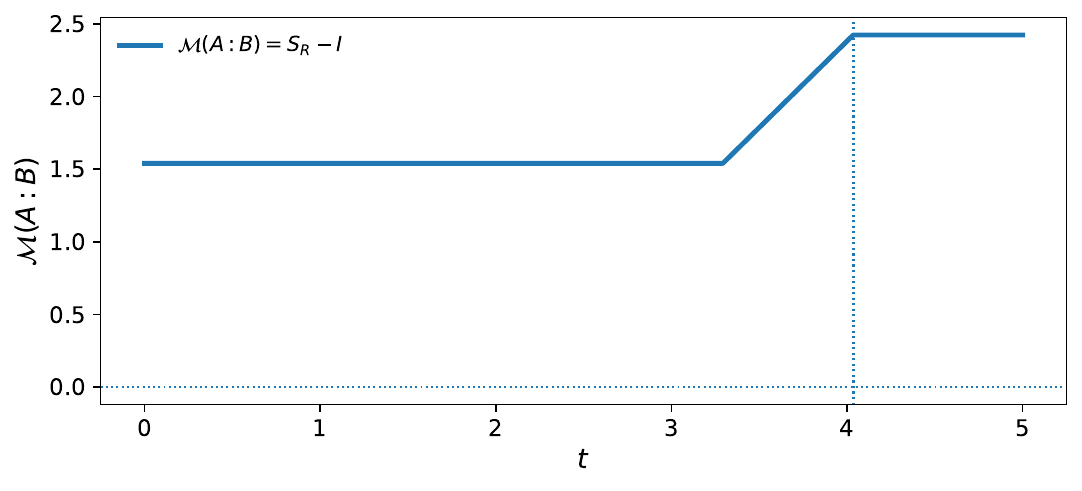}
\caption{Markov gap $\mathcal M=S_R-I$, shown on a separate axis to avoid obscuring the reflected-entropy curve.}
\end{subfigure}
\caption{Adjacent intervals: transition from a single surviving HM bridge to an island-supported saddle.  Panel (a) displays the competing union entropies, panel (b) compares $S_R$ with $I$, and panel (c) isolates the Markov gap.  The post-transition plateau is controlled by the brane/island sector rather than by wormhole stretching.  A nonzero late-time $\mathcal M$ is consistent with the island phase remaining non-Markovian and retaining multipartite organization involving $A$, $B$ and the purifying degrees of freedom.}
\label{fig:adj-page-bridge-island}
\end{figure}
The curves in \cref{fig:adj-page-bridge-island} make the island takeover visible simultaneously in the union entropy and in the internal purification geometry.  Panel (c) provides an independent diagnostic of this rearrangement.  Before the transition, the natural third party in the tripartite split is largely associated with the opposite exterior/interior sector of the TFD state; afterwards, part of that purifying role is transferred to the gravitating island degrees of freedom.  The nonzero late-time Markov gap is therefore consistent with the interpretation that the island does not simply remove correlations, but reorganizes the purification among $A$, $B$ and the island/complement sector \cite{Chandrasekaran:2020qtn,Lu:2022cgq}.

The wormhole-to-island transition is obtained by replacing the remaining HM contribution in $S_2$ by the corresponding island pair in $S_4$, namely by solving $H(L_3,R_3;t)=J(L_3,R_3)$.  This gives
\begin{align}\label{t-adj-2-4}
 t_{E_{2\to4}}=\frac{\ell^2}{2r_h}\operatorname{arccosh}\!\left[
 \frac{2\ell^2\mathcal B_0^2}{\epsilon_y^2}
 \sinh\frac{r_hL_3}{\ell^2}\sinh\frac{r_hR_3}{\ell^2}\left(1+\nu\mathcal Q_3\right)-\mathcal C_3\right].
\end{align}
Equation~\eqref{t-adj-2-4} is the elementary HM--island crossing for the remaining outer pair.  The subsequent choice among island-surface and brane-ending cross sections is an independent minimization inside the island-supported wedge, as displayed in \cref{fig:adj-page-bridge-island}.

\subsubsection*{Sequential loss of one HM bridge}
\begin{figure}[ht]
\centering
\begin{subfigure}[b]{0.49\textwidth}
\centering
\includegraphics[width=\linewidth]{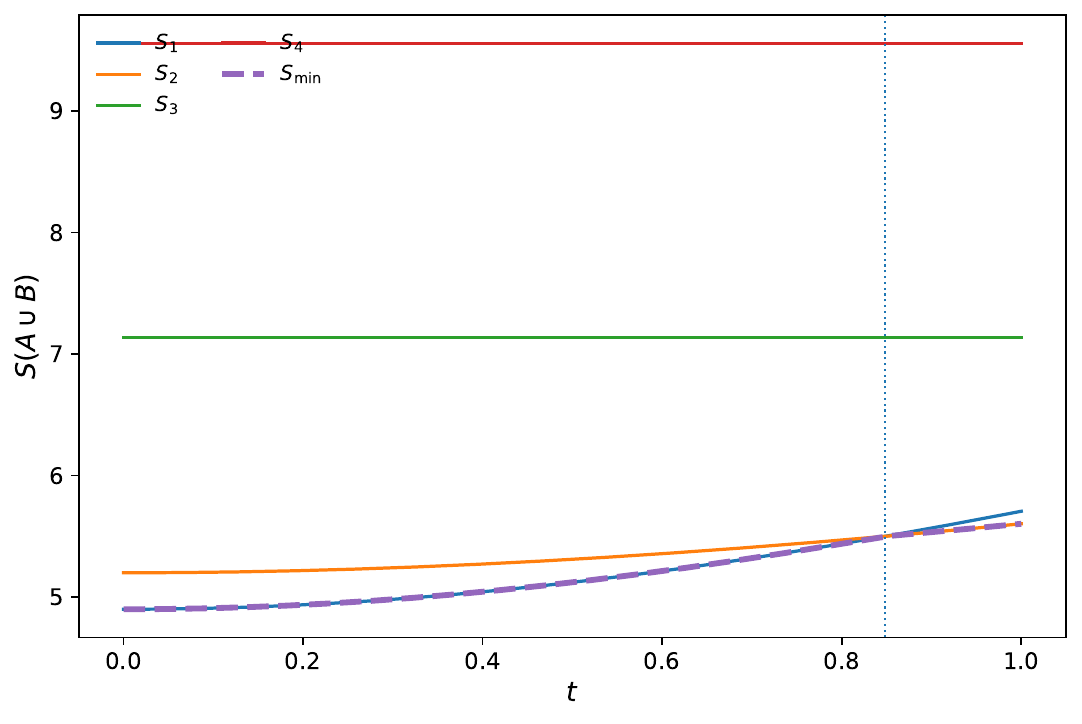}
\caption{Competing union-entropy saddles and their lower envelope.}
\end{subfigure}
\hfill
\begin{subfigure}[b]{0.49\textwidth}
\centering
\includegraphics[width=\linewidth]{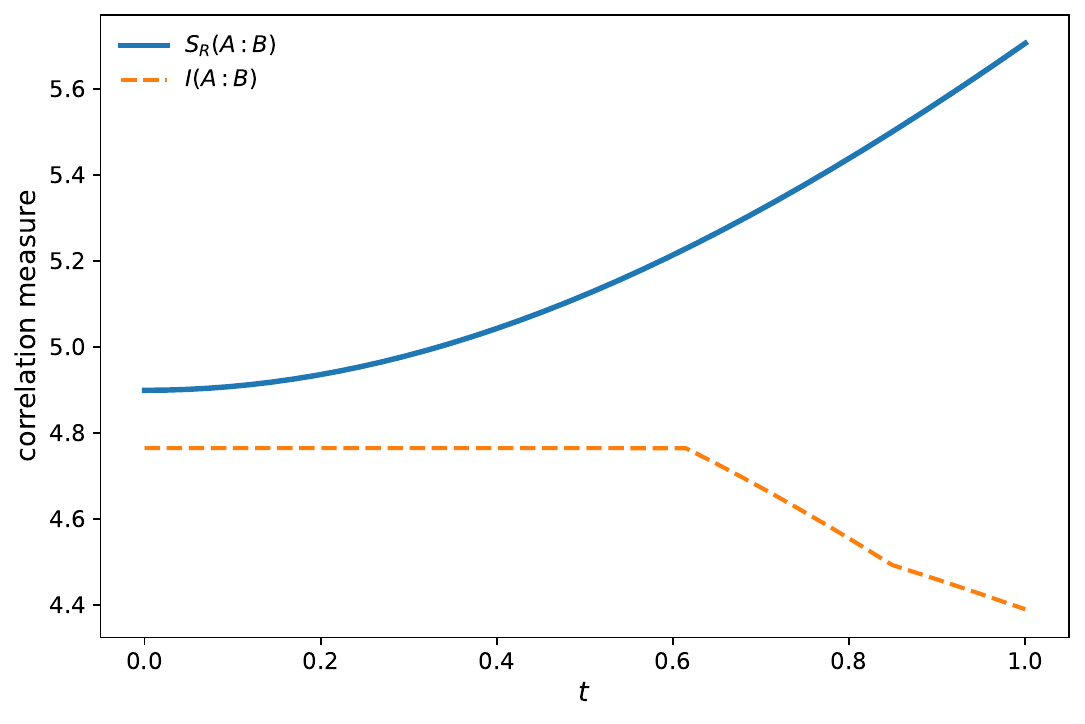}
\caption{Physical reflected entropy $S_R$ and mutual information $I$.}
\end{subfigure}
\par\medskip
\begin{subfigure}[b]{0.62\textwidth}
\centering
\includegraphics[width=\linewidth]{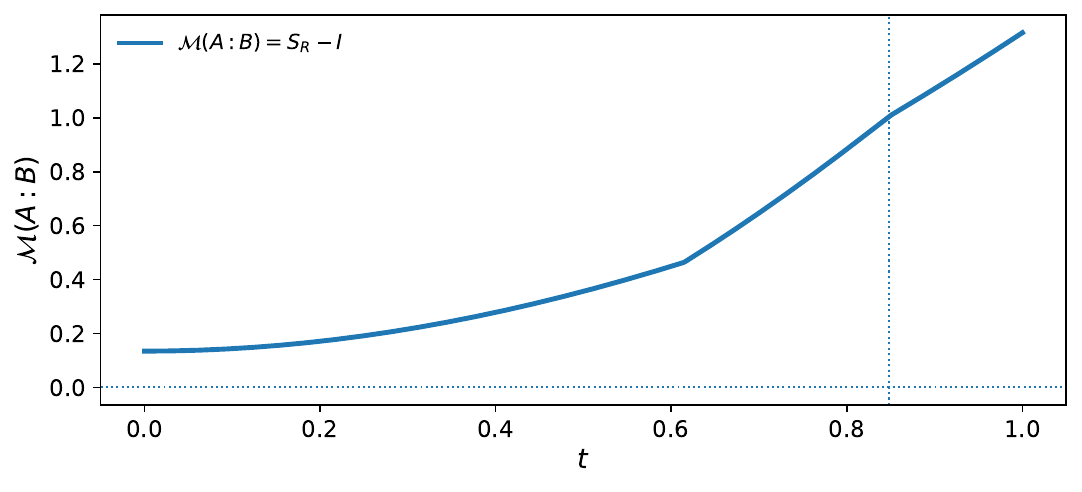}
\caption{Markov gap $\mathcal M=S_R-I$, shown on a separate axis to avoid obscuring the reflected-entropy curve.}
\end{subfigure}
\caption{Adjacent intervals: sequential loss of one of two HM bridges.  Panel (a) shows the union-entropy competition and the intermediate asymmetric wedge, panel (b) shows the minimized $S_R$ and $I$, and panel (c) displays the Markov gap.  The staged change of $\mathcal M$ resolves the corresponding reorganization of the non-Markovian/multipartite diagnostic one wormhole-supported component at a time.}
\label{fig:adj-page-sequential}
\end{figure}
The sequential rearrangement shown in \cref{fig:adj-page-sequential} separates the loss of the two wormhole-crossing channels into distinct stages, allowing the reflected entropy to resolve an asymmetry that is much less pronounced in the union entropy.  The Markov gap gives this staging an information-theoretic interpretation: its endpoint-sensitive diagnostic changes when one part of the purifier becomes island supported while the second still uses the wormhole.  Only after the later rearrangement is the wormhole-supported part fully removed, so the two kinks in the correlation diagnostics need not coincide with a single Page time.

The first step in the sequential wormhole-to-island evolution replaces the inner HM bridge by an island pair.  Because the second HM term is common to both saddles, the crossing is simply $H(L_1,R_1;t)=J(L_1,R_1)$, and hence
\begin{align}\label{t-adj-1-2}
 t_{E_{1\to2}}=\frac{\ell^2}{2r_h}\operatorname{arccosh}\!\left[
 \frac{2\ell^2\mathcal B_0^2}{\epsilon_y^2}
 \sinh\frac{r_hL_1}{\ell^2}\sinh\frac{r_hR_1}{\ell^2}\left(1+\nu\mathcal Q_1\right)-\mathcal C_1\right].
\end{align}
The first crossing \eqref{t-adj-1-2} removes only one HM contribution, leaving an asymmetric mixed wedge.  This separation between the outer transition and the later internal cross-section exchange is the origin of the two-stage structure in \cref{fig:adj-page-sequential}.

\subsubsection*{Direct transition from two HM bridges to dome saddles}
\begin{figure}[ht]
\centering
\begin{subfigure}[b]{0.49\textwidth}
\centering
\includegraphics[width=\linewidth]{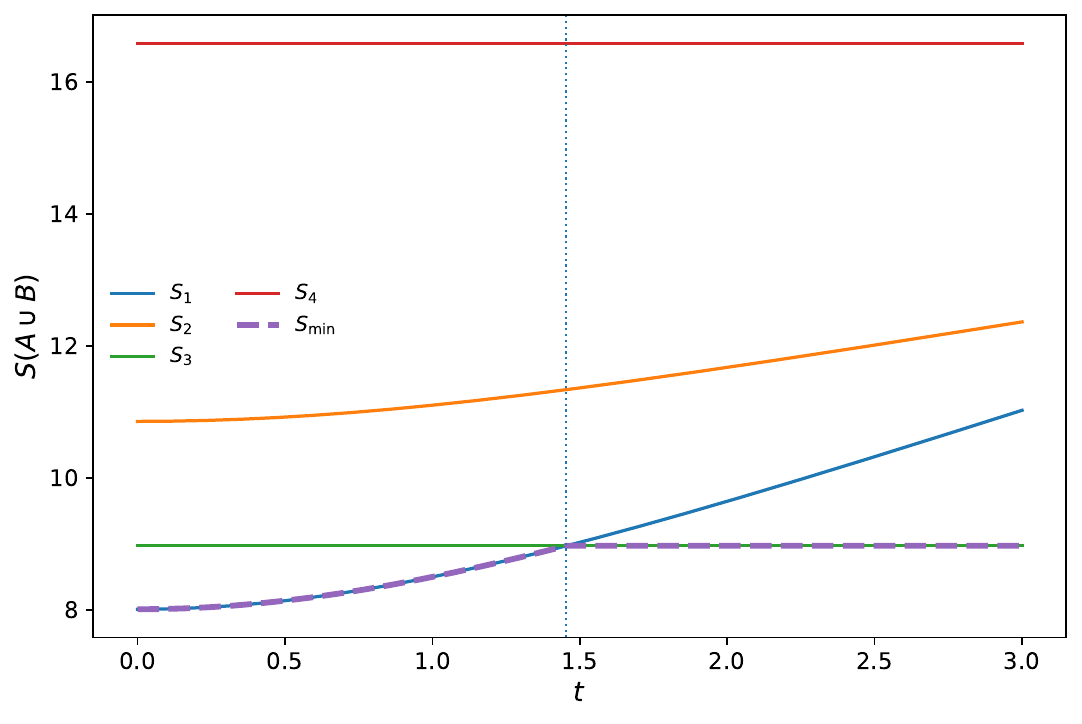}
\caption{Competing union-entropy saddles and their lower envelope.}
\end{subfigure}
\hfill
\begin{subfigure}[b]{0.49\textwidth}
\centering
\includegraphics[width=\linewidth]{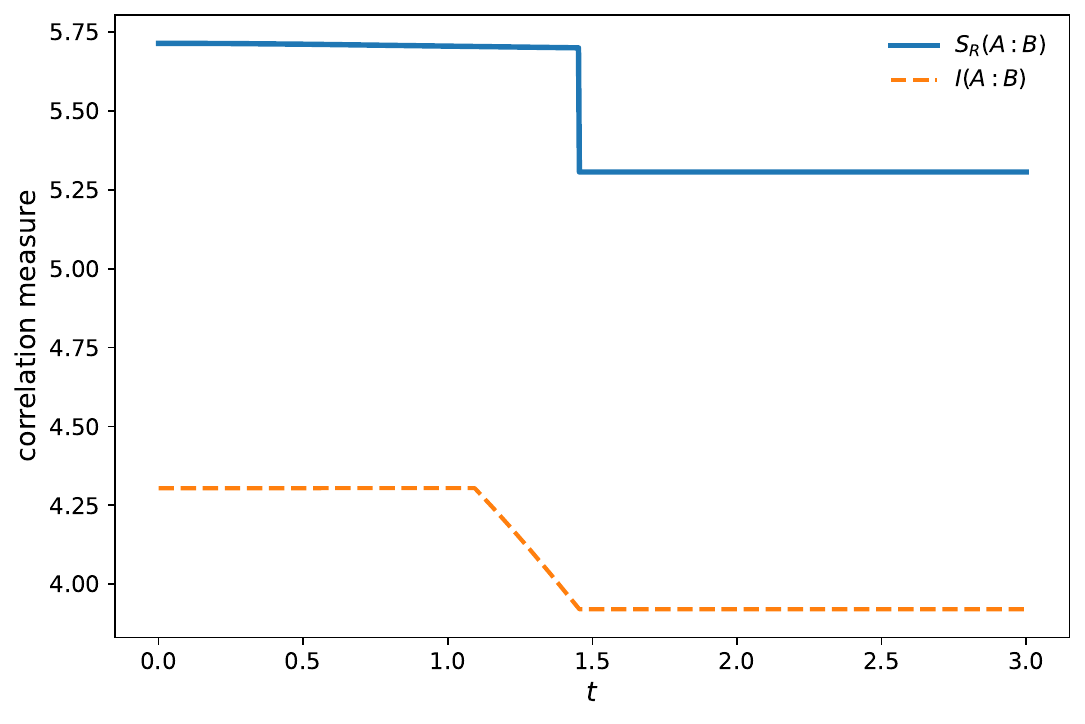}
\caption{Physical reflected entropy $S_R$ and mutual information $I$.}
\end{subfigure}
\par\medskip
\begin{subfigure}[b]{0.62\textwidth}
\centering
\includegraphics[width=\linewidth]{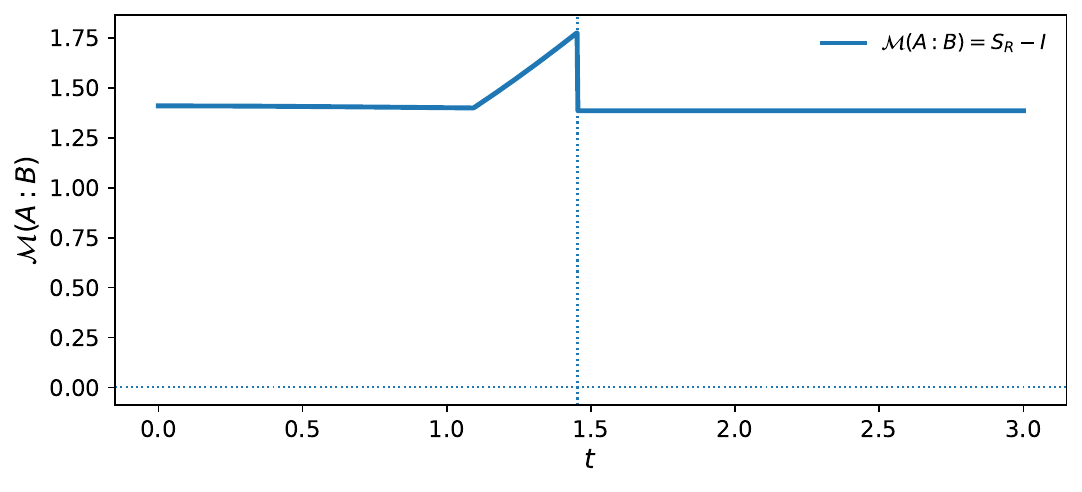}
\caption{Markov gap $\mathcal M=S_R-I$, shown on a separate axis to avoid obscuring the reflected-entropy curve.}
\end{subfigure}
\caption{Adjacent intervals: direct replacement of two HM bridges by same-boundary domes.  Panel (a) shows the direct crossing of the time-growing and static union-entropy saddles, panel (b) shows the corresponding reflected entropy and mutual information, and panel (c) isolates the Markov gap.  The abrupt reduction of $\mathcal M$ tracks the simultaneous reorganization away from the wormhole-mediated non-Markovian structure, while its nonzero late-time value remains sensitive to the nontrivial endpoint structure of the remaining cross section.}
\label{fig:adj-page-direct-dome}
\end{figure}
The direct transition in \cref{fig:adj-page-direct-dome} removes both wormhole-crossing contributions at once; the flattening of the late-time curves is therefore a direct geometric consequence of replacing time-growing bridges by static same-boundary geodesics.  The Markov-gap panel adds a sharper interpretation: the abrupt drop corresponds to a simultaneous removal of the two wormhole-supported components from the endpoint-sensitive purification geometry.  The residual constant gap is not evidence for surviving wormhole entanglement; rather, it is the endpoint contribution associated with the static geometric purification and therefore reflects the fact that $A$, $B$ and their complement are still not arranged as an exact quantum Markov chain.

The direct simultaneous wormhole-to-dome transition follows from the degeneracy of $S_1=H(L_1,R_1;t)+H(L_3,R_3;t)$ with the time-independent saddle $S_3=D(L_1,L_3)+D(R_1,R_3)$.  Solving $S_1(t)=S_3$ yields
\begin{align}\label{t-adj-1-3}
 t_{E_{1\to3}}=\frac{\ell^2}{2r_h}\operatorname{arccosh}\!\left[\frac12\left(
 \sqrt{(\mathcal C_1-\mathcal C_3)^2+16\sinh^2\frac{r_hL_{31}}{2\ell^2}\sinh^2\frac{r_hR_{31}}{2\ell^2}}-\mathcal C_1-\mathcal C_3\right)\right].
\end{align}
This direct crossing replaces both time-growing HM contributions at once by the static dome saddle.  The reflected-entropy and Markov-gap panels therefore show the most abrupt loss of wormhole-supported purification among the adjacent examples.

 \subsection{Disjoint subsystems}
We now separate the two parties by a finite gap and take $A=(L_1,L_2)\cup(R_1,R_2)$ and $B=(L_3,L_4)\cup(R_3,R_4)$ with $L_1<L_2<L_3<L_4$ and the same ordering on the right.  The inner endpoints $(L_2,L_3)$ and $(R_2,R_3)$ introduce central dome surfaces that survive in several saddles.  This additional scale makes the disjoint problem richer than the adjacent one: the union may retain a connected wedge while the optimal reflected-entropy cross section moves between the central domes, a wormhole-crossing surface, an island surface and the EOW brane.

Before analyzing connected EWCS topologies, we must also include the disconnected outer wedge.  For a two-sided interval with ordered endpoint labels $i<j$, define its independently minimized entropy by
\begin{align}
S_{[ij]}(t)\equiv\min\Big\{&H(L_i,R_i;t)+H(L_j,R_j;t),\notag\\
&J(L_i,R_i)+H(L_j,R_j;t),\notag\\
&D(L_i,L_j)+D(R_i,R_j),\notag\\
&J(L_i,R_i)+J(L_j,R_j)\Big\}.
\label{thermal-single-interval-min}
\end{align}
These are respectively the two-HM, mixed island--HM, same-boundary dome and fully island-supported homology classes.  The ordering relative to the EOW brane makes the displayed mixed class the admissible one in this phase catalogue: if the farther endpoint has already switched to an island support, the nearer endpoint has done so as well.  The disconnected competitor for the disjoint union is therefore
\begin{align}
S_{\rm disc}^{\rm th}(A\cup B;t)=S_{[12]}(t)+S_{[34]}(t).
\label{thermal-disconnected-union}
\end{align}
The connected saddles below are admissible as physical reflected-entropy phases only when
\begin{align}
\min_{k=1,\ldots,4}S_k(A\cup B;t)<S_{\rm disc}^{\rm th}(A\cup B;t).
\label{thermal-connected-domain}
\end{align}
If the inequality is reversed, the leading entanglement wedge is disconnected and there is no connected EWCS; consequently $I(A:B)=S_R(A:B)=0+O(c^0)$.  The numerical minimization used for the disjoint Page curves includes \cref{thermal-disconnected-union}, so a connected $S_R$ branch is never continued outside its outer-wedge domain.

We organize the four outer entanglement saddles with the elementary blocks $D$, $H$ and $J$ defined in \cref{thermal-D,thermal-H,thermal-J}.  This notation fixes the endpoint pairing and keeps the time-growing wormhole contributions separate from the static dome and island terms.  On the boundary side, the reflected-entropy saddles arise from different semiclassical channels of a common eight-point replicated correlator.  To keep the replica structure legible, define
\begin{align}
\mathcal G_{AB}^{(m,n)}\equiv&\,
\big\langle \sigma_{g_A}(P_1)\sigma_{g_A^{-1}}(P_2)
\sigma_{g_B}(P_3)\sigma_{g_B^{-1}}(P_4)\notag\\
&\qquad\times\sigma_{g_B}(Q_4)\sigma_{g_B^{-1}}(Q_3)
\sigma_{g_A}(Q_2)\sigma_{g_A^{-1}}(Q_1)\big\rangle,\\
\mathcal G_{0}^{(m)}\equiv&\,
\big\langle \sigma_{g_m}(P_1)\sigma_{g_m^{-1}}(P_2)
\sigma_{g_m}(P_3)\sigma_{g_m^{-1}}(P_4)\notag\\
&\qquad\times\sigma_{g_m}(Q_4)\sigma_{g_m^{-1}}(Q_3)
\sigma_{g_m}(Q_2)\sigma_{g_m^{-1}}(Q_1)\big\rangle .
\end{align}
Before choosing a semiclassical OPE channel, the effective reflected entropy is then
\begin{align}\label{dis-SR-def}
S_R^{\text{bdy}}(A:B)=S_R^{\rm eff}(A:B)
=\lim_{m,n\to1}\frac{1}{1-n}
\log\frac{\mathcal G_{AB}^{(m,n)}}{\big(\mathcal G_0^{(m)}\big)^n}.
\end{align}
Here $P_i=(L_i,t_0,\ell^2/\epsilon)$ and $Q_i=(R_i,-t_0+i\beta/2,\ell^2/\epsilon)$.  The eight-point function makes the new feature of the disjoint problem explicit: the four inner twists associated with the finite gap can remain in a nontrivial block even when the outer twists factorize.  Bulk-wise, this is precisely the possibility that the EWCS runs between the two central domes rather than following an outer HM or island surface.

\subsubsection{Two wormhole-crossing outer surfaces with central domes}
\begin{figure}[ht]
    \centering
    \includegraphics[width=0.55\linewidth]{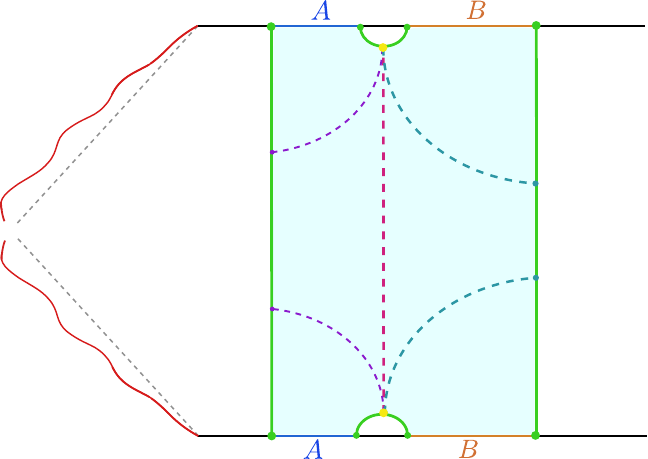}
    \caption{Schematic depiction of the distinct reflected-entropy, or equivalently bulk-EWCS, saddles between $A$ and $B$ when the entropy of $A\cup B$ contains two wormhole-crossing outer surfaces and central domes.  Candidate cross sections are indicated by dashed curves.
}
    \label{fig:-FT-dis-2HM}
\end{figure}
When both disjoint regions extend far from the brane, the outer endpoint pairs $(L_1,R_1)$ and $(L_4,R_4)$ are connected through the black-hole interior, while the inner endpoint pairs are joined by same-boundary domes. These extremal surfaces are shown as solid curves in \cref{fig:-FT-dis-2HM}.  The union entropy therefore contains two time-growing HM contributions and two static central-dome contributions,
\begin{align}\label{dis-HM-EE}
S_1=H(L_1,R_1;t_0)+H(L_4,R_4;t_0)+D(L_2,L_3)+D(R_2,R_3).
\end{align}
The central domes do not grow with $t_0$, so the eventual loss of this saddle is driven entirely by the two wormhole-crossing pieces.  Nevertheless, those domes strongly affect reflected entropy because the internal cross section can pass between them rather than through the wormhole.  Thus even at early times the dominant mixed-state purification need not follow the same geometry as the outer entanglement surface.

This outer saddle contains no island and therefore no lower-dimensional reflected-island generalized-area contribution.  Nevertheless, three inequivalent separators are available: a common perpendicular between the central domes, a dome-to-HM separator controlled by $A$, and its mirror controlled by $B$.  The finite central gap is crucial here.  It supplies a static geometric channel that can compete with the wormhole-supported channels even while the union entropy is still dominated by two HM surfaces.

\subsubsection*{Cross section between the two dome surfaces}

\paragraph{Boundary description.}
For the cross section running between the two central domes, the outer twists pair across the two TFD copies while the four inner twists remain in a nontrivial block.  The numerator of \cref{dis-SR-def} is consequently, in the dominant large-$c$ channel, approximated by two two-point functions and one four-point function,
\begin{align}
    &\langle \sigma_{g^{}_A}({P_1})\sigma_{g_A^{-1}}({P_{2}}) \sigma_{g_B} ({P_{3}})\sigma_{g_B^{-1}}({P_{4}}) \sigma_{g^{}_B}({Q_{4}}) \sigma_{g^{-1}_B}({Q_{3}}) \sigma_{g^{}_A  }({Q_{2}})\sigma_{g_A^{-1}}({Q_{1}}) \rangle\notag\\
    &\simeq \langle \sigma_{g^{}_A}({P_1}) \sigma_{g_A^{-1}}({Q_{1}}) \rangle ~\langle \sigma_{g_B^{-1}}({P_{4}})\sigma_{g^{}_B}({Q_{4}}) \rangle ~\langle \sigma_{g_A^{-1}}({P_{2}}) \sigma_{g_B} ({P_{3}})\sigma_{g^{-1}_B}({Q_{3}}) \sigma_{g^{}_A  }({Q_{2}}) \rangle.
\end{align}
The denominator has the same leading factorization in that channel, so all outer spectator factors cancel.  The reflected entropy is reduced to the normalized four-point function of the inner twists,
\begin{align}
    S_R^{\text{bdy}}(A:B)=  \lim_{m,n \to 1}\frac{1}{1-n}\log \frac{\langle \sigma_{g_A^{-1}}({P_{2}}) \sigma_{g_B} ({P_{3}})\sigma_{g^{-1}_B}({Q_{3}}) \sigma_{g^{}_A  }({Q_{2}})\rangle}{\langle \sigma_{g_m^{-1}}({P_{2}}) \sigma_{g_m} ({P_{3}})\sigma_{g^{-1}_m}({Q_{3}}) \sigma_{g^{}_m  }({Q_{2}})\rangle^n}.
\end{align}
The semiclassical four-point block is the channel dual to the common perpendicular between the two dome surfaces.  Writing the bulk invariant that appears below as $\mathcal Z_{\rm dd}(t_0)$, the boundary answer is
\begin{align}\label{dis-sr1-i-bdy}
S_R^{\rm bdy}(A:B)=\frac{c}{3}\operatorname{arccosh}\mathcal Z_{\rm dd}(t_0),
\end{align}
where the entire dependence on the two-copy dome geometry is captured by the dimensionless invariant
\begin{align}
\mathcal Z_{\rm dd}(t_0)=&\frac{\sqrt{[\cosh\kappa(L_2-R_3)+\cosh(2\kappa t_0)]
[\cosh\kappa(L_3-R_2)+\cosh(2\kappa t_0)]}}{2\sinh\frac{\kappa L_{32}}2\sinh\frac{\kappa R_{32}}2}\notag\\
&+\frac{\sqrt{[\cosh\kappa(L_2-R_2)+\cosh(2\kappa t_0)]
[\cosh\kappa(L_3-R_3)+\cosh(2\kappa t_0)]}}{2\sinh\frac{\kappa L_{32}}2\sinh\frac{\kappa R_{32}}2}.
\end{align}
This is the same semiclassical block that appears in the rigid-brane phase analysis \cite{Banerjee:2016qca,Basu:2023jtf,Dey:2025dlj}.  The invariant $\mathcal Z_{\rm dd}$ measures the hyperbolic separation of the two central dome geodesics.  Its time dependence is inherited indirectly from embedding the two domes in the two-sided TFD geometry; the cross section can therefore evolve even though each same-boundary dome length is individually time independent.

\paragraph{Bulk description.}
In the bulk description, the extremal curve ($\Sigma_{AB}$) is the common perpendicular between the two central domes shown as a dashed curve in \cref{fig:-FT-dis-2HM}.
 Let $(x_1,t_0,r_1)$ and $(x_2,-t_0+i\beta/2,r_2)$ denote arbitrary points on the left- and right-copy domes.  Before imposing orthogonality, their BTZ embedding invariant gives the trial length 
{\small\begin{align}
    \mathcal{L}=\ell \operatorname{arccosh}\left[\frac{\sqrt{\left(r_1^2-r_h^2\right) \left(r_2^2-r_h^2\right)} \cosh \frac{2 r_h t_0}{\ell^2}+r_1 r_2 \cosh \frac{r_h (x_1-x_2)}{\ell^2}}{r_h^2}\right].
\end{align}}
The dome-type RT surface anchored on $(L_2,L_3)$ and $(R_2,R_3)$ can be parametrized by signed proper distances $\sigma_L$ and $\sigma_R$ measured from their respective midpoints:
    \begin{align}\label{dome-RT-para}
        \sqrt{\frac{r_1^2}{r_h^2}-1}&=\cosh\sigma_L\,\operatorname{csch}\frac{r_h L_{32}}{2\ell^2},\notag\\
        \tanh\frac{r_h x_1}{\ell^2}&=
        \frac{\cosh\sigma_L\left(\sinh\frac{r_hL_2}{\ell^2}+\sinh\frac{r_hL_3}{\ell^2}\right)
        +\sinh\sigma_L\left(\sinh\frac{r_hL_3}{\ell^2}-\sinh\frac{r_hL_2}{\ell^2}\right)}
        {\cosh\sigma_L\left(\cosh\frac{r_hL_2}{\ell^2}+\cosh\frac{r_hL_3}{\ell^2}\right)
        +\sinh\sigma_L\left(\cosh\frac{r_hL_3}{\ell^2}-\cosh\frac{r_hL_2}{\ell^2}\right)},\notag\\
        \sqrt{\frac{r_2^2}{r_h^2}-1}&=\cosh\sigma_R\,\operatorname{csch}\frac{r_h R_{32}}{2\ell^2},\notag\\
        \tanh\frac{r_h x_2}{\ell^2}&=
        \frac{\cosh\sigma_R\left(\sinh\frac{r_hR_2}{\ell^2}+\sinh\frac{r_hR_3}{\ell^2}\right)
        +\sinh\sigma_R\left(\sinh\frac{r_hR_3}{\ell^2}-\sinh\frac{r_hR_2}{\ell^2}\right)}
        {\cosh\sigma_R\left(\cosh\frac{r_hR_2}{\ell^2}+\cosh\frac{r_hR_3}{\ell^2}\right)
        +\sinh\sigma_R\left(\cosh\frac{r_hR_3}{\ell^2}-\cosh\frac{r_hR_2}{\ell^2}\right)}.
    \end{align}
    This form is equivalent to the usual exponential proper-length parametrization but makes the reflection symmetry about the midpoint of each dome explicit.  In particular, $\sigma_L=0$ and $\sigma_R=0$ select the respective geodesic midpoints.
Substituting these intrinsic dome coordinates into the BTZ invariant packages the entire trial cross section into one dimensionless chordal quantity $\zeta$:
\begin{align}
    \mathcal{L}= \ell \operatorname{arccosh}\,\zeta .
\end{align}
All dependence on the four left--right endpoint separations and on time can be isolated into four positive hyperbolic combinations.  We define
\begin{align}
A_{ij}\equiv \cosh\frac{r_h(L_i-R_j)}{\ell^2}+\cosh\frac{2r_ht_0}{\ell^2},\qquad i,j\in\{2,3\}.
\end{align}
In terms of these combinations the trial invariant takes the manifestly hyperbolic form
{\footnotesize\begin{align}
  \zeta=\frac{\operatorname{csch} \frac{r_h R_{32}}{2\ell^2}}{4\sinh\frac{r_h L_{32}}{2\ell^2}}\Big[&
  \cosh(\sigma_R+\sigma_L)(A_{22}+A_{33})+\sinh(\sigma_R+\sigma_L)(A_{33}-A_{22})\notag\\
  &+\cosh(\sigma_R-\sigma_L)(A_{23}+A_{32})+\sinh(\sigma_R-\sigma_L)(A_{23}-A_{32})\Big].\label{zeta-dome-dome}
\end{align}}
The two lines separate the contributions even and odd under interchange of the two endpoints.  This representation is useful for asymmetric left/right intervals because the loss of reflection symmetry is carried entirely by the terms proportional to $\sinh(\sigma_R\pm\sigma_L)$.
The physical EWCS is the unique common perpendicular, so its endpoint on each dome must satisfy an orthogonality condition.  Extremizing with respect to $\sigma_L$ and $\sigma_R$ imposes these two conditions and fixes
{\small\begin{align}
    &\sigma_L=\frac{1}{4} \log \left[\frac{\left(\cosh \frac{r_h (L_2-R_2)}{\ell^2}+ \cosh \frac{2 r_h t_0}{\ell^2} \right) \left(\cosh \frac{r_h (L_2-R_3)}{\ell^2} +\cosh \frac{2 r_h t_0}{\ell^2}\right) } { \left( \cosh \frac{r_h (L_3-R_2)}{\ell^2} + \cosh \frac{2 r_h t_0}{\ell^2} \right) \left(\cosh \frac{r_h (L_3-R_3)}{\ell^2}+\cosh \frac{2 r_h t_0}{\ell^2}\right)}\right]\notag\\
    &\sigma_R= \frac{1}{4} \log \left[\frac{\left(\cosh \frac{r_h (L_2-R_2)}{\ell^2}+\cosh \frac{2 r_h t_0}{\ell^2}\right) \left(\cosh \frac{r_h (L_3-R_2)}{\ell^2}+\cosh \frac{2 r_h t_0}{\ell^2}\right)}{\left(\cosh \frac{r_h (L_2-R_3)}{\ell^2}+\cosh \frac{2 r_h t_0}{\ell^2}\right) \left(\cosh \frac{r_h (L_3-R_3)}{\ell^2}+\cosh \frac{2 r_h t_0}{\ell^2}\right)}\right].
\end{align}}
Evaluating the invariant at this common perpendicular and converting the geodesic length to reflected entropy gives
{\small\begin{align}\label{sr-thermal-dis-dome-dome}
    S_R^{\text{bulk}}(A:B)= \frac{\ell}{2 G_N} \operatorname{arccosh}\bigg[&\frac{\sqrt{ \left(\cosh \frac{r_h (L_2-R_3)}{\ell^2}+\cosh \frac{2 r_h t_0}{\ell^2}\right) \left(\cosh \frac{r_h (L_3-R_2)}{\ell^2}+\cosh \frac{2 r_h t_0}{\ell^2}\right)}}{2 \sinh \frac{r_h L_{32} }{2 \ell^2} \sinh \frac{r_h R_{32} }{2 \ell^2}}\notag\\ 
    &+\frac{\sqrt{\left(\cosh \frac{r_h (L_2-R_2)}{\ell^2}+\cosh \frac{2 r_h t_0}{\ell^2}\right) \left(\cosh \frac{r_h (L_3-R_3)}{\ell^2}+\cosh \frac{2 r_h t_0}{\ell^2}\right)}}{2 \sinh \frac{r_h L_{32}}{2 \ell^2} \sinh \frac{r_h R_{32} }{2 \ell^2} }\bigg].
\end{align}}
After the Brown--Henneaux relation this is exactly \cref{dis-sr1-i-bdy}.  The equality has a useful physical consequence: although the two supporting domes are individually static, their relative embedding across the two TFD copies retains a time dependence.  Thus $S_R$ can evolve inside an outer entanglement saddle whose central RT components themselves do not grow.  This is one of the clearest examples of reflected entropy detecting internal wedge geometry that is invisible to the total RT length.

\subsubsection*{Cross section ending on the inner HM bridge}

\paragraph{Boundary description.}
When $A$ is the smaller disjoint component, the cross section leaves the central dome and terminates on the inner HM support.  The boundary OPE channel isolates the spectator associated with the far outer endpoint of $B$, so the dominant large-$c$ channel approximates the eight-point numerator of \cref{dis-SR-def} by one spectator two-point block and a six-point block:
\begin{align}
     &\langle \sigma_{g^{}_A}({P_1})\sigma_{g_A^{-1}}({P_{2}}) \sigma_{g_B} ({P_{3}})\sigma_{g_B^{-1}}({P_{4}}) \sigma_{g^{}_B}({Q_{4}}) \sigma_{g^{-1}_B}({Q_{3}}) \sigma_{g^{}_A  }({Q_{2}})\sigma_{g_A^{-1}}({Q_{1}}) \rangle\notag\\
    &\simeq \langle \sigma_{g_B^{-1}}({P_{4}})\sigma_{g^{}_B}({Q_{4}}) \rangle ~\langle \sigma_{g^{}_A}({P_1}) \sigma_{g_A^{-1}}({P_{2}}) \sigma_{g_B} ({P_{3}})\sigma_{g^{-1}_B}({Q_{3}}) \sigma_{g^{}_A  }({Q_{2}})\sigma_{g_A^{-1}}({Q_{1}}) \rangle.
\end{align}
In the semiclassical large-$c$ channel relevant here, the six-point block further resolves into two four-point blocks \cite{Banerjee:2016qca,Basu:2023jtf}.  This boundary factorization mirrors the fact that the bulk cross section has one common-perpendicular problem on each TFD copy:
\begin{align}
    \mathcal{F}_6(P_1,P_2,P_3,Q_3,Q_2,Q_1;h,h_{AB})\simeq{}& \mathcal{F}_4 (P_1,P_2,P_3,Q_1;h,h_{AB}) \notag\\&\times \mathcal{F}_4 (Q_1,Q_2,Q_3,P_1;h,h_{AB}).
\end{align}
In the large-central-charge limit the exchanged primary is $\sigma_{g_Ag_B^{-1}}$.  Evaluating the two resulting four-point blocks gives
\begin{align}\label{dis-sr1-ii-bdy}
S_R^{\rm bdy}(A:B)=\frac{c}{3}\operatorname{arccosh}\mathcal Z_L+\frac{c}{3}\operatorname{arccosh}\mathcal Z_R,
\end{align}
where the left-copy contribution is conveniently summarized by the invariant
\begin{align}
\mathcal Z_L=&\frac{\sinh\frac{\kappa(L_3-R_1)}2\sqrt{\cosh(\kappa L_{21})+\cosh(2\kappa t_0)}
+\sinh\frac{\kappa(L_2-R_1)}2\sqrt{\cosh(\kappa L_{31})+\cosh(2\kappa t_0)}}
{\sinh\frac{\kappa L_{32}}2\sqrt{\cosh\kappa(L_1-R_1)+\cosh(2\kappa t_0)}},\\
\mathcal Z_R=&\mathcal Z_L\big|_{L_i\leftrightarrow R_i}.
\end{align}
The bulk calculation below reproduces these invariants directly.  This equality is exact at the rigid-brane order.

\paragraph{Bulk description.}
In the bulk description, the separator joins the central dome to the inner HM surface, depicted as dashed curves in \cref{fig:-FT-dis-2HM}.  This geometry interpolates between a purely exterior dome--dome separator and the fully wormhole-crossing branch: one endpoint remains tied to the central gap, while the other follows the time-dependent bridge.  Before restricting either endpoint to its supporting geodesic, let the trial points be $(x_1,t_1,r_1)$ and $(x_2,t_2,r_2)$.  Their invariant BTZ separation is
\begin{align}
    \mathcal{L}=\ell\,\operatorname{arccosh}\left[\frac{r_1 r_2}{r_h^2}\cosh\frac{r_h(x_1-x_2)}{\ell^2}-\frac{\sqrt{(r_1^2-r_h^2)(r_2^2-r_h^2)}}{r_h^2}\cosh\frac{r_h(t_1-t_2)}{\ell^2}\right].
\end{align}
We now express both supporting geodesics intrinsically.  The central dome anchored on $(L_2,L_3)$ is parametrized by the signed proper distance $\sigma_D$ of \cref{dome-RT-para}, while the HM support is parametrized by the signed affine coordinate $\sigma_H$ of \cref{HM-parametrization}.  This choice is preferable to coordinate endpoint variables because orthogonality of the cross section becomes ordinary stationarity with respect to $\sigma_D$ and $\sigma_H$.  Substituting the two parametrizations into the BTZ invariant gives
    \begin{align*}
        \mathcal L=\ell\,\operatorname{arccosh}\,\zeta
    \end{align*}
The trial invariant $\zeta$ contains four recurring thermal chords.  To expose their geometric origin and keep the stationarity equations readable, define
\begin{align}
B_1&\equiv\sinh^2\frac{r_hL_{21}}{2\ell^2}+\cosh^2\frac{r_ht_0}{\ell^2},\qquad
B_2\equiv\sinh^2\frac{r_hL_{31}}{2\ell^2}+\cosh^2\frac{r_ht_0}{\ell^2},\notag\\
B_3&\equiv\sinh^2\frac{r_h(L_2-R_1)}{2\ell^2},\qquad
B_4\equiv\sinh^2\frac{r_h(L_3-R_1)}{2\ell^2}.
\end{align}
The chordal invariant is then
    {\small\begin{align}
\zeta&=\frac{\mathcal N_{DH}}
{2\sinh\frac{r_hL_{32}}{2\ell^2}
\sqrt{\sinh^2\frac{r_h(L_1-R_1)}{2\ell^2}+\cosh^2\frac{r_ht_0}{\ell^2}}},\label{zeta-dome-HM}\\
\mathcal N_{DH}&\equiv
\cosh(\sigma_H+\sigma_D)(B_1+B_4)+\sinh(\sigma_H+\sigma_D)(B_4-B_1)\notag\\
&\quad+\cosh(\sigma_H-\sigma_D)(B_2+B_3)+\sinh(\sigma_H-\sigma_D)(B_3-B_2).\notag
\end{align}}
The physical separator is the unique geodesic orthogonal to both supports.  Thus $\partial_{\sigma_H}\zeta=\partial_{\sigma_D}\zeta=0$.  Solving these two equations fixes the signed positions of the two contact points as
    \begin{align}
        \sigma_H&=\frac{1}{4} \log \left[\frac{\left(\cosh\frac{r_h L_{21}}{\ell^2}+\cosh\frac{2r_ht_0}{\ell^2}\right) \left(\cosh \frac{r_h L_{31}}{\ell^2}+\cosh \frac{2r_ht_0}{\ell^2}\right)}{4 \sinh ^2\frac{r_h \left(L_2-R_1\right)}{2 \ell^2} \sinh ^2\frac{r_h \left(L_3-R_1\right)}{2 \ell^2}}\right]\,,\notag\\
        \sigma_D&=\frac{1}{4}\log\left[\frac{\sinh ^2\frac{r_h \left(L_2-R_1\right) }{2 \ell^2} \left(\cosh\frac{r_h L_{21} }{\ell^2}+\cosh\frac{2 r_ht_0}{\ell^2}\right)}{\sinh ^2\frac{r_h \left(L_3-R_1\right)}{2 \ell^2} \left(\cosh \frac{r_h L_{31} }{\ell^2}+\cosh\frac{2 r_ht_0}{\ell^2}\right)}\right]
    \end{align}
The logarithms locate the common perpendicular relative to the midpoints of the two supporting geodesics; their signs therefore record which side of each midpoint the separator reaches in an asymmetric configuration.  Evaluating the chordal invariant at these orthogonal intersection points gives the left-copy extremal length
    {\footnotesize\begin{align}
        \mathcal L_{\rm ext}=\ell\,\operatorname{arccosh}\Bigg[\frac{\sinh\frac{r_h \left(L_3-R_1\right)}{2 \ell^2} \sqrt{\cosh\frac{r_h L_{21} }{\ell^2}+\cosh\frac{2 r_ht_0}{\ell^2}}+\sinh \frac{r_h \left(L_2-R_1\right)}{2 \ell^2} \sqrt{\cosh \frac{r_h L_{31} }{\ell^2}+\cosh\frac{2r_ht_0}{\ell^2}}}{\sinh\frac{r_h L_{32}}{2 \ell^2}\sqrt{\cosh\frac{r_h \left(L_1-R_1\right)}{\ell^2}+\cosh\frac{2r_h t_0}{\ell^2}}}\Bigg].
    \end{align}}
The extremal left-copy invariant is precisely $\mathcal Z_L$ in the boundary block.  Adding its right-copy image therefore gives
\begin{align}
    S_R^{\text{bulk}}(A:B)= \frac{\ell}{2 G_N}  \operatorname{arccosh}\Bigg[&\frac{\sinh\frac{r_h \left(L_3-R_1\right)}{2 \ell^2} \sqrt{\cosh\frac{r_h L_{21}}{\ell^2}+\cosh\frac{2 r_ht_0}{\ell^2}}}{\sinh\frac{r_h L_{32}}{2 \ell^2}\sqrt{\cosh\frac{r_h \left(L_1-R_1\right)}{\ell^2}+\cosh\frac{2r_h t_0}{\ell^2}}}\notag\\    
    &+\frac{\sinh \frac{r_h \left(L_2-R_1\right)}{2 \ell^2} \sqrt{\cosh \frac{r_h L_{31}}{\ell^2}+\cosh\frac{2r_ht_0}{\ell^2}}}{ \sinh\frac{r_h L_{32}}{2 \ell^2}\sqrt{\cosh\frac{r_h \left(L_1-R_1\right)}{\ell^2}+\cosh\frac{2r_h t_0}{\ell^2}} }\Bigg]+ \left(L_{i} \rightarrow R_{i}\right).
\end{align}
After the Brown--Henneaux relation, this bulk expression reproduces the two boundary invariants exactly.  This agreement makes the role of the finite gap transparent.  The denominator contains the central-dome chord $\sinh(r_hL_{32}/2\ell^2)$, whereas the square-root factors measure the distance to the time-dependent HM support.  As the wormhole stretches, the common perpendicular is driven away from the bridge and the competing dome--dome channel can become favorable before the outer entanglement saddle itself changes.

\subsubsection*{Mirror cross section ending on the outer HM bridge}

\paragraph{Boundary description.}
When $B$ is smaller, the preferred separator is the mirror common perpendicular ending on the outer HM support.  The replica channel is obtained by exchanging the two parties, and the eight-point numerator is correspondingly, in the dominant large-$c$ channel, approximated by a spectator two-point function and a six-point block:
\begin{align}
     &\langle \sigma_{g^{}_A}({P_1})\sigma_{g_A^{-1}}({P_{2}}) \sigma_{g_B} ({P_{3}})\sigma_{g_B^{-1}}({P_{4}}) \sigma_{g^{}_B}({Q_{4}}) \sigma_{g^{-1}_B}({Q_{3}}) \sigma_{g^{}_A  }({Q_{2}})\sigma_{g_A^{-1}}({Q_{1}}) \rangle\notag\\
    &\simeq \langle \sigma_{g_A^{}}({P_{1}})\sigma_{g^{-1}_A}({Q_{4}}) \rangle ~\langle  \sigma_{g_A^{-1}}({P_{2}}) \sigma_{g_B} ({P_{3}}) \sigma_{g^{-1}_B} ({P_4}) \sigma_{g_B^{}}({Q_{4}}) \sigma_{g^{-1}_B}({Q_{3}}) \sigma_{g^{}_A  }({Q_{2}})\rangle.
\end{align}
In the dominant large-$c$ channel, the six-point block is approximated by the two four-point blocks appropriate to this mirror OPE channel \cite{Banerjee:2016qca,Basu:2023jtf}:
\begin{align}
      \mathcal{F}_6(P_2,P_3,P_4,Q_4,Q_3,Q_2;h,h_{AB})=& \mathcal{F}_4 (P_2,P_3,P_4,Q_4;h,h_{AB}) \notag\\&\times \mathcal{F}_4 (Q_2,Q_3,Q_4,P_4;h,h_{AB}).
\end{align}
The continuation is the mirror of the inner HM cross-section channel.  Consequently
\begin{align}\label{dis-sr1-iii-bdy}
S_R^{\rm bdy}(A:B)=\frac{c}{3}\operatorname{arccosh}\widetilde{\mathcal Z}_L+\frac{c}{3}\operatorname{arccosh}\widetilde{\mathcal Z}_R,
\end{align}
with $\widetilde{\mathcal Z}_{L,R}$ obtained from $\mathcal Z_{L,R}$ in \cref{dis-sr1-ii-bdy} by the endpoint replacement $(L_1,R_1)\to(L_4,R_4)$ and reversal of the corresponding interval orientation.  The explicit expression is displayed in the bulk result below and agrees after $c=3\ell/(2G_N)$.

\paragraph{Bulk description.}
The mirror bulk saddle is obtained by attaching the common perpendicular to the outer HM surface instead, shown as dashed curves in \cref{fig:-FT-dis-2HM}.  No new local extremization is required: exchanging the outer endpoint pair maps the previous invariant to the present one.  The resulting bulk reflected entropy is
\begin{align}
     S_R^{\text{bulk}}(A:B)= \frac{\ell}{2 G_N}  \operatorname{arccosh}\Bigg[&\frac{\sinh\frac{r_h \left(L_3-R_4\right)}{2 \ell^2} \sqrt{\cosh\frac{r_h L_{42}}{\ell^2}+\cosh\frac{2 r_ht_0}{\ell^2}}}{\sinh\frac{r_h L_{32}}{2 \ell^2}\sqrt{\cosh\frac{r_h \left(L_4-R_4\right)}{\ell^2}+\cosh\frac{2r_h t_0}{\ell^2}}}\notag\\    
    &+\frac{\sinh \frac{r_h \left(L_2-R_4\right)}{2 \ell^2} \sqrt{\cosh \frac{r_h L_{43}}{\ell^2}+\cosh\frac{2r_ht_0}{\ell^2}}}{ \sinh\frac{r_h L_{32}}{2 \ell^2}\sqrt{\cosh\frac{r_h \left(L_4-R_4\right)}{\ell^2}+\cosh\frac{2r_h t_0}{\ell^2}} }\Bigg]+ \left(L_{i} \rightarrow R_{i}\right).
\end{align}
Under exchange of the two disjoint parties, the supporting HM geodesic is swapped while the central-dome denominator remains unchanged, as required by the mirror symmetry.  Thus the finite gap is a common geometric reference, while the numerator identifies which outer wormhole support controls the purification.  The boundary and bulk formulas again coincide by $c=3\ell/(2G_N)$.

\subsubsection{One island-supported outer pair with a remaining wormhole bridge} 
\begin{figure}[ht]
    \centering
    \includegraphics[width=0.55\linewidth]{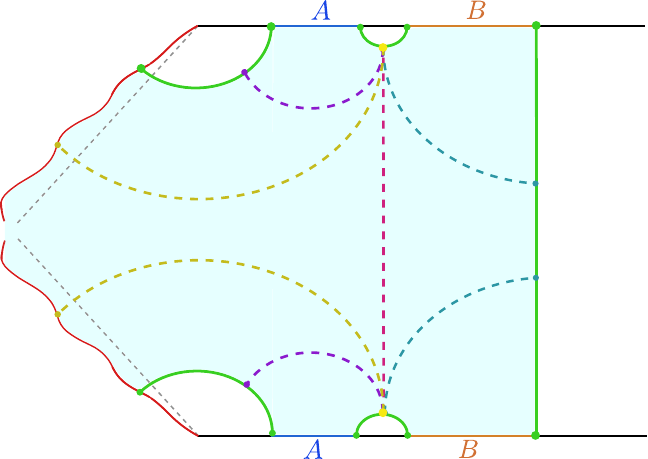}
    \caption{Schematic depiction of the distinct reflected-entropy, or equivalently bulk-EWCS, saddles between $A$ and $B$ when the entropy of $A\cup B$ contains one island-supported outer pair and one remaining wormhole bridge.  Candidate cross sections are shown by dashed curves.}
    \label{fig:-FT-dis-1HM}
\end{figure}
A second class of wedges appears when the endpoint pair closest to the brane has already undergone an island transition but the outer pair still prefers a wormhole-crossing surface.  The central gap continues to be bounded by two domes.  The resulting union entropy is therefore a sum of one island pair, one HM contribution and the two central domes, shown as solid curves in \cref{fig:-FT-dis-1HM}, which may be computed as
\begin{align}\label{dis-EE-S2}
S_2=J(L_1,R_1)+H(L_4,R_4;t_0)+D(L_2,L_3)+D(R_2,R_3).
\end{align}
Only the outer HM term carries time dependence.  The wedge is therefore geometrically asymmetric: one outer pair is already purified through the brane while the other still probes the wormhole interior.  Reflected entropy can distinguish four internal arrangements inside this same outer saddle: the central dome--dome separator, a dome--island separator, a dome--brane separator, and the mirror separator tied to the remaining outer support.  Their competition determines whether the canonical purification is localized near the central gap, shared with the island, or still partly stored in the black-hole interior.

\subsubsection*{Dome--dome cross section}
One possibility is that the internal cross section remains entirely between the central dome surfaces.  The local geometry of those domes is unchanged by replacing an outer wormhole-crossing surface with an island pair, so the common-perpendicular problem is identical to the dome--dome saddle derived above.  Its reflected entropy is therefore still given by \cref{sr-thermal-dis-dome-dome}.  This is an instructive example in which the union has undergone an island transition while the local mixed-state cross section has not yet changed topology.

\subsubsection*{Cross section ending on an island surface}

\paragraph{Boundary description.}
If $A$ is the smaller component, the common perpendicular leaves the central dome and ends on the island surface associated with the inner island pair.  There is no independent reflected-island separator on the brane, so the lower-dimensional generalized-area contribution in \cref{SR-bdy} is absent; the higher-dimensional EWCS area remains nonzero, and the gravitational dependence enters through the location of the supporting island geodesic.  The corresponding boundary contribution is the dominant large-$c$ four-point block in the dome--island channel, whose endpoint pairing is dual to this common perpendicular.  It is the surviving block after the spectator factors associated with the mixed outer wedge are removed, rather than an exact factorization identity:
\begin{align}
    S_R^{\text{bdy}}(A:B)= & \lim_{{m,n} \to 1}\frac{1}{1-n}\log \frac{   \langle \sigma_{g^{}_A}(P_1)\sigma_{ g_A^{-1}}(P_2) \sigma_{ g_B^{}}(P_3) \sigma_{g^{-1}_A}( a_{P_{1}} )\rangle}{  \langle \sigma_{g^{}_m}(P_1)\sigma_{ g_m^{-1}}(P_2) \sigma_{ g_m^{}}(P_3) \sigma_{g^{-1}_m}( a_{P_{1}} )\rangle^n}\notag\\
    & +\lim_{{m,n} \to 1}\frac{1}{1-n}\log \frac{      \langle \sigma_{g^{-1}_A}(Q_1)\sigma_{g_A^{}}(Q_2) \sigma_{g_B^{-1}}(Q_3) \sigma_{g^{}_A}( a_{Q_{1}} )\rangle}{\langle \sigma_{g^{}_m}(Q_1)\sigma_{ g_m^{-1}}(Q_2) \sigma_{ g_m^{}}(Q_3) \sigma_{g^{-1}_m}( a_{Q_{1}}  )\rangle^n},
\end{align}
The large-$c$ four-point block exponentiates into the geodesic invariant between the central dome and the supporting island surface.  Keeping the left and right copies distinct, the boundary result is
\begin{align}\label{sr-dis2-ii-bdy}
S_R^{\rm bdy}(A:B)=&\frac{c}{3}\operatorname{arccosh}\!\left[
\frac{1-m_{1L}m_{2L}\cosh\frac{r_h(x_{c2,L}-x_{c1,L})}{\ell^2}}
{\sqrt{(1-m_{1L}^2)(1-m_{2L}^2)}}\right]+(L\rightarrow R).
\end{align}
The brane fluctuation enters $m_{1L,R}$ through the DES endpoint.  At first order its on-shell contribution can be evaluated on the unperturbed cross section by \cref{first-order-extremization}; the result agrees with the bulk expression through $\mathcal O(\nu)$.

\paragraph{Bulk description.}

The corresponding dashed curves in \cref{fig:-FT-dis-1HM} display the dome--island saddle explicitly.  On each copy the EWCS is the common perpendicular between the central dome and the island surface.  Both endpoints are therefore variational, and the reflected entropy is sensitive to the brane only through the displacement of the supporting island geodesic.  Because both supporting surfaces are static on the chosen TFD copy, the endpoints share the boundary time $t_0$.  The geodesic length between arbitrary points $(x_1,t_0,r_1)$ and $(x_2,t_0,r_2)$ therefore follows from \cref{BTZ-length} as
\begin{align}
    \mathcal{L}=\ell\,\operatorname{arccosh}\left[\frac{r_1 r_2}{r_h^2}\cosh\frac{r_h(x_1-x_2)}{\ell^2}-\frac{\sqrt{(r_1^2-r_h^2)(r_2^2-r_h^2)}}{r_h^2}\right].
\end{align}
Substituting the two geodesic profiles converts the radial coordinates into intrinsic parameters of the supporting surfaces.  The equal-time distance then becomes
\begin{align}
\mathcal{L}=\ell\,\operatorname{arccosh}\!\left[
\frac{\cosh\frac{r_h(x_1-x_2)}{\ell^2}-m_1m_2
\cosh\frac{r_h(x_1-x_{c_1})}{\ell^2}\cosh\frac{r_h(x_2-x_{c_2})}{\ell^2}}
{\sqrt{\left(1-m_1^2\cosh^2\frac{r_h(x_1-x_{c_1})}{\ell^2}\right)
\left(1-m_2^2\cosh^2\frac{r_h(x_2-x_{c_2})}{\ell^2}\right)}}\right].
\end{align}
where $(x_{c_{1}},m_1)$ is defined in \cref{center_radius_island} and
\begin{align}\label{radius-center-dome}
    x_{c_{2}}=\frac{L_2+L_3}{2} ~, ~ m_2= \operatorname{sech}\frac{r_h(L_3-L_2)}{2\ell^2}.
\end{align}
The two endpoint coordinates are auxiliary: what is physically invariant is the common perpendicular between the island geodesic and the central dome.  It is therefore preferable not to solve the two stationarity equations in exponential coordinates.  Denoting the dimensionless displacements from the geodesic centers by
\begin{align}
 y_1\equiv\frac{r_h(x_1-x_{c1})}{\ell^2},\qquad
 y_2\equiv\frac{r_h(x_2-x_{c2})}{\ell^2},\qquad
 \Delta_c\equiv\frac{r_h(x_{c2}-x_{c1})}{\ell^2},
\end{align}
the extremal cross section is equivalently characterized by
\begin{align}
 \partial_{y_1}\mathcal L=0,\qquad \partial_{y_2}\mathcal L=0.
 \label{dome-island-orthogonality}
\end{align}
Geometrically, these conditions state that the cross section meets both supporting geodesics orthogonally.  Eliminating $y_1$ and $y_2$ from \cref{dome-island-orthogonality} produces a compact answer that depends only on the two geodesic invariants $m_1,m_2$ and on the separation of their centers.  Adding the left- and right-copy contributions, the reflected entropy is
   \begin{align}\label{sr-thermal-dis-dome-island}
S_R^{\rm bulk}(A:B)=&\frac{\ell}{2G_N}\operatorname{arccosh}\!\left[
\frac{1-m_{1L}m_{2L}\cosh\frac{r_h(x_{c2,L}-x_{c1,L})}{\ell^2}}
{\sqrt{(1-m_{1L}^2)(1-m_{2L}^2)}}\right]\notag\\
&+\frac{\ell}{2G_N}\operatorname{arccosh}\!\left[
\frac{1-m_{1R}m_{2R}\cosh\frac{r_h(x_{c2,R}-x_{c1,R})}{\ell^2}}
{\sqrt{(1-m_{1R}^2)(1-m_{2R}^2)}}\right],
\end{align}
where $(m_{1L},x_{c1,L})$ and $(m_{1R},x_{c1,R})$ are the left/right island-geodesic parameters, while $(m_{2L},x_{c2,L})$ and $(m_{2R},x_{c2,R})$ are the corresponding dome parameters.  The two terms must not be artificially identified when the left and right endpoints are asymmetric.  The answer depends only on the invariant separation of the two supporting geodesics, which makes its physical content clear: the brane fluctuation modifies reflected entropy by moving the island support of the cross section, while the central dome acts as a fixed thermal reference surface.

\subsubsection*{Cross section ending directly on the EOW brane}

\paragraph{Boundary description.}
For this reflected-entropy saddle the cross section leaves the central dome and reaches the EOW brane itself.  The generalized reflected entropy contains the defect contribution associated with the brane endpoint and the effective reflected entropy of the two enlarged regions.  Denoting the left and right cross-section endpoints by $a_L,a_R$, we write
\begin{align}\label{dis-eff}
S_R^{\rm bdy}(A:B)=\underset{a_L,a_R}{\rm ext}\Bigg\{\frac{c}{3}\frac{\rho_0}{\ell}\left[2+\Phi(a_L)+\Phi(a_R)\right]
+S_{R,\rm eff}^{(L)}(a_L)+S_{R,\rm eff}^{(R)}(a_R)\Bigg\}.
\end{align}
In the semiclassically dominant brane-ending channel, the effective terms are the thermal three-point blocks obtained by fusing the twists at the two endpoints of the central dome.  This conformal-block choice is the boundary image of a common perpendicular from that dome to the EOW brane and is used only while the corresponding outer wedge and internal channel dominate.  This form makes the island/DES matching manifest and is sufficient for the first-order expansion, because the $\mathcal O(\nu)$ displacement of $a_{L,R}$ does not enter the on-shell value at this order; cf. \cref{first-order-extremization}.

\paragraph{Bulk description.}
In the bulk, $\Sigma_{AB}$ is the common perpendicular from the central dome to the EOW brane.  Because it intersects the brane, it directly partitions the reflected island and must be varied together with the defect entropy. These extremal geodesics are shown as dashed curves in \cref{fig:-FT-dis-1HM}.  The effective term in \cref{SR-bulk} is the defect/matter contribution already encoded in \cref{dis-eff}; the geometric term is the equal-time distance from a point $(x_1,t_0,r_1)$ on the central dome to a point $(x_2,t_0,r_2)$ on the EOW brane.  Using \cref{BTZ-length}, the trial distance is
\begin{align}
    \mathcal{L}=&\ell\,\operatorname{arccosh}\left[\frac{r_1 r_2}{r_h^2}\cosh\frac{r_h(x_1-x_2)}{\ell^2}-\frac{\sqrt{(r_1^2-r_h^2)(r_2^2-r_h^2)}}{r_h^2}\right],\notag\\
    =& \ell \operatorname{arccosh} \left[ \frac{r_2 \cosh\frac{r_h(x_1-x_2)}{\ell^2}- m_2 \sqrt{r_2^2-r_h^2} \cosh\frac{r_h(x_1-x_{c_{2}})}{\ell^2} }{r_h \sqrt{1- m_2^2 \cosh^2 \frac{r_h(x_1-x_{c_{2}})}{\ell^2}}}  \right],
\end{align}
where we have used the geodesic equation of the dome-type RT surface to rewrite $r_1$ in terms of $x_1$.  The coordinate value of the contact point on the dome is not illuminating and, in exponential coordinates, obscures the symmetry of the result.  We instead impose the invariant common-perpendicular condition
\begin{align}
 \partial_{x_1}\mathcal L=0,
 \label{dome-brane-orthogonality}
\end{align}
which states that the cross section is orthogonal to the dome at its endpoint.  Using the brane point \cref{EOW-point} to eliminate $(x_2,r_2)$, the trial distance can be expressed entirely in hyperbolic functions,
\begin{align}
\frac{\mathcal L(a,\rho)}{\ell}
=\operatorname{arccosh}\!\left[
\sqrt{1+\frac{\left(\operatorname{csch}(\kappa a)-m_2\left[\coth(\kappa a)\cosh(\kappa x_{c2})+\tanh(\rho/\ell)\sinh(\kappa x_{c2})\right]\right)^2}
{(1-m_2^2)\operatorname{sech}^2(\rho/\ell)}}\right].
\label{dome-brane-hyperbolic}
\end{align}
The rigid-brane stationarity condition is surprisingly simple.  With $m_2=\operatorname{sech}d_L$, $d_L\equiv\kappa(L_3-L_2)/2$ and $U_L\equiv\kappa(L_2+L_3)/2$, it fixes
\begin{align}
\kappa a_L^{(0)}=\operatorname{arccosh}\!\left(m_2\cosh U_L\right).
\label{D2b-LO}
\end{align}
Although the coordinate location $a_L^{(0)}$ is independent of the rigid brane position $\rho_0$, the proper distance to the brane is not.  Evaluating \cref{dome-brane-hyperbolic} at the stationary point and using
\begin{align}
\sinh y_L=\frac{\sqrt{\sinh(\kappa L_2)\sinh(\kappa L_3)}}{\sinh d_L},
\qquad
\cosh y_L=\frac{\sinh U_L}{\sinh d_L},
\label{dome-brane-y}
\end{align}
reduces the chordal combination to $\sinh(\rho_0/\ell+y_L)$.  Hence the rigid common-perpendicular length is simply
\begin{align}
\mathcal L_L^{(0)}
=\rho_0+\ell y_L
=\rho_0+\ell\operatorname{arccosh}\!\left[\frac{\sinh U_L}{\sinh d_L}\right].
\label{dome-brane-rigid-length}
\end{align}
Moving the constant-tension brane outward shifts the common-perpendicular length by exactly the same proper distance, while the remaining term depends only on the thermal geometry of the central dome.

We now turn on the JT mode by writing $\rho=\rho_0+\Delta\rho$.  At the rigid saddle one finds $\partial\mathcal L/\partial\rho=1$, so the explicit first-order response is simply the local brane displacement.  By \cref{first-order-extremization}, the first-order shift of $a_L$ does not contribute to the on-shell length.  Using the thermal dilaton relation $\Delta\rho(a)=\ell\nu\coth(\kappa a)$ therefore gives
\begin{align}
\mathcal L_L
=\rho_0+\ell\operatorname{arccosh}\!\left[\frac{\sinh U_L}{\sinh d_L}\right]
+\ell\nu\coth\!\left(\kappa a_L^{(0)}\right)+\mathcal O(\nu^2).
\label{BTZ-dis-ext-length}
\end{align}
The corresponding left-copy contribution to the reflected entropy, including the effective defect term, is
\begin{align}
S_{R,L}^{\rm bulk}=&\frac{1}{2G_N}\left\{\rho_0+\ell\operatorname{arccosh}\!\left[\frac{\sinh\frac{\kappa(L_2+L_3)}{2}}{\sinh\frac{\kappa(L_3-L_2)}{2}}\right]
+\ell\nu\coth\!\left(\kappa a_L^{(0)}\right)\right\}
+S_{R,\rm defect}^{(L)}+\mathcal O(\nu^2),
\label{dis-eow-left}
\end{align}
with $a_L^{(0)}$ given in \cref{D2b-LO}.  The right-copy term follows by $L_i\to R_i$.  Equations \cref{dis-eff,dis-eow-left} are dimensionally homogeneous: every geometric length is divided by $2G_N$, whereas the brane fluctuation enters as $\Delta\rho/(2G_N)$.  This dimensional bookkeeping is essential because the geodesic length and the entropy carry different normalizations before division by $2G_N$.  Evaluating the boundary thermal blocks gives the same result through first order in $\nu$.  We keep the answer in this invariant form because it remains compact and is numerically stable in the limits $L_3\to L_2$ and $r_h\to0$.  This brane-ending saddle is also the disjoint analogue of the most genuinely gravitational adjacent configuration: the cross section itself intersects the EOW brane and partitions the reflected island.  Consequently the linearized displacement $\Delta\rho$ contributes directly to the generalized cross-section cost.  The fact that its first-order endpoint shift drops out of the on-shell value is a concrete application of \cref{first-order-extremization}.

\subsubsection*{Mirror cross section ending on the outer island surface}
\paragraph{Boundary description.}
When subsystem $B$ is the smaller component, the relevant channel is the mirror of the island-surface channel: the EWCS terminates on the island surface associated with the outer endpoint $L_4$ (and $R_4$ on the second copy).  The twist-block result is obtained from \cref{sr-dis2-ii-bdy} by the endpoint exchange
\begin{align}
(L_1,R_1;L_2,L_3;R_2,R_3)\longleftrightarrow(L_4,R_4;L_3,L_2;R_3,R_2).
\label{dis-phaseIV-map}
\end{align}
\paragraph{Bulk description.}
The same replacement maps the island--dome common perpendicular in \cref{sr-thermal-dis-dome-island} to the required bulk cross section.  Since both descriptions are related by the same endpoint map, the equality through $\mathcal O(\nu)$ follows without a new independent extremization.  We nevertheless include this saddle in the numerical minimization used for the Page curves.

\subsubsection{Fully same-boundary dome saddle}
\begin{figure}[ht]
    \centering
    \includegraphics[width=0.5\linewidth]{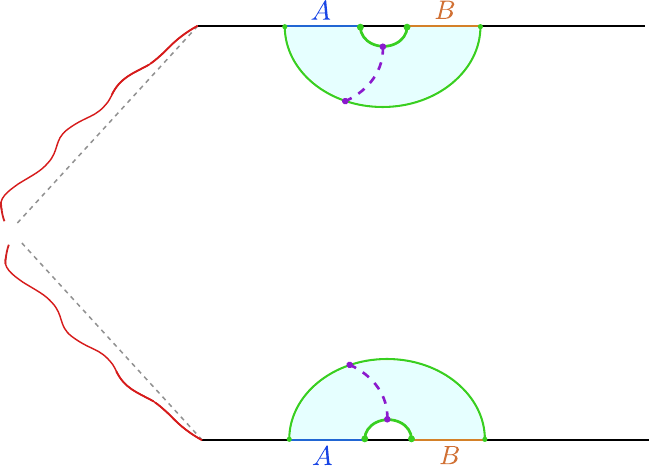}
    \caption{Schematic illustration of the reflected-entropy, or equivalently bulk-EWCS, saddle between $A$ and $B$ when the entropy of $A\cup B$ is supported by same-boundary domes.  The cross section is shown by a dashed curve.}
    \label{fig:-FT-dis-dome}
\end{figure}
When both disjoint regions are sufficiently small, the complete union is reconstructed on the individual asymptotic boundaries.  The entanglement surface consists of four same-boundary domes and is independent of $t_0$.  In building-block notation the union entropy is
\begin{align}\label{dis-EE-S3}
 S_3=D(L_1,L_4)+D(R_1,R_4)+D(L_2,L_3)+D(R_2,R_3).
\end{align}
There is no island term in this saddle.  Consequently all time dependence drops out of the outer entropy, and the reflected entropy is determined entirely by common perpendiculars between same-boundary dome surfaces.  This branch is the thermal analogue of the vacuum disjoint-interval cross-ratio result and represents a purification localized completely in the exterior geometry.

\paragraph{Boundary description.}
In the boundary description no reflected island is present, so $S_R$ is purely the effective CFT reflected entropy.  The left and right TFD copies decouple in the dominant same-boundary channel, and the eight-point numerator of \cref{dis-SR-def} is, in the dominant large-$c$ channel, approximated by one four-point block on each copy:
\begin{align}
 &\langle \sigma_{g^{}_A}({P_1})\sigma_{g_A^{-1}}({P_{2}}) \sigma_{g_B} ({P_{3}})\sigma_{g_B^{-1}}({P_{4}}) \sigma_{g^{}_B}({Q_{4}}) \sigma_{g^{-1}_B}({Q_{3}}) \sigma_{g^{}_A  }({Q_{2}})\sigma_{g_A^{-1}}({Q_{1}}) \rangle\notag\\
    &\simeq \langle \sigma_{g^{}_A}({P_1})\sigma_{g_A^{-1}}({P_{2}}) \sigma_{g_B} ({P_{3}})\sigma_{g_B^{-1}}({P_{4}})\rangle 
     \times \langle  \sigma_{g^{}_B}({Q_{4}}) \sigma_{g^{-1}_B}({Q_{3}}) \sigma_{g^{}_A  }({Q_{2}})\sigma_{g_A^{-1}}({Q_{1}})\rangle.
\end{align}
The normalization correlator has the same dominant-channel decomposition.  Dividing by it removes the ordinary R\'enyi normalization and leaves one normalized reflected-entropy four-point block on each TFD copy.  Therefore
 \begin{align}
     S_R^{\rm bdy}(A:B)=& \lim_{{m,n} \to 1}\frac{1}{1-n}\log \frac{      \langle \sigma_{g^{}_A}({P_1})\sigma_{g_A^{-1}}({P_{2}}) \sigma_{g_B} ({P_{3}})\sigma_{g_B^{-1}}({P_{4}})\rangle}{ \langle \sigma_{g^{}_m}({P_1})\sigma_{g_m^{-1}}({P_{2}}) \sigma_{g_m} ({P_{3}})\sigma_{g_m^{-1}}({P_{4}})\rangle^n} \notag\\
        &+\lim_{{m,n} \to 1}\frac{1}{1-n}\log \frac{   \langle  \sigma_{g^{}_B}({Q_{4}}) \sigma_{g^{-1}_B}({Q_{3}}) \sigma_{g^{}_A  }({Q_{2}})\sigma_{g_A^{-1}}({Q_{1}})\rangle}{ \langle  \sigma_{g^{}_m}({Q_{4}}) \sigma_{g^{-1}_m}({Q_{3}}) \sigma_{g^{}_m  }({Q_{2}})\sigma_{g_m^{-1}}({Q_{1}})\rangle^n}.
 \end{align}
Using the large-$c$ four-point block on each copy, the boundary answer is
\begin{align}\label{dis-sr3-bdy}
S_R^{\rm bdy}(A:B)=&\frac{c}{3}\operatorname{arccosh}\!\left[
\frac{\sinh\frac{r_hL_{31}}{2\ell^2}\sinh\frac{r_hL_{42}}{2\ell^2}+\sinh\frac{r_hL_{21}}{2\ell^2}\sinh\frac{r_hL_{43}}{2\ell^2}}
{\sinh\frac{r_hL_{41}}{2\ell^2}\sinh\frac{r_hL_{32}}{2\ell^2}}\right]\notag\\
&+\frac{c}{3}\operatorname{arccosh}\!\left[
\frac{\sinh\frac{r_hR_{31}}{2\ell^2}\sinh\frac{r_hR_{42}}{2\ell^2}+\sinh\frac{r_hR_{21}}{2\ell^2}\sinh\frac{r_hR_{43}}{2\ell^2}}
{\sinh\frac{r_hR_{41}}{2\ell^2}\sinh\frac{r_hR_{32}}{2\ell^2}}\right].
\end{align}
The formula is the thermal-cylinder image of the familiar common-perpendicular result for two disjoint intervals \cite{Dutta:2019gen,Jeong:2019xdr,Tamaoka:2018ned}.  On each copy define, for example,
\begin{align}
X_\beta^{(L)}=\frac{\sinh\frac{\kappa L_{21}}2\,\sinh\frac{\kappa L_{43}}2}
{\sinh\frac{\kappa L_{31}}2\,\sinh\frac{\kappa L_{42}}2},
\qquad X_\beta^{(R)}=X_\beta^{(L)}\big|_{L_i\to R_i}.
\label{thermal-cross-ratios}
\end{align}
For ordered endpoints $0<X_\beta^{(L,R)}<1$, so the corresponding common perpendiculars exist as real geodesics.  This does not by itself establish dominance: in the ordinary same-boundary large-$c$ competition the connected channel on a copy dominates only for $X_\beta>1/2$, and in the full two-sided problem the decisive condition is the outer-wedge inequality \cref{thermal-connected-domain}.  Thus \cref{dis-sr3-bdy} is used only in the dominant connected channel; if the disconnected union wins, the leading reflected entropy is zero.  Within its domain, the late-time reflected entropy is controlled by local interval geometry rather than by wormhole growth or brane dynamics.

\paragraph{Bulk description.}
The bulk calculation is the common-perpendicular problem between the inner and outer dome geodesics on each copy.  Choosing arbitrary points $(x_1,t_0,r_1)$ and $(x_2,t_0,r_2)$ on the two left-copy domes, the BTZ invariant gives the trial length
\begin{align}
    \mathcal{L}=&\ell\,\operatorname{arccosh}\left[\frac{r_1 r_2}{r_h^2}\cosh\frac{r_h(x_1-x_2)}{\ell^2}-\frac{\sqrt{(r_1^2-r_h^2)(r_2^2-r_h^2)}}{r_h^2}\right].
\end{align}
Using the parametrization of dome-type RT surfaces from \cref{dome-RT-para}, it is convenient to introduce the four positive chord factors
\begin{align}
C_1&\equiv\sinh^2\frac{r_hL_{21}}{2\ell^2},\qquad C_2\equiv\sinh^2\frac{r_hL_{31}}{2\ell^2},\notag\\
C_3&\equiv\sinh^2\frac{r_hL_{42}}{2\ell^2},\qquad C_4\equiv\sinh^2\frac{r_hL_{43}}{2\ell^2}.
\end{align}
The length can then be written without exponential proper-length weights.  Define the hyperbolic numerator
\begin{align}
\mathcal N_{\rm dd}\equiv{}&
\cosh(\sigma_{\rm out}+\sigma_{\rm in})(C_1+C_4)
+\sinh(\sigma_{\rm out}+\sigma_{\rm in})(C_4-C_1)\notag\\
&+\cosh(\sigma_{\rm out}-\sigma_{\rm in})(C_2+C_3)
+\sinh(\sigma_{\rm out}-\sigma_{\rm in})(C_3-C_2).
\end{align}
The common-perpendicular length is therefore
\begin{align}
\mathcal L=\ell\,\operatorname{arccosh}\!\left[
\frac{\mathcal N_{\rm dd}}
{2\sinh(\kappa L_{32}/2)\sinh(\kappa L_{41}/2)}\right].\label{four-dome-length}
\end{align}
Here $\sigma_{\rm in}$ and $\sigma_{\rm out}$ measure signed proper distance from the midpoints of the inner and outer dome geodesics.  The terms even in $\sigma_{\rm in},\sigma_{\rm out}$ describe the symmetric part of the geometry, whereas the $\sinh$ terms encode the displacement of the common perpendicular away from the midpoints.  Extremizing \cref{four-dome-length} over $\sigma_{\rm in}$ and $\sigma_{\rm out}$ gives
\begin{align}
    &\sigma_{\rm in}= \frac{1}{2} \log \left[\sinh \frac{ r_h L_{21} }{2 \ell^2}  \operatorname{csch}\frac{r_h L_{31}}{2 \ell^2} \sinh \frac{r_h L_{42}}{2 \ell^2} \operatorname{csch} \frac{r_h L_{43}}{2 \ell^2}\right],\notag\\
    &\sigma_{\rm out}=\frac{1}{2} \log \left[\sinh \frac{ r_h L_{21} }{2 \ell^2}   \sinh\frac{r_h L_{31}}{2 \ell^2} \operatorname{csch} \frac{r_h L_{42}}{2 \ell^2} \operatorname{csch} \frac{r_h L_{43}}{2 \ell^2}\right].
\end{align}
Evaluating the common-perpendicular invariant at these stationary points and adding the right-copy contribution gives
\begin{align}
    S_R^{\text{bulk}}(A:B) =& \frac{\ell}{2 G_N} \operatorname{arccosh} \left[\frac{\sinh  \frac{r_h L_{31} }{2 \ell^2}  \sinh \frac{r_h L_{42}}{2 \ell^2}+\sinh \frac{r_h L_{21} }{2 \ell^2}  \sinh  \frac{r_h L_{43}}{2 \ell^2}}{\sinh \frac{r_h L_{41} }{2 \ell^2} \sinh \frac{r_h L_{32} }{2 \ell^2}}\right]\notag\\
    &+\frac{\ell}{2 G_N} \operatorname{arccosh} \left[\frac{\sinh  \frac{r_h R_{31} }{2 \ell^2}  \sinh \frac{r_h R_{42}}{2 \ell^2}+\sinh \frac{r_h R_{21} }{2 \ell^2}  \sinh  \frac{r_h R_{43}}{2 \ell^2}}{\sinh \frac{r_h R_{41} }{2 \ell^2} \sinh \frac{r_h R_{32} }{2 \ell^2}}\right].
\end{align}
The Brown--Henneaux relation reproduces \cref{dis-sr3-bdy}.  The absence of any $\rho_0$, $\nu$, or $t_0$ dependence is consistent with the fact that this channel contains neither a brane endpoint nor a wormhole-supported cross section.

\subsubsection{Both outer pairs island-supported with central domes}
\begin{figure}[ht]
    \centering
    \includegraphics[width=0.55\linewidth]{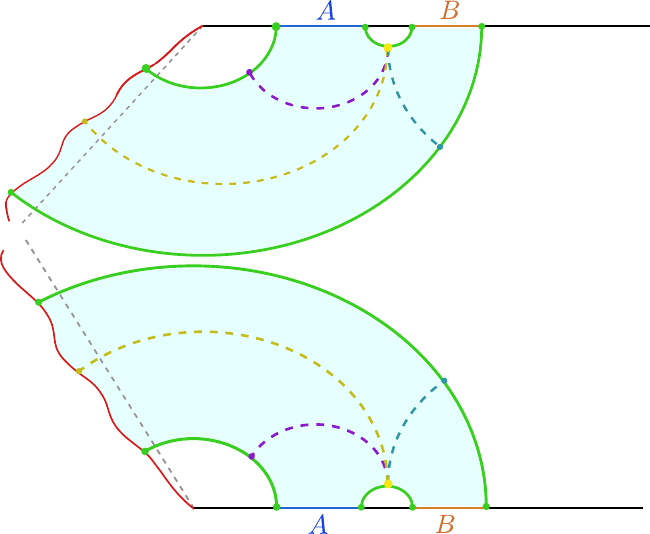}
    \caption{Schematic illustration of the distinct reflected-entropy, or equivalently bulk-EWCS, saddles between $A$ and $B$ when the entropy of $A\cup B$ is supported by two island pairs together with central domes.  Candidate cross sections are shown by dashed curves.}
    \label{fig:-FT-dis-Island}
\end{figure}
The final outer saddle occurs when both endpoint pairs nearest the interface prefer islands while the two inner pairs still support the central domes.  All wormhole-crossing components have disappeared, leaving a time-independent geometry containing two island pairs and two central same-boundary domes.  Its entropy is
\begin{align}\label{dis-EE-S4}
S_4=J(L_1,R_1)+J(L_4,R_4)+D(L_2,L_3)+D(R_2,R_3).
\end{align}
Although the union entropy is now completely static, reflected entropy still distinguishes how the island sector is partitioned.  The separator can end on the inner island surface, end on the outer island surface, or reach the EOW brane between them.  The first two options probe how the supporting island geodesics divide the canonical purification; the brane-ending option instead carries a direct generalized-area cost.  Hence even after all wormhole growth has disappeared, the mixed-state correlations retain a nontrivial gravitational phase structure.

\subsubsection*{Cross section ending on the inner island surface}
If the cross section terminates on the island surface associated with the inner endpoint $L_1$ (and $R_1$), the result is the same island--dome invariant as the island-surface cross section of the mixed entanglement saddle, namely \cref{sr-thermal-dis-dome-island}.  For this outer saddle both endpoint pairs are already island-supported, so only the spectator homology assignment changes; the local common-perpendicular problem and its leading dome--island conformal block are unchanged.  The formula is again conditional on that block being the minimal reflected-entropy channel.

\subsubsection*{Central cross section ending on the EOW brane}
\paragraph{Boundary description.}
For the central saddle the EWCS reaches the EOW brane, so the generalized reflected entropy is the two-sided version of \cref{dis-eff}.  Both outer island pairs belong to the enlarged reflected regions, while the brane endpoint separates their canonical-purification partners.  Unlike either island-surface saddle, this topology carries a direct generalized-area contribution from the reflected-island separator.
\paragraph{Bulk description.}
The bulk cross section is the common perpendicular from the central dome to the EOW brane on each copy.  Its regulated value is obtained from \cref{dis-eow-left} and its $L_i\to R_i$ partner.  The first-order shift of the contact point again drops out of the on-shell value, so the DES and island answers agree through $\mathcal O(\nu)$.  Physically, the remaining first-order term is the normal displacement of the brane evaluated at the rigid common perpendicular, which is precisely the geometric image of the JT dilaton contribution.

\subsubsection*{Cross section ending on the outer island surface}
The final saddle is the mirror of the inner-island-surface saddle with the cross section ending on the island surface associated with $B$.  It is obtained from \cref{sr-thermal-dis-dome-island} by the endpoint map \cref{dis-phaseIV-map}, with the same dominant-channel qualification.

\subsubsection{Page-like transitions for disjoint intervals}
The disjoint case has the same four basic replacement patterns as the adjacent case, but the central separation region introduces an additional pair of dome surfaces that survives in several saddles.  Consequently the transition of the union entropy and the transition of the optimal reflected-entropy cross section need not coincide.  In each figure below, panel (a) shows the competing entanglement saddles and their lower envelope, panel (b) displays the physical reflected entropy together with mutual information, and panel (c) shows the Markov gap on a separate axis.  The separation region is part of the purifying complement $C$ of $A\cup B$, so the disjoint configurations admit a particularly concrete tripartite interpretation: $\mathcal M$ is geometrically sensitive to whether the central complement and, in island phases, the gravitating sector participate in the organization of the $A$--$B$ correlations.

\subsubsection*{From one HM bridge to a dome saddle}

\begin{figure}[ht]
\centering
\begin{subfigure}[b]{0.49\textwidth}
\centering
\includegraphics[width=\linewidth]{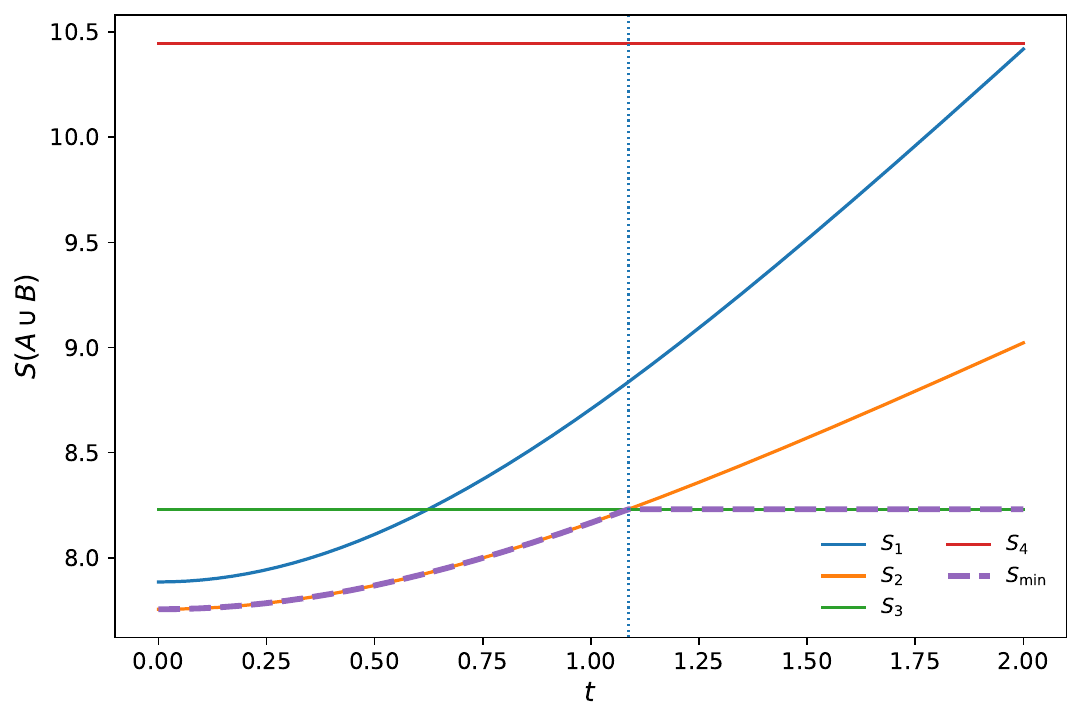}
\caption{Competing union-entropy saddles and their lower envelope.}
\end{subfigure}
\hfill
\begin{subfigure}[b]{0.49\textwidth}
\centering
\includegraphics[width=\linewidth]{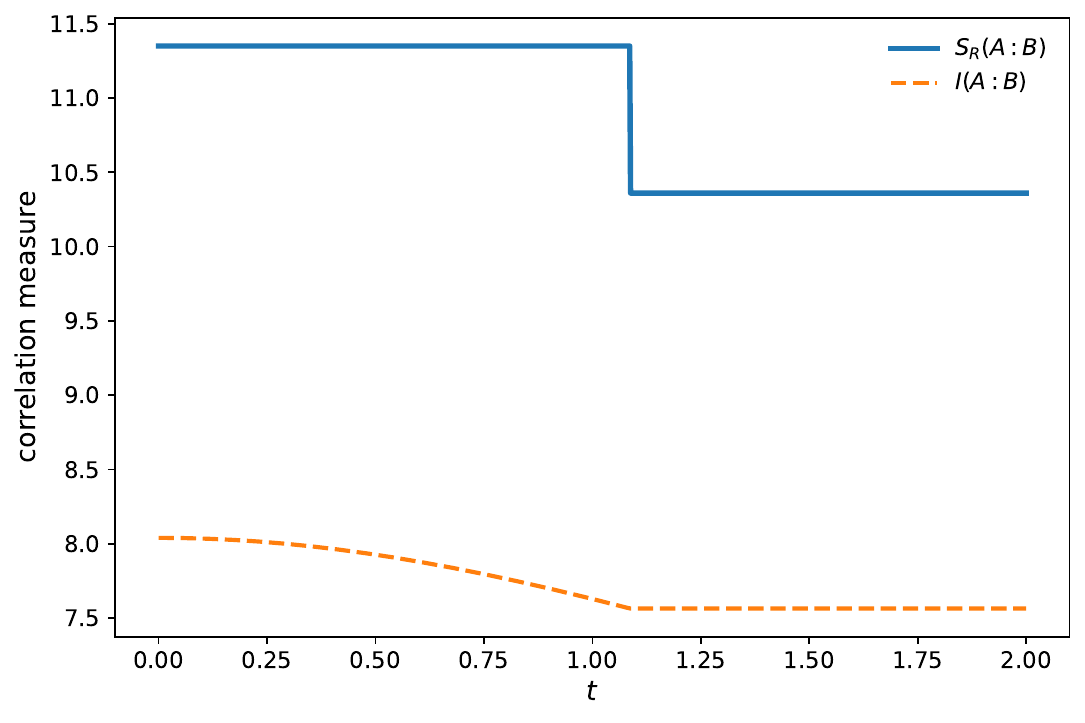}
\caption{Physical reflected entropy $S_R$ and mutual information $I$.}
\end{subfigure}
\par\medskip
\begin{subfigure}[b]{0.62\textwidth}
\centering
\includegraphics[width=\linewidth]{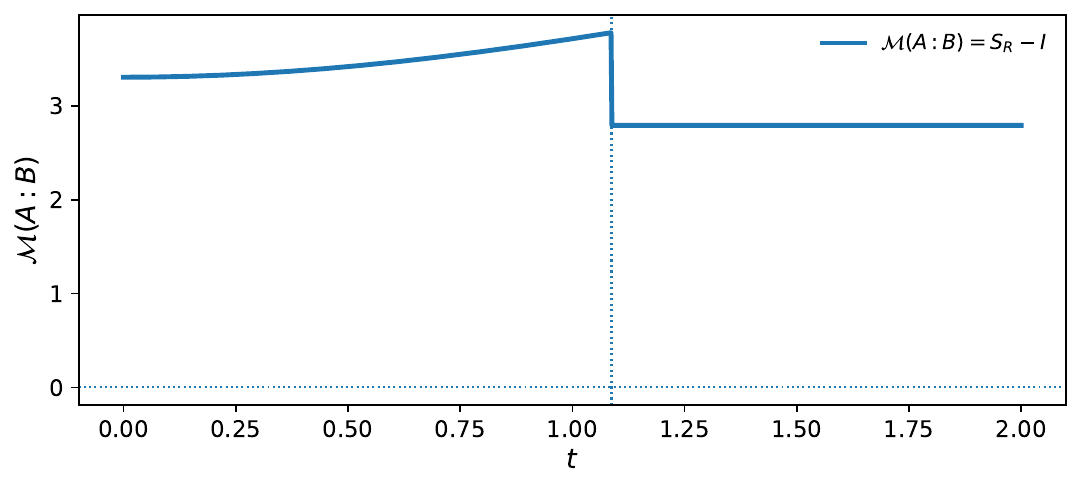}
\caption{Markov gap $\mathcal M=S_R-I$, shown on a separate axis to avoid obscuring the reflected-entropy curve.}
\end{subfigure}
\caption{Disjoint intervals: transition from the remaining outer HM bridge to a same-boundary dome while the central dome pair remains.  Panel (a) shows the union-entropy lower envelope, panel (b) displays $S_R$ and $I$ on unobstructed axes, and panel (c) shows the Markov gap.  Because the separation region between $A$ and $B$ is part of the purifying complement, the nontrivial evolution of $\mathcal M$ is also sensitive to tripartite organization involving this central gap.}
\label{fig:dis-page-bridge-dome}
\end{figure}
The disjoint-interval curves in \cref{fig:dis-page-bridge-dome} demonstrate how the central interval gap introduces an additional geometric scale, so that the optimal cross section can respond more sharply than the entropy of the union.  From the tripartite perspective, this central region belongs to the purifier $C$ and therefore provides an additional sector through which the $A$--$B$ correlations can be organized.  The cusp in panel (c) diagnoses a change in that organization even though the central dome pair itself survives.

For disjoint intervals, the wormhole-to-same-boundary transition is defined by the crossing of $S_2=J(L_1,R_1)+H(L_4,R_4;t)+D(L_2,L_3)+D(R_2,R_3)$ with the four-dome saddle $S_3=D(L_1,L_4)+D(R_1,R_4)+D(L_2,L_3)+D(R_2,R_3)$.  Solving $S_2(t)=S_3$ gives
\begin{align}\label{t-dis-2-3}
 t_{E_{2\to3}}=\frac{\ell^2}{2r_h}\operatorname{arccosh}\!\left[
 \frac{2\epsilon_y^2\,\sinh^2\!\frac{r_hL_{41}}{2\ell^2}\,\sinh^2\!\frac{r_hR_{41}}{2\ell^2}}
 {\ell^2\mathcal B_0^2\sinh\frac{r_hL_1}{\ell^2}\sinh\frac{r_hR_1}{\ell^2}}
 \left(1-\nu \mathcal Q_1\right)-\mathcal C_4\right].
\end{align}
The crossing \cref{t-dis-2-3} replaces the remaining HM contribution by a same-boundary dome while the central domes are spectators.  The additional structure in \cref{fig:dis-page-bridge-dome} comes from the independent internal cross-section competition associated with the finite central gap.

\subsubsection*{From one HM bridge to an island saddle}
\begin{figure}[ht]
\centering
\begin{subfigure}[b]{0.49\textwidth}
\centering
\includegraphics[width=\linewidth]{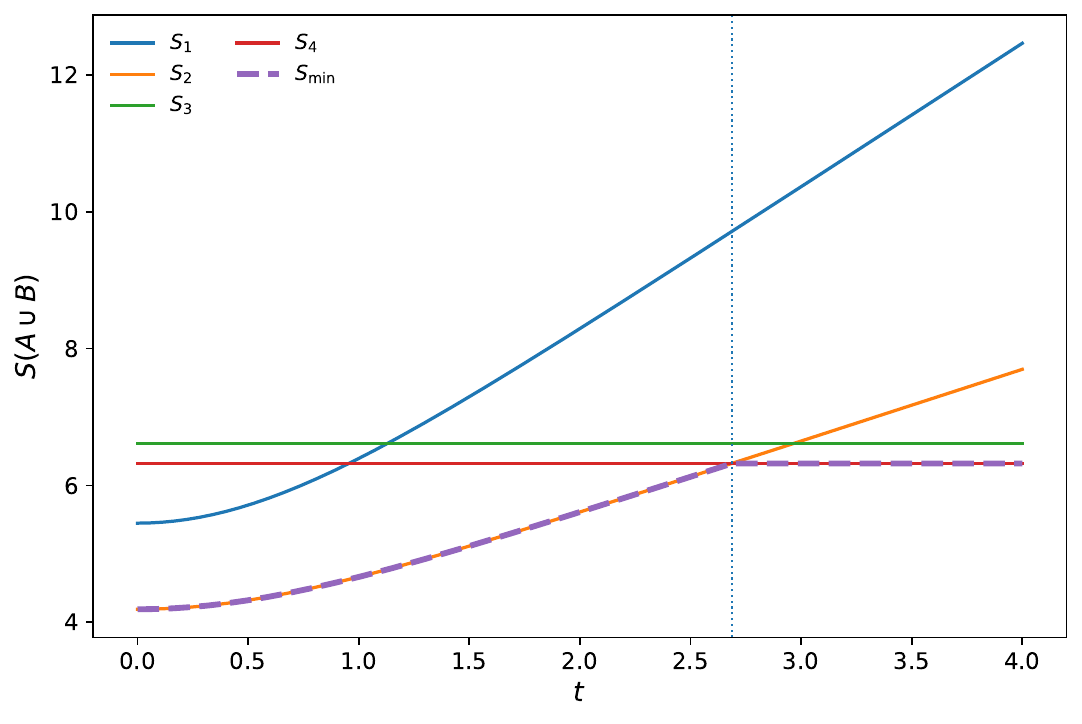}
\caption{Competing union-entropy saddles and their lower envelope.}
\end{subfigure}
\hfill
\begin{subfigure}[b]{0.49\textwidth}
\centering
\includegraphics[width=\linewidth]{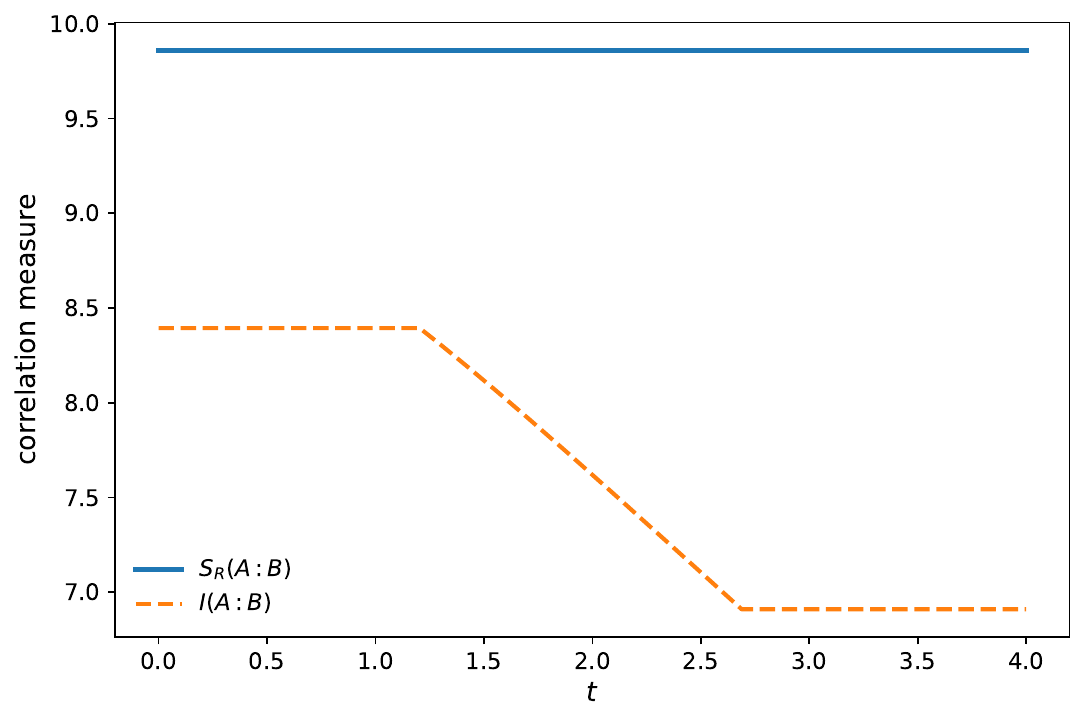}
\caption{Physical reflected entropy $S_R$ and mutual information $I$.}
\end{subfigure}
\par\medskip
\begin{subfigure}[b]{0.62\textwidth}
\centering
\includegraphics[width=\linewidth]{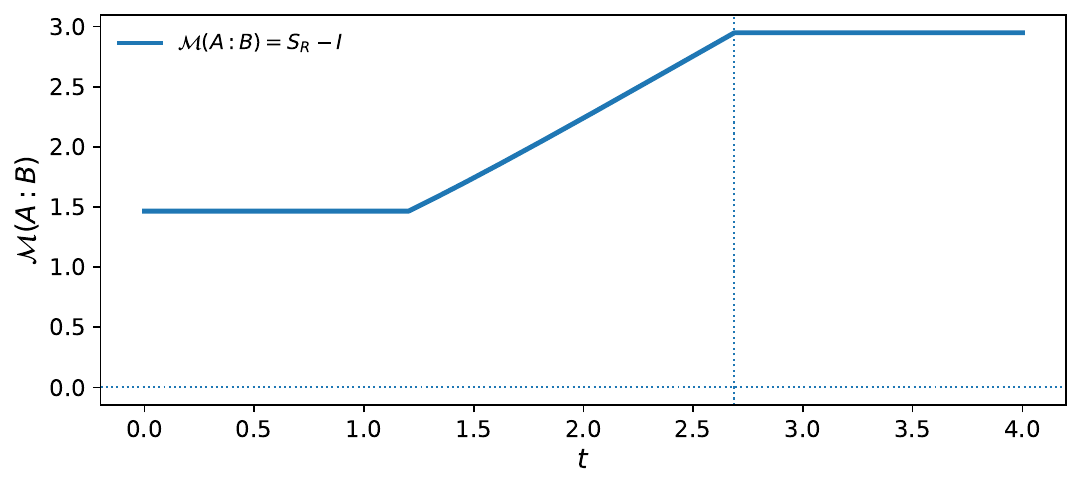}
\caption{Markov gap $\mathcal M=S_R-I$, shown on a separate axis to avoid obscuring the reflected-entropy curve.}
\end{subfigure}
\caption{Disjoint intervals: transition from the remaining outer HM bridge to an island-supported pair, with the central dome pair unchanged.  Panels (a) and (b) show the union entropy and the pair $S_R,I$, respectively, while panel (c) isolates $\mathcal M$.  The persistence of a positive Markov gap after the wormhole branch disappears is consistent with the island and the central complement continuing to participate in the non-Markovian/multipartite organization of the purification.}
\label{fig:dis-page-bridge-island}
\end{figure}
The island takeover displayed in \cref{fig:dis-page-bridge-island} shows that the late-time island phase preserves a nonzero mixed-state correlation between the separated intervals even after the wormhole-supported channel is no longer dominant.  More specifically, the nonzero post-transition gap shows that exact Markov recovery remains obstructed.  Geometrically, the central separation region and the island form a natural purifying sector, so the endpoint structure is sensitive to the multipartite organization of $A$, $B$ and their complement.

The disjoint wormhole-to-island transition replaces the remaining outer HM bridge in $S_2$ by the corresponding island pair in $S_4$.  Equating $H(L_4,R_4;t)$ with $J(L_4,R_4)$ gives
\begin{align}\label{t-dis-2-4}
 t_{E_{2\to4}}=\frac{\ell^2}{2r_h}\operatorname{arccosh}\!\left[\frac{2\ell^2\mathcal B_0^2}{\epsilon_y^2}\sinh\frac{r_hL_4}{\ell^2}\sinh\frac{r_hR_4}{\ell^2}\left(1+\nu\mathcal Q_4\right)-\mathcal C_4\right].
\end{align}
Equation~\eqref{t-dis-2-4} replaces the last HM bridge by an island pair while the central domes remain fixed.  The nonzero post-transition Markov gap in \cref{fig:dis-page-bridge-island} is therefore naturally interpreted as a diagnostic of the resulting non-Markovian organization involving the island and central complement, rather than as evidence for a surviving wormhole bridge.

\subsubsection*{Sequential loss of one HM bridge}

\begin{figure}[ht]
\centering
\begin{subfigure}[b]{0.49\textwidth}
\centering
\includegraphics[width=\linewidth]{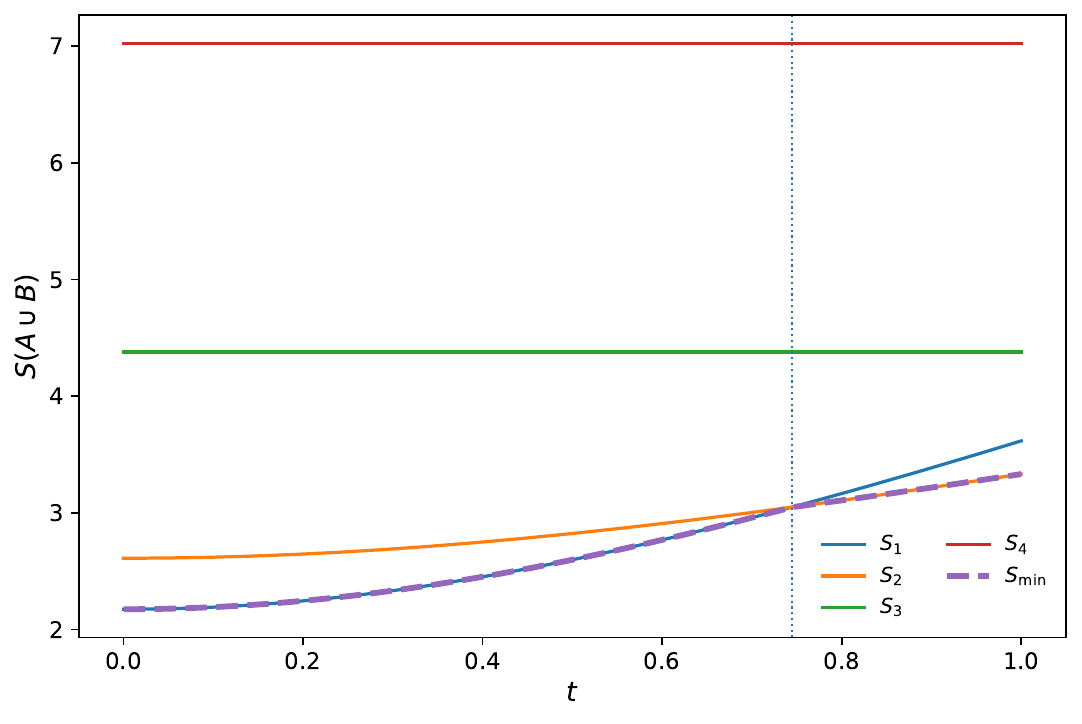}
\caption{Competing union-entropy saddles and their lower envelope.}
\end{subfigure}
\hfill
\begin{subfigure}[b]{0.49\textwidth}
\centering
\includegraphics[width=\linewidth]{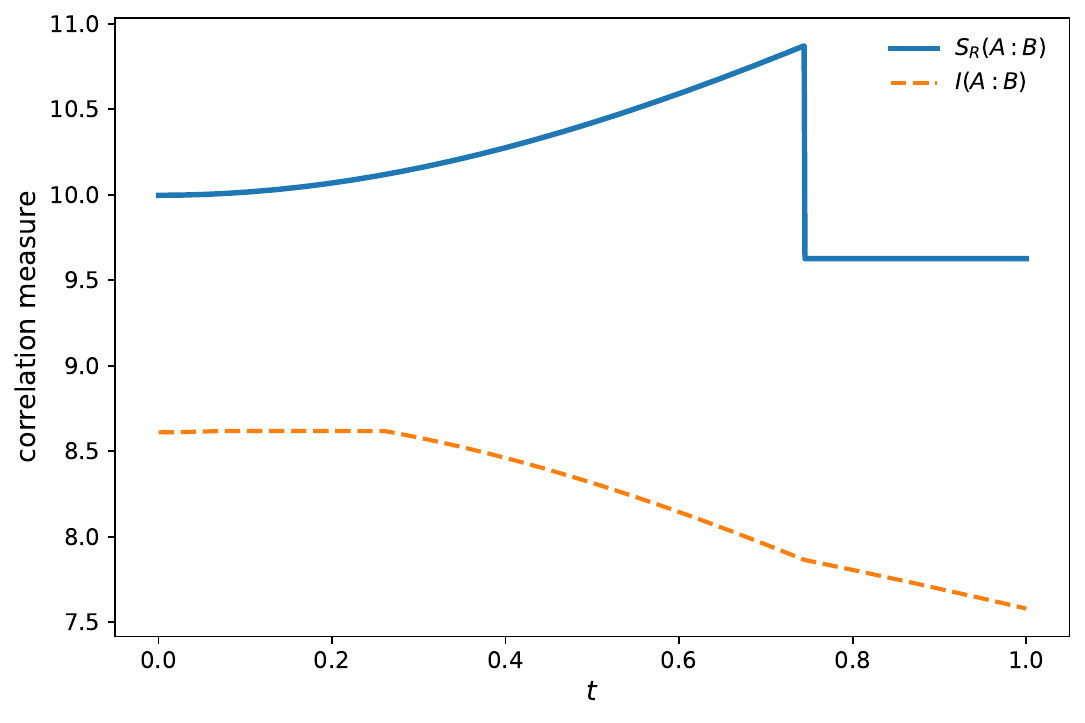}
\caption{Physical reflected entropy $S_R$ and mutual information $I$.}
\end{subfigure}
\par\medskip
\begin{subfigure}[b]{0.62\textwidth}
\centering
\includegraphics[width=\linewidth]{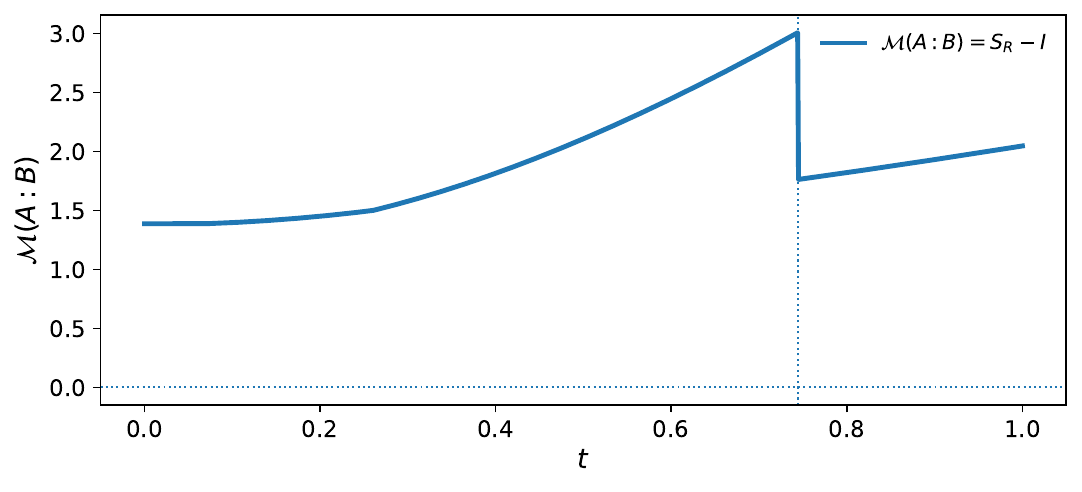}
\caption{Markov gap $\mathcal M=S_R-I$, shown on a separate axis to avoid obscuring the reflected-entropy curve.}
\end{subfigure}
\caption{Disjoint intervals: sequential removal of one of two outer HM bridges.  The central dome pair survives through the transition.  Panel (a) shows the outer-saddle competition, panel (b) shows the physical reflected entropy and mutual information, and panel (c) displays the Markov gap.  The two-stage behavior of $\mathcal M$ makes the asymmetric reorganization of the non-Markovian/multipartite diagnostic more apparent than in the union entropy alone.}
\label{fig:dis-page-sequential}
\end{figure}
The two-stage evolution in \cref{fig:dis-page-sequential} makes the geometrical asymmetry of the intermediate wedge explicit.  The reflected entropy is sensitive to which of the two wormhole-crossing components survives, while panel (c) shows that the obstruction to Markov recovery changes in the same staged fashion, consistently with the changing participation of the purifying sector.  This provides an example in which the diagnostic of tripartite/non-Markovian organization evolves nontrivially even though the central geometric purifier is unchanged.

The first transition in the disjoint sequential wormhole-to-island sequence again comes from the elementary degeneracy $H(L_1,R_1;t)=J(L_1,R_1)$, because the common outer HM contribution and the central domes cancel between $S_1$ and $S_2$.  Thus
\begin{align}\label{t-dis-1-2}
 t_{E_{1\to2}}=\frac{\ell^2}{2r_h}\operatorname{arccosh}\!\left[\frac{2\ell^2\mathcal B_0^2}{\epsilon_y^2}\sinh\frac{r_hL_1}{\ell^2}\sinh\frac{r_hR_1}{\ell^2}\left(1+\nu\mathcal Q_1\right)-\mathcal C_1\right].
\end{align}
The first crossing \eqref{t-dis-1-2} produces the asymmetric connected wedge in which only one outer pair remains HM supported.  The later internal change of the reflected-entropy saddle accounts for the staged behavior in \cref{fig:dis-page-sequential}.

\subsubsection*{Direct transition from two HM bridges to dome saddles}
\begin{figure}[ht]
\centering
\begin{subfigure}[b]{0.49\textwidth}
\centering
\includegraphics[width=\linewidth]{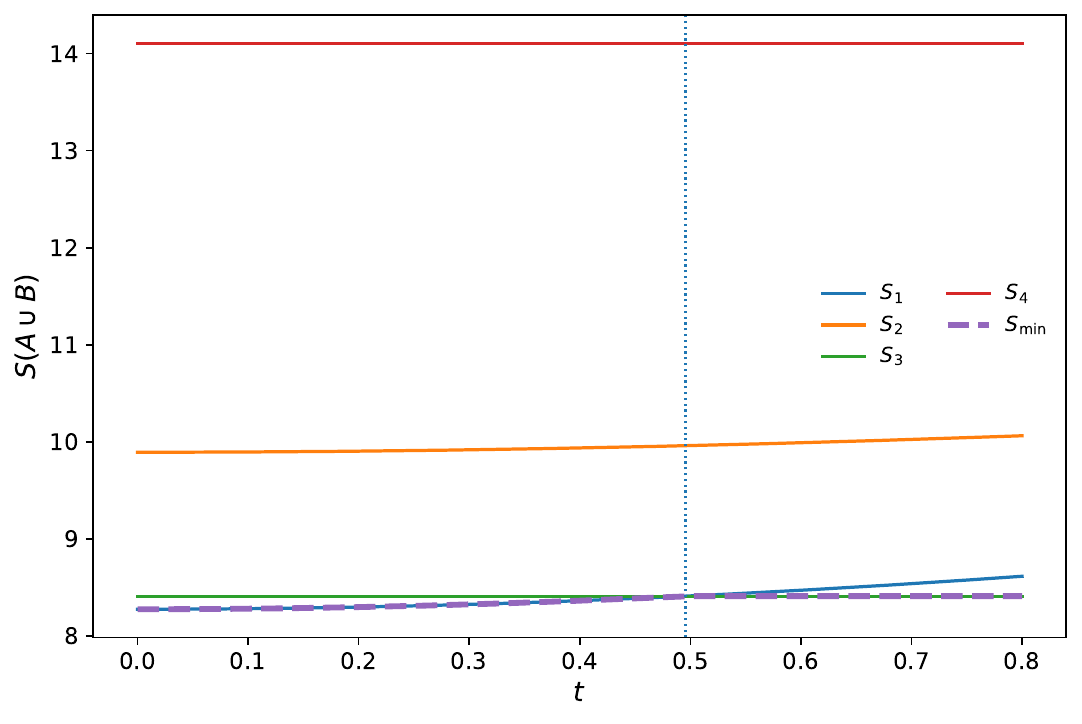}
\caption{Competing union-entropy saddles and their lower envelope.}
\end{subfigure}
\hfill
\begin{subfigure}[b]{0.49\textwidth}
\centering
\includegraphics[width=\linewidth]{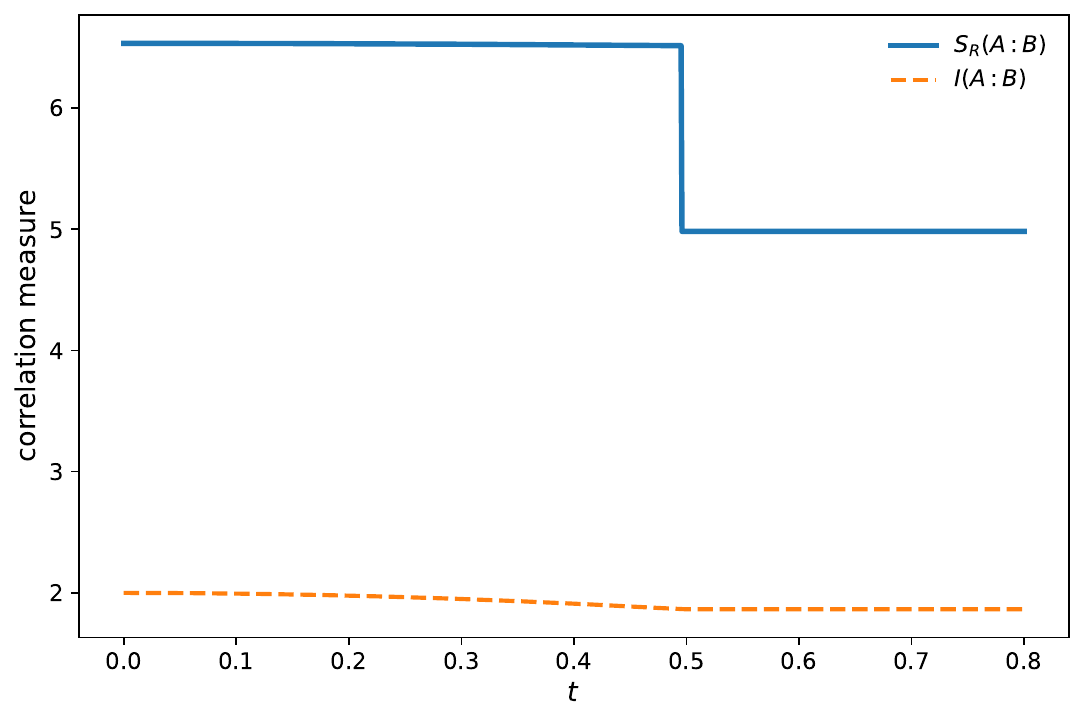}
\caption{Physical reflected entropy $S_R$ and mutual information $I$.}
\end{subfigure}
\par\medskip
\begin{subfigure}[b]{0.62\textwidth}
\centering
\includegraphics[width=\linewidth]{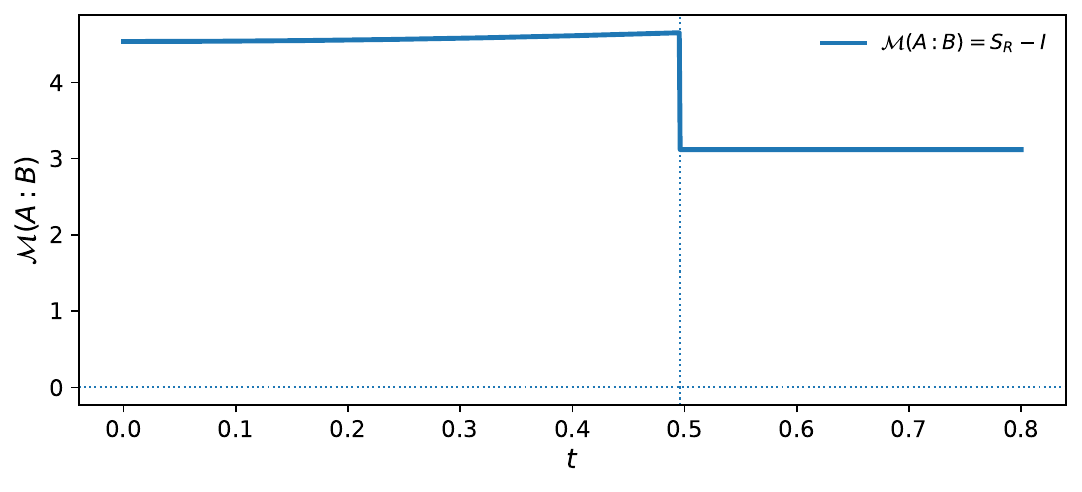}
\caption{Markov gap $\mathcal M=S_R-I$, shown on a separate axis to avoid obscuring the reflected-entropy curve.}
\end{subfigure}
\caption{Disjoint intervals: direct replacement of two outer HM bridges by a fully same-boundary dome configuration.  Panel (a) displays the entropy crossing, panel (b) shows the unobstructed $S_R$ and $I$ curves, and panel (c) isolates the Markov gap.  The sharp change in $\mathcal M$ reflects the disappearance of both wormhole-supported channels, whereas the residual late-time gap is tied to the remaining geometric endpoints and to correlations with the central purifying region.}
\label{fig:dis-page-direct-dome}
\end{figure}
The direct rearrangement in \cref{fig:dis-page-direct-dome} illustrates the most abrupt loss of wormhole-supported purification in the disjoint setup; after the crossing, both the reflected entropy and the Markov gap are governed by static near-boundary data.  Because the central separation region remains part of the purifier, the late-time positive value of $\mathcal M$ is naturally interpreted as a residual obstruction to a purely pairwise/Markov organization involving $A$, $B$ and that complement, rather than as a remnant of the black-hole bridge.  This distinction is precisely why the Markov gap carries information beyond the ordinary Page curve.

The direct disjoint simultaneous wormhole-to-dome transition is obtained by equating the two-wormhole-plus-central-dome saddle $S_1$ with the four-dome saddle $S_3$.  After the common central-dome contribution is cancelled, the resulting quadratic equation for $\cosh(2r_ht/\ell^2)$ gives
\begin{align}\label{t-dis-1-3}
 t_{E_{1\to3}}=\frac{\ell^2}{2r_h}\operatorname{arccosh}\!\left[\frac12\left(\sqrt{(\mathcal C_1-\mathcal C_4)^2+16\sinh^2\frac{r_hL_{41}}{2\ell^2}\sinh^2\frac{r_hR_{41}}{2\ell^2}}-\mathcal C_1-\mathcal C_4\right)\right].
\end{align}
The direct crossing \eqref{t-dis-1-3} removes both HM pieces at once, after which the union and reflected-entropy branches are supported entirely by static same-boundary geometry, explaining the rapid late-time saturation in \cref{fig:dis-page-direct-dome}.

    \section{Summary and discussion}\label{sec:summary}
    In this work we have studied reflected entropy in the partial-dimensional-reduction realization of JT gravity, retaining both connected and disconnected entanglement-wedge phases of the island/defect-extremal-surface construction rather than restricting attention to a single symmetric configuration.  The central advantage of this framework is that the two descriptions are related geometrically: in the lower-dimensional theory the generalized entropy contains a dilaton evaluated at a quantum extremal endpoint, whereas in the AdS$_3$ description the same correction is the first-order change of a geodesic ending on a fluctuating EOW brane.  This makes reflected entropy a particularly sharp mixed-state test of the partial-reduction dictionary \cite{Deng:2020ent,Deng:2022yll,Li:2021dmf,Ling:2021vxe}.

    At zero temperature we considered adjacent and disjoint intervals in the bath.  For adjacent intervals the no-island entanglement saddle contains the familiar boundary-to-RT cross section and reproduces the large-$c$ three-point twist correlator.  In the island-supported entanglement saddle there are three inequivalent reflected-entropy saddles: the EWCS can end on the island surface associated with $A$, can reach the EOW brane and split the reflected island, or can end on the island surface associated with $B$.  We retained the direct geodesic extremizations for each topology.  For the island-surface saddle, the relevant invariant is the distance from the junction point $b_2$ to the supporting island geodesic, proportional to $(b_2-x_c)^2-r^2$.  This geometric projection fixes the sign of the first-order response and agrees with the twist-field expansion.  For disjoint intervals the connected dome--dome phase is naturally written in terms of the conformal cross ratio, while the island phases involve dome--island and dome--brane common perpendiculars.  We distinguish the existence of these common perpendiculars from dominance of the connected outer wedge: in the vacuum no-island channel the RT transition occurs at $X=1/2$, and in any disconnected-wedge regime the leading holographic reflected entropy vanishes.  The mirror phase in which the $B$-island controls the cross section follows from the endpoint exchange symmetry and has been written explicitly.

    The finite-temperature analysis was performed directly in the eternal BTZ geometry.  The elementary functions $D$, $H$ and $J$ in \cref{thermal-D,thermal-H,thermal-J} keep the same-boundary, wormhole-crossing and island-supported contributions distinct, including the thermal-cylinder half-arguments and the HM normalization.  Adjacent and disjoint union entropies are obtained by the corresponding lower envelopes; for disjoint regions this minimization also includes the factorized competitor \cref{thermal-disconnected-union}.  This avoids repeating the full saddle catalogue here and makes explicit that connected reflected-entropy formulas are conditional on the dominance of their outer entanglement wedge.

    A technically important point concerns asymmetric HM surfaces.  The constant-$x$ profile often used in symmetric TFD calculations is valid only when the left and right spatial endpoints coincide.  Applying it when $L_i\neq R_i$ produces an incorrect sign or normalization in the boundary-to-wormhole common perpendicular.  Using the general embedding-space parametrization instead gives the invariant boundary-to-HM distance \cref{point-to-HM-invariant} and gives exact agreement in the rigid-brane sector.  The same invariant method can be used for the dome--HM and dome--dome common perpendiculars appearing in the disjoint phases.  This invariant construction is independent of the CFT block representation and fixes the geometric endpoint assignments for the corresponding dominant channels \cite{Hartman:2013qma,Dutta:2019gen,Basu:2023jtf,Dey:2025dlj}.

    The JT correction was kept to first order in the transverse brane fluctuation.  At finite brane angle the minimal QES and DES functionals agree in value and first variation at the rigid saddle but differ in their off-shell curvature, with coefficient $q_\theta=e^{-2\rho_0/\ell}$ in the vacuum and the analogous thermal expression in \cref{thermal-finite-angle-offshell}.  This off-shell difference is quadratic around the rigid extremum and therefore first affects the on-shell entropy at $O(q_\theta\Phi_r^2)$.

    The first-order endpoint comparison requires an additional distinction.  The raw AdS$_2$-slicing labels give different QES and DES shifts, but the slicing label on a bent brane is not automatically the dimensionally reduced coordinate.  The background projection $Y=-x+e^{-\rho_0/\ell}z$ produces the universal shift \cref{vacuum-projection-shift} and restores the QES/DES endpoint equality through $O(\Phi_r)$ at arbitrary finite angle, with the same mechanism for the disjoint endpoints.  In the thermal geometry the uniformizing-coordinate prescription \cref{thermal-projection-F,thermal-projection-shift} gives the corresponding result.  We interpret these calculations as evidence for a natural projection-covariant, or St\"uckelberg-completed, endpoint dictionary at linear order; a derivation of its nonlinear completion from the reduced action remains open.  The large-angle limit suppresses the genuine higher-order finite-angle corrections, but is not established here as a necessary condition for the leading projected endpoint location.

    The transition times between the four entanglement saddles follow directly from the same building blocks.  The elementary HM-to-island transition is obtained by solving $H(L_i,R_i;t)=J(L_i,R_i)$, which gives \cref{t-adj-2-4,t-adj-1-2} and their disjoint analogues.  Transitions in which a single wormhole-crossing contribution is replaced by a same-boundary dome follow by equating the appropriate mixed island--wormhole saddle with the corresponding dome saddle, while the simultaneous replacement of two wormhole-crossing contributions reduces to a quadratic equation for $\cosh(2r_ht/\ell^2)$.  The representative Page curves are generated directly from these analytic saddle expressions.  For each representative choice of endpoints, the physical curve is the lower envelope of the allowed saddles.  The companion reflected-entropy plots perform the analogous minimization over the EWCS topologies.

    In addition to $S_R$, the plots display the Markov gap $\mathcal M=S_R-I$ in a dedicated third panel.  Mutual information is evaluated from the same RT/HRT/island minimization used for $S(A)$, $S(B)$ and $S(A\cup B)$, so the Markov-gap curve never mixes inconsistent saddles.  For the parameter sets in \cref{app:numerics}, $\mathcal M\ge0$ throughout the displayed time intervals without positivity clipping.  Kinks in $\mathcal M$ occur either when the entanglement saddle entering $I$ changes or when the minimal reflected-entropy cross section changes.  This separates a Page transition in the total correlation from a rearrangement of the non-Markovian/multipartite organization of the purification.  In disjoint configurations the central gap belongs to the purifying complement, so its geometry participates directly in this tripartite interpretation.  The interpretation is consistent with the geometric Markov-gap analysis and its extension to islands \cite{Hayden:2021gno,Lu:2022cgq,Basu:2025sqk}.

    Several consistency limits are worth emphasizing.  First, $\Phi_r\to0$ removes the dynamical JT correction and reduces the formulas to the corresponding rigid-brane/DES results.  Second, the $\beta\to\infty$ limit of the same-boundary thermal chords, the QES/DES endpoint shifts and the finite-angle functional difference reproduces the corresponding vacuum formulas after matching the regulators; the two-sided HM branch instead has its own zero-temperature degeneration and is not identified with a same-boundary vacuum chord.  Third, setting $L_i=R_i$ reduces the general HM invariants to the familiar constant-$x$ expressions.  Fourth, exchanging $A\leftrightarrow B$ maps the pairs of island-surface phases into one another, as required by the endpoint-exchange symmetry of the higher-point block channels.  Finally, the Brown--Henneaux relation converts the CFT coefficient $c/3$ precisely into the geometric factor $\ell/(2G_N)$ for reflected entropy.

    The present calculation remains semiclassical and perturbative in the transverse brane fluctuation.  It would be interesting to go beyond first order, where the shift of the extremal cross section begins to contribute explicitly and the simple identity \cref{first-order-extremization} is no longer sufficient.  The finite-angle analysis suggests a more specific extension. At linear order, the background projection gives a natural St\"uckelberg-completed endpoint identification and removes the apparent QES/DES displacement in the projected coordinate. Beyond linear order, however, the off-shell differences \cref{finite-angle-offshell,thermal-finite-angle-offshell} remain. Deriving the projection map and the accompanying higher-order terms directly from the reduced action would determine whether they arise from finite-cutoff, nonlocal or additional St\"uckelberg operators generated when the transverse AdS wedge is integrated out. The large-angle limit suppresses these effects, but the linear analysis does not establish it as a prerequisite for finite-angle endpoint matching.  A further direction is to derive the reflected-island formula directly from the gravitational replica path integral in the partially reduced theory, including the backreaction of the reflected-entropy cosmic branes.  Other natural extensions include finite-interval baths, higher-dimensional braneworlds, covariant time-dependent partial reduction, symmetry-resolved mixed-state measures, and multipartite generalizations.  The detailed phase catalogue obtained here should provide a useful benchmark for those developments and for future comparisons between lower-dimensional quantum extremal cross sections and their higher-dimensional defect counterparts \cite{Dutta:2019gen,Akers:2022max,Basu:2024xjq,Basu:2025sqk,Dey:2025dlj}.

    \appendix

\section{Numerical implementation and representative Page curves}\label{app:numerics}
The Page-curve plots in the Page-curve subsections of section~\ref{sec:thermal-reflected} are intended to illustrate the different geometric replacement patterns allowed by the analytic saddle formulas rather than to single out a preferred microscopic parameter choice.  For each data set we evaluate every admissible entropy saddle at the same boundary time, take the lower envelope for $S(A\cup B)$, and independently minimize the saddles entering $S(A)$ and $S(B)$.  Only after this entropy minimization do we compare the reflected-entropy cross sections that are compatible with the dominant entanglement wedge.  This ordering is important: minimizing $S_R$ over a cross section that belongs to a non-dominant outer wedge would mix distinct semiclassical saddles and can spuriously violate $S_R\ge I$.

The third panel in every Page-curve figure shows the Markov gap.  With the mutual information evaluated from independently minimized entropies, the quantity plotted is
\begin{equation}
 \mathcal M(A:B)=S_R(A:B)-I(A:B),\qquad
 I(A:B)=S(A)+S(B)-S(A\cup B).
\end{equation}
No positivity prescription or clipping is imposed on $\mathcal M$; the parameter sets are retained only when the independently minimized saddles give $\mathcal M\ge0$ over the displayed range.  A cusp in the Markov-gap panel can have two distinct origins.  It can coincide with a Page transition of the outer entanglement wedge, or it can arise from an internal change of the minimal reflected-entropy cross section while the outer wedge remains unchanged.  The latter possibility is one of the main reasons for displaying the Markov gap together with $S_R$.

The eight representative data sets used in the Page-curve figures realize four transition patterns for adjacent intervals and the corresponding four patterns for disjoint intervals. We set $\ell=1$ and $G_N=1/2$ only for plotting; all analytic formulas in the main text retain the full dimensionful parameters. Perturbative control must be imposed on the local brane displacement, not merely on $\nu$. At a thermal brane endpoint,
\begin{align}
\frac{|\Delta\rho(a)|}{\rho_0}=\frac{\ell\nu}{\rho_0}\coth(\kappa a),\qquad \kappa=r_h/\ell^2.
\label{local-fluctuation-control}
\end{align}
Since the physical island endpoints move outward for positive $\nu$, evaluating \cref{local-fluctuation-control} at the smallest relevant undeformed endpoint gives a conservative first-order bound. Across the eight plotted examples the maximal local fluctuation lies between approximately $0.023$ and $0.075$ and is always smaller than $0.076$. The values of $\rho_0$ and $\epsilon_y$ were chosen along curves of fixed
\begin{align}
\mathcal C_Q\equiv\frac{\rho_0}{\ell}+\log\frac{2\ell\cosh(\rho_0/\ell)}{\epsilon_y},
\end{align}
so the first-order island building block $J$, and likewise the first-order reflected-entropy combinations entering the plots, are unchanged. Indeed, using
$\sec\theta_0+\tan\theta_0=e^{\rho_0/\ell}$ and
$B_0=2\sec\theta_0(\sec\theta_0+\tan\theta_0)$, one has
$\mathcal C_Q=\log(\ell B_0/\epsilon_y)$. Thus the combinations $B_0/\epsilon_y$ that enter the thermal transition formulas are fixed as well. At the order retained here, the island-surface branches depend on the fluctuation through $\nu$ and the boundary endpoints, while the brane-ending branches contain $(\rho_0,\epsilon_y)$ through this same fixed combination. Consequently the displayed curves are unchanged to first order under this perturbative-control reparametrization.
For all eight examples the lower envelope realizes the intended transition sequence, remains inside the connected-wedge domain whenever a connected reflected-entropy formula is used, and the Markov gap remains non-negative over the displayed time interval without clipping.  The numerical choices are representative geometric laboratories rather than fine tuning: their purpose is to separate the allowed replacement patterns cleanly enough that the origin of every kink can be identified.  The four adjacent examples isolate, respectively, the replacement of a single wormhole bridge by a same-boundary dome, replacement by an island pair, sequential loss of one of two bridges, and the simultaneous loss of both bridges.  The disjoint examples realize the same four outer transitions while retaining the central dome pair associated with the finite separation between $A$ and $B$.  Comparing these two sets makes clear which features of the curves come from the Page transition of the union and which are caused specifically by the additional internal geometry of a disjoint entanglement wedge.

\section{Geodesic invariants in Poincar\'e AdS$_3$ and BTZ}\label{app:embedding}
Many of the reflected-entropy saddles in the main text are common-perpendicular problems between geodesics.  Embedding-space methods are useful because they separate the invariant content of these problems from coordinate choices and, in particular, prevent the symmetric HM parametrization from being used outside its domain of validity.  We collect here the formulas used repeatedly in \cref{sec:vacuum-reflected,sec:BTZ-review,sec:thermal-reflected} and explain their geometric meaning.

AdS$_3$ is the one-sheeted hyperboloid $X\cdot X=-\ell^2$ embedded in $\mathbb R^{2,2}$ with ambient metric $\eta_{AB}=\operatorname{diag}(-1,-1,1,1)$.  We use $X\cdot Y\equiv\eta_{AB}X^AY^B$, so the ambient line element and the hyperboloid constraint are
\begin{align}
\d s_{\mathbb R^{2,2}}^2=\eta_{AB}\d X^A\d X^B,
\qquad X\cdot X=-\ell^2.
\end{align}
Differentiating the constraint twice shows that the acceleration of an affinely parametrized AdS geodesic is normal to the hyperboloid.  For the dimensionless affine parameter $\lambda$ used below, the embedding-space geodesic equation is
\begin{align}
\ddot X^A=\frac{\dot X^2}{\ell^2}X^A.
\end{align}
For a spacelike geodesic we choose a point $P^A$ on the hyperboloid and a unit tangent $V^A$ satisfying $P\cdot V=0$.  The solution then takes the boost-like form
\begin{align}
X^A(\lambda)=P^A\cosh\lambda+\ell V^A\sinh\lambda,
\qquad P^2=-\ell^2,
\quad V^2=1,
\quad P\cdot V=0.
\label{embedding-geodesic-hyperbolic}
\end{align}
The hyperbolic dependence is simply the statement that translation along a spacelike AdS geodesic is a Lorentz boost in the two-dimensional plane spanned by $P$ and $V$.  Since $\lambda$ is dimensionless, proper length is $\d s=\ell\,\d\lambda$.  Consequently the geodesic distance between two spacelike-separated points $X_a$ and $X_b$ is determined completely by their ambient inner product,
\begin{align}
\mathcal L_{ab}=\ell\,\Delta\lambda
=\ell\,\operatorname{arccosh}\left(-\frac{X_a\cdot X_b}{\ell^2}\right).
\label{embedding-geod}
\end{align}
This invariant formula is the starting point for all of the coordinate expressions below.  For reflected entropy, its practical advantage is especially important: a common perpendicular is found by varying invariant distances along the supporting RT/HRT surfaces, so orthogonality can be imposed without committing to a potentially misleading coordinate profile.

\subsection*{Poincar\'e AdS$_3$}
For the Poincar\'e coordinates used in the vacuum analysis, a convenient embedding is
\begin{align}
X^0&=\frac{z}{2}+\frac{1}{2z}\left(\ell^2+x^2-t^2\right),\notag\\
X^1&=\frac{\ell t}{z},\notag\\
X^2&=\frac{\ell x}{z},\notag\\
X^3&=\frac{z}{2}-\frac{1}{2z}\left(\ell^2-x^2+t^2\right).
\end{align}
Substituting two such points into \cref{embedding-geod} gives the bulk-to-bulk Poincar\'e distance used throughout \cref{sec:vacuum-reflected},
\begin{align}
\mathcal L_{ab}=\ell\,\operatorname{arccosh}\left[
\frac{-(t_a-t_b)^2+(x_a-x_b)^2+z_a^2+z_b^2}{2z_az_b}
\right].
\label{Poincare-length}
\end{align}
At equal time this reduces to the standard distance in the hyperbolic plane.  The vacuum calculations in \cref{sec:vacuum-reflected} exploit this fact repeatedly: RT and island surfaces are semicircles, and extremizing a distance to one of them is simply the construction of a hyperbolic common perpendicular.  Sending either endpoint to $z=\epsilon$ produces the familiar logarithmic UV divergence; the finite reflected-entropy quantity is obtained after the corresponding replica or RT/HRT normalization is included.

\subsection*{Planar BTZ black hole}
For the planar BTZ exterior, the coordinates $(t,x,r)$ are embedded as
\begin{align}
X^0&=\ell\frac{\sqrt{r^2-r_h^2}}{r_h}\sinh\frac{r_ht}{\ell^2},\notag\\
X^1&=\ell\frac{r}{r_h}\cosh\frac{r_h x}{\ell^2},\notag\\
X^2&=\ell\frac{r}{r_h}\sinh\frac{r_h x}{\ell^2},\notag\\
X^3&=\ell\frac{\sqrt{r^2-r_h^2}}{r_h}\cosh\frac{r_ht}{\ell^2}.
\end{align}
The invariant distance between arbitrary exterior points then becomes
{\small\begin{align}
\mathcal L_{ab}=\ell\,\operatorname{arccosh}\left[
\frac{r_a r_b}{r_h^2}\cosh\frac{r_h(x_a-x_b)}{\ell^2}
-\frac{\sqrt{(r_a^2-r_h^2)(r_b^2-r_h^2)}}{r_h^2}
\cosh\frac{r_h(t_a-t_b)}{\ell^2}
\right].
\label{BTZ-length}
\end{align}}
The relative minus sign between the spatial and temporal invariants is important.  On an equal-time slice it gives the common-perpendicular formulas used for dome and island surfaces.  After analytic continuation by half the Euclidean thermal circle to the second TFD boundary, the temporal term changes sign and produces the sum of hyperbolic cosines characteristic of a HM geodesic.  This single invariant therefore unifies the apparently different thermal geometries used throughout \cref{sec:thermal-reflected}.

\subsection*{Island geodesics in BTZ}
A static island surface is an equal-time BTZ geodesic that begins at a regulated bath endpoint and terminates on the EOW brane.  It is convenient to describe it by a center $x_c$ and a dimensionless parameter $m$,
\begin{align}
r(x)=\frac{r_h}{\sqrt{1-m^2\cosh^2\!\left[\frac{r_h(x-x_c)}{\ell^2}\right]}}.
\label{island-profile-appendix}
\end{align}
The denominator measures the radial depth of the geodesic: it vanishes at an asymptotic boundary endpoint and is finite at the brane.  For the island surface anchored at $L_1$, the constants $(x_{c1},m_1)$ are fixed by imposing these two endpoint conditions.  In the boundary limit and at the brane point $(x_b,r_b)$ they read
\begin{align}\label{center_radius_island}
m_1\cosh\!\left[\frac{r_h(L_1-x_{c1})}{\ell^2}\right]&=1,\\
m_1\cosh\!\left[\frac{r_h(x_b-x_{c1})}{\ell^2}\right]&=\sqrt{1-\frac{r_h^2}{r_b^2}},
\qquad (x_b,r_b)\ \text{given by \cref{EOW-point}}.\notag
\end{align}
These two endpoint conditions determine the physical island branch uniquely.  More importantly for the calculations in \cref{sec:thermal-reflected}, they separate the geometric data into the center $x_c$ and the depth parameter $m$.  A dome--island or boundary--island common perpendicular can then be characterized by $(x_c,m)$ and an ordinary orthogonality condition, avoiding the exponentially weighted affine coordinates that obscure both the zero-temperature limit and the effect of a small brane displacement.

\subsection*{HM geodesics}
The wormhole-crossing HM surface is most cleanly described by boundary null rays rather than by pretending that the regulated cutoff points are null embedding vectors.  Introduce
\begin{align}
\tau_0\equiv\frac{r_ht_0}{\ell^2},\qquad
u_L\equiv\frac{r_hx_L}{\ell^2},\qquad
u_R\equiv\frac{r_hx_R}{\ell^2},
\end{align}
and choose the dimensionless null representatives
\begin{align}
p_L&=(\sinh\tau_0,\,\cosh u_L,\,\sinh u_L,\,\cosh\tau_0),\notag\\
p_R&=(\sinh\tau_0,\,\cosh u_R,\,\sinh u_R,\,-\cosh\tau_0).
\label{HM-null-rays}
\end{align}
The second line incorporates the half-thermal-circle continuation to the opposite TFD boundary.  With the ambient metric used above, $p_L^2=p_R^2=0$ and
\begin{align}
\mathcal Z_{LR}\equiv-p_L\!\cdot p_R
=\cosh(2\tau_0)+\cosh(u_L-u_R)>0.
\label{HM-null-invariant}
\end{align}
The exact unregulated geodesic on the AdS$_3$ hyperboloid is then
\begin{align}
X_{LR}(\lambda)=\frac{\ell}{\sqrt{2\mathcal Z_{LR}}}
\left[(p_L+p_R)\cosh\lambda+(p_L-p_R)\sinh\lambda\right].
\label{A8}
\end{align}
Because the inputs in \cref{A8} are null boundary representatives, $X_{LR}(\lambda)^2=-\ell^2$ identically.  The limit $\lambda\to+\infty$ approaches the ray $p_L$, whereas $\lambda\to-\infty$ approaches $p_R$.  A radial regulator $r_c=\ell^2/\epsilon$ simply truncates this infinite geodesic at the corresponding large values of $|\lambda|$; it does not replace $p_L,p_R$ by regulated bulk endpoints.

The regulated HM length follows by matching the asymptotic coefficient of the null rays to $r_c$.  To leading order in the cutoff,
\begin{align}
\mathcal L_{\rm HM}(x_L,x_R;t_0)
=\ell\log\!\left[2\left(\frac{\ell^2}{r_h\epsilon}\right)^2
\left(\cosh\frac{2r_ht_0}{\ell^2}
+\cosh\frac{r_h(x_L-x_R)}{\ell^2}\right)\right]
+\mathcal O(\epsilon^2).
\label{HM-length}
\end{align}
This is the expression used in the main text.  The time-dependent term measures the wormhole growth, while the second term retains the spatial mismatch of asymmetric anchors.

The BTZ coordinates along the geodesic are reconstructed from
\begin{align}
\frac{X^2}{X^1}=\tanh\frac{r_hx}{\ell^2},\qquad
\frac{X^0}{X^3}=\tanh\frac{r_ht}{\ell^2},\qquad
(X^1)^2-(X^2)^2=\ell^2\frac{r^2}{r_h^2}.
\label{A10}
\end{align}
Substitution of \cref{A8} gives the fully asymmetric parametrization
\begin{align}\label{HM-parametrization}
\tanh\frac{r_h t}{\ell^2}&=\coth\lambda\,\tanh\tau_0,\notag\\
\tanh\frac{r_h x}{\ell^2}&=
\frac{\cosh\lambda(\sinh u_L+\sinh u_R)+\sinh\lambda(\sinh u_L-\sinh u_R)}
{\cosh\lambda(\cosh u_L+\cosh u_R)+\sinh\lambda(\cosh u_L-\cosh u_R)},\\
r^2&=r_h^2\frac{\cosh 2\lambda+\cosh(u_L-u_R)}
{\cosh(2\tau_0)+\cosh(u_L-u_R)}.
\label{HM-parametric}
\end{align}
The terms odd in $\lambda$ carry the left--right spatial asymmetry.  For $x_L=x_R=x_0$ they cancel and one recovers
\begin{align}
\sqrt{1-\frac{r_h^2}{r^2}}=
\frac{\sinh(r_ht_0/\ell^2)}{\sinh(r_ht/\ell^2)},
\qquad x=x_0.
\end{align}
Thus the constant-$x$ curve is a symmetric special case, not a general HM profile.  The boundary-null-ray convention in \cref{HM-null-rays,A8} and the regulated-bulk-point convention used in \cref{point-to-HM-invariant} are complementary but distinct: the former defines the supporting geodesic exactly, while the latter supplies the UV cutoff only when a boundary-to-geodesic distance is evaluated.

	\bibliographystyle{JHEP}
	\bibliography{main}
\end{document}